\documentclass{aa}  

\usepackage{nameref}
\usepackage{graphbox}
\usepackage{graphics}
\usepackage{caption}
\usepackage{multirow}
\usepackage{txfonts}
\usepackage[dvipsnames]{xcolor}
\usepackage{placeins}
\usepackage{longtable}
\usepackage{subfig}
\usepackage{multirow}
\usepackage[normalem]{ulem}
\usepackage{verbatim}
\usepackage{wrapfig}
\usepackage{graphicx}
\usepackage{txfonts}
\usepackage{lipsum}
\usepackage{float}

\usepackage{subcaption}      	
\usepackage{lscape}             	
\usepackage{placeins}           	
                                
\usepackage{hyperref}
\newcommand\T{\rule{0pt}{2.6ex}}       	
\newcommand\B{\rule[-1.2ex]{0pt}{0pt}} 	

\begin{document}

  \title{Infrared spectroscopy of gas-phase hydrogenated and methylated pyrenes: From laboratory spectra to the simulated 3.4~$\mu$m emission band}

\titlerunning{High-temperature IR spectrum of hydrogenated and methylated pyrene in the gaseous phase}

   \author{Karine Demyk\inst{1}
          \and Christine Joblin\inst{1}
          \and Louan de Bentzmann\inst{1}
          \and Dominique Toublanc\inst{2}
         \and Giacomo Mulas\inst{3}
          }

 \institute{Institut de Recherche en Astrophysique et Plan\'etologie (IRAP), Universit\'e de Toulouse, CNRS, CNES, 9 Avenue du Colonel Roche, F-31028 Toulouse, France \\ \email{karine.demyk@cnrs.fr, christine.joblin@cnrs.fr} 
 \and
Laboratoire Collisions Agrégats Réactivité (LCAR/FeRMI), Université de Toulouse, CNRS, 118 Route de Narbonne, F-31062 Toulouse, France 
 \and
Istituto Nazionale di Astrofisica (INAF), Osservatorio Astronomico di Cagliari, via della Scienza 5, 09047 Selargius (CA), Italy  \\%
}

    \date{Received 30 September 2026 / Accepted 07 October 2026}

 
  \abstract
   {Observations of the aromatic infrared (IR) emission band at 3.3~$\mu$m often reveal satellite emission features in the 3.4 -- 3.6~$\mu$m range. While the 3.3~$\mu$m band is attributed to the CH stretching vibration of polycylic aromatic hydrocarbons (PAHs), the satellite bands, particularly its prominent 3.4~$\mu$m component, is assigned to aliphatic CH stretching vibrations in hydrogenated and methylated PAH-like species. }
   {We derive state-of-the-art IR emission spectra for aliphatic-containing pyrene derivatives and compare them with astronomical observations. This will help us to refine our understanding of the contribution of these species to the 3.4~$\mu$m emission band.} 
   {Infrared  spectra (1.4--25~$\mu$m) of gas-phase hydrogenated pyrenes (dihydropyrene, tetrahydropyrene, and hexahydropyrene), methylated pyrene, and pyrene were recorded at temperatures ranging from 373 to 673~K, depending on the species. The band profiles were analysed using a multi-component fitting tool, and empirical anharmonicity laws were derived to quantify the evolution of the band positions and widths with temperature. The obtained spectral data were combined with the results of a Monte Carlo emission model to simulate the emission spectra following UV-photon absorption, up to the dissociation limit ($\lesssim$6~eV). The resulting synthetic spectra were then compared with astronomical spectra including James Webb Space Telescope observations of the Orion Bar region (PDRs4All program).} 
   {Based on these state-of-the-art simulated spectra, we show that 1,2,3,6,7,8-hexahydropyrene matches all the constrains to fit the red component of the 3.4 $\mu$m band observed at 3.403 $\mu$m. While 1-methylpyrene may also contribute to the underlying emission plateau, its lack of a strong IR band complicates detection in observed spectra, unlike hexahydropyrene. All experimental spectra and their temperature-dependent analyses are available in the new cosmicPAH-IRDB database of anharmonic IR spectra.}   
    {}

\maketitle
\nolinenumbers

\section{Introduction}

Aromatic infrared bands (AIBs) are emission features detected in a wide range of environments, particularly those linked to star formation and evolution \citep{peeters2002, li2020}. In these regions, ultraviolet (UV) radiation shapes photodissociation regions (PDRs), which play a crucial role in the interstellar medium of the Milky Way, and in nearby and early-type galaxies \citep{hollenbach1999}. In addition to the main AIBs at 3.3, 6.2, 7.7, 8.6, 11.2, and 12.7~$\mu$m, many other tightly correlated weaker bands exist up to $\sim$ 20~$\mu$m \citep[eg.,][]{ chown2024}. The leading candidates for AIB carriers are polycyclic aromatic hydrocarbons (PAHs), which are molecules that are composed of multiple aromatic rings that emit in the infrared (IR) as they radiatively cool after excitation by UV photons \citep{Sellgren1984, Leger1984, leger1989,allamandola1989}. The main AIBs are thought to originate from the vibrations of aromatic C-H and C-C bonds. As the emitting molecules are hot, anharmonic effects give rise to numerous combination bands and hot bands, which in turn affect the band positions and shapes  \citep{allamandola1989, joblin1995, pech2002}. The AIB spectrum is known to change depending on excitation conditions, particularly the UV radiation field. This variation is often attributed to changes in the chemical and physical properties of the PAH population, such as size, charge state, and composition \citep{joblin1996a, sloan1999, berne2007, shannon2016, peeters2017}. 
The James Webb Space Telescope (JWST) now observes AIBs at high angular and spectral resolution with unprecedented detail. These observations provide access to accurate band positions and shapes, including minor features and band shoulders, together with information on their spatial evolution, as demonstrated in the observations of the Orion Bar \citep{chown2024, peeters2024}.

In addition to the 3.3~$\mu$m band that arises from the CH stretching vibration of aromatic CH, several emission bands are observed in the 3.4 -- 3.6~$\mu$m range at 3.40, 3.46, 3.51 and 3.57~$\mu$m, on top of a broad plateau extending from 3.2 to 3.6~$\mu$m. The 3.4~$\mu$m/3.3~$\mu$m ratio is found to be anti-correlated with the intensity of the UV radiation field, showing that their carriers are easily photodissociated by UV photons \citep{joblin1996b,pilleri2015,schroetter2024}. Several carriers were proposed to explain the 3.4 -- 3.6~$\mu$m bands: hot bands of the CH stretching modes, although these alone cannot explain the observations \citep{geballe1994}; aliphatic CH stretching vibrations from methyl sidegroups (-CH$_3$) in methylated PAHs (Me-PAHs) \citep{joblin1996b}; and aliphatic CH stretching vibrations from methylene groups (-CH$_2$) in hydrogenated PAHs (H-PAHs) \citep{bernstein1996}. \\
The identification of the carriers of spectroscopic features observed in astronomical spectra relies on experimental and theoretical works that provide spectra of dust and molecules to compare with the observations. Some of this work has focused on the identification of the 3.4 -- 3.6~$\mu$m emission features. The IR spectra of various hydrogenated and methylated PAHs have been measured using multiple techniques: absorption spectroscopy at room temperature in the solid state \citep{steglich2013}, in argon matrix at low temperature \citep{bernstein1996, sandford2013}, and in the gas phase at high temperature \citep{joblin1996b}; using infrared resonance-enhanced multiphoton ionization (IR-REMPI) spectroscopy in jets \citep{maltseva2018}; and emission spectroscopy of gas-phase UV laser-excited PAHs \citep{wagner2000}. Density functional theory (DFT) calculations have complemented these studies with a focus on the 3.4~$\mu$m/3.3~$\mu$m band ratio for methyl-substituted PAHs \citep{pauzat1999,yang2013} and superhydrogenated PAHs \citep{pauzat2001,yang2020}. All these experimental and theoretical studies focused on small PAHs up to the size of coronene (24 carbon atoms).\\
Because the emitting PAHs are hot, anharmonic effects should be considered in generating synthetic infared spectra that can be compared with astronomical AIB spectra. Two studies \citep{verstraete2001,pech2002} made use of the empirical anharmonic factors derived from gas-phase experiments by \citet{joblin1995}. The latter study included a few neutral small- and medium-sized PAHs (N$_{C} \leq 32$) in the gas phase. Further theoretical studies were published and reported empirical anharmonic factors for the band positions and widths for a number of species  \citep{joalland2010,calvo2011,simon2011,chakraborty2021}. However, no relevant data are available for PAHs containing aliphatic CH bonds.\\

We present new absorption spectroscopic measurements of gas-phase pyrene, methylated pyrene (Me-pyrene), and hydrogenated pyrene (nH-pyrene with n=2, 4, and 6) in the 1.4-25~$\mu$m wavelength range at various temperatures between 373 and 673~K. Detailed anharmonic calculations at 0~K were also performed to support the band analysis. The analysis of the temperature dependence of the experimental IR spectra allowed us to derive empirical anharmonicity factors. These factors were then incorporated into an emission model to generate synthetic spectra that can be compared with JWST observations. Experimental methods and data analysis, including supporting theoretical calculations, are detailed in Sect.~\ref{sect:methods}, along with the emission model. The temperature-dependent spectral evolution and derived anharmonicity parameters are presented in Sect.~\ref{sect:res}.  Section~\ref{sect:astro} presents the synthetic spectra, their comparison with astronomical observations, and a discussion of the method accuracy. Finally, astrophysical implications are discussed in Sect.~\ref{Sect:res:astro}, followed by the conclusions in Sect.~\ref{Sect:concl}.

\section{Methods}
\label{sect:methods}

\subsection{Experiments}
\label{sect:exp}

Pyrene, 1-methylpyrene, (Me-pyrene), 4,5-dihydropyrene (hereafter 2H-pyrene), 4,5,9,10-tetrahydropyrene (hereafter 4H-pyrene), and 1,2,3,6,7,8-hexahydropyrene (hereafter 6H-pyrene) were purchased from Aldrich and used without further purification. 
The measurement of the IR spectra was performed on the ESPOIRS (Etude Spectroscopique des Propriétés Optiques dans l'InfraRouge et le Submillimétrique) setup described in detail in \cite{demyk2017a}. For this study the IR Fourier Transform spectrometer (Bruker Vertex 70V) was equipped with a globar source, a KBr beamsplitter, and a DTLaTGS detector, enabling measurements in the 7000 -- 400~cm$^{-1}$ range (1.4 -- 25~$\mu$m). Measurements were made in the gas  phase with a spectral resolution of 2~cm$^{-1}$.\\
To measure the spectrum of PAHs in the gas phase, we developed a dedicated high-temperature cell. It consists of a stainless steel cell (18.4 cm long and of 3.4 cm diameter) equipped with a valve and two wedged diamond windows that allow us to access a wide range of temperatures and wavelengths. The cell was inserted into a custom-made oven that was insulated so that it can be positioned in the sample compartment of the spectrometer. The temperature of the cell was monitored by three probes, two of which were regulated by two controllers. The temperature of the cell was homogeneous within 1~K except on the two windows, which are expected to be colder. The samples were introduced into the cell from one side when the windows were dismounted and after the cell and windows were cleaned. The cell was then closed and evacuated through the valve to a typical pressure of a few 10$^{-6}$ mbar to remove at best atmospheric H$_2$O and CO$_2$. The cell can be baked to improve the degassing of the wall depending on the vapour pressure of the studied species. To optimise the uniformity of the species temperature during the experiments, about 200 mbar of N$_2$ was introduced in the cell. The spectra of the cell containing the samples and 200 mbar of N$_2$ were recorded at various temperatures, depending on the species, with temperature steps of 50~K. At each step, we measured the IR spectrum of the cell every 15 minutes to monitor its evolution with time until the maximum intensity of the bands was reached, which indicates that the cell reached equilibrium in terms of vapour pressure and temperature. We repeated the measurements with the same conditions (temperature steps and time for stabilisation) for the cell containing only N$_2$. These spectra were used as a reference to construct the absorbance spectrum of the samples at each temperature. \\
All experimental spectra and anharmonicity parameters (see Sect~\ref{sect:analysis}) are made available on the new the \href{https://cosmicpah-irdb.irap.omp.eu/}{CosmicPAH-IRDB database}\footnote{https://cosmicpah-irdb.irap.omp.eu/}. The id numbers of the data used in this study are \href{https://cosmicpah-irdb.irap.omp.eu/science/Pyrene/Gas/22/sample/}{id:22} and \href{https://cosmicpah-irdb.irap.omp.eu/science/Pyrene/Gas/2/sample/}{id:2} for pyrene, \href{https://cosmicpah-irdb.irap.omp.eu/science/1,2,3,6,7,8-hexahydropyrene/Gas/3/sample/}{id:3} and \href{https://cosmicpah-irdb.irap.omp.eu/science/1,2,3,6,7,8-hexahydropyrene/Gas/4/sample/}{id:4} for 6H-pyrene, \href{https://cosmicpah-irdb.irap.omp.eu/science/1-methylpyrene/Gas/7/sample/}{id:7} and \href{https://cosmicpah-irdb.irap.omp.eu/science/1-methylpyrene/Gas/17/sample/}{id:17} for Me-pyrene, \href{https://cosmicpah-irdb.irap.omp.eu/science/4,5-dihydropyrene/Gas/9/sample/}{id:9} for 2H-pyrene, and \href{https://cosmicpah-irdb.irap.omp.eu/science/4,5,9,10-tetrahydropyrene/Gas/12/sample/}{id:12} for 4H-pyrene.

\subsection{Theoretical calculations}
\label{sect:theory}
Theoretical spectra for all molecules considered were obtained in the framework of the DFT using the B3LYP exchange-correlation functional \citep{becke1993, lee1988} together with Grimme's D3 dispersion correction \citep{grimme2010}. We used several basis sets, namely Pople's 6-311G(d,p) \citep{krishnan1980} and the N-07D Gaussian basis set specifically optimised by Barone and co-workers also for GVPT2 anharmonic calculations \citep{barone2008}. The \textsc{Gaussian16} code \citep{g16} was employed to obtain the minimum energy geometries and perform a harmonic vibrational analysis with the above basis sets, and we then obtained semi-quartic force fields and first and second derivatives of the electric dipole moment, only with N-07D. The \textsc{AnharmoniCaOs} code \citep{mulas2018} was then employed with these force fields and dipole moment first and second derivatives to obtain anharmonic absorption spectra from the ground-vibrational state. 
Harmonic analyses at the B3LYP-D3/6-311G(d,p) level are known to yield rather accurate vibrational frequencies, when properly scaled, and acceptable intensities, except for those of CH stretches. 
B3LYP-D3/N-07D yields more accurate intensities, but poorer frequencies at the harmonic level, being instead optimised for anharmonic calculations. Hence, the B3LYP-D3/N-07D level was used as a basis for anharmonic calculations with \textsc{AnharmoniCaOs}. When harmonic calculations were used, since we only needed frequencies, we employed those at the B3LYP-D3/6-311G(d,p) level, scaled with the factors 0.965 for CH stretching modes and 0.982 for all other modes at lower frequency, following typical values provided by \citet{bauschlicher2018}. \\
In all \textsc{Gaussian16} calculations, we used the largest built-in grids for numerical integration and the tightest convergence criteria. The parameters for \textsc{AnharmoniCaOs} calculations were tuned to maximise accuracy while keeping the maximum sizes of polyads of resonating harmonic states within $\sim$40000. Hydrogenating pyrene removes some of the electrons from the aromatic delocalised $\pi$ orbitals, using them to form aliphatic bonds. This in turn breaks the planarity of the molecule and creates a staggered structure of the carbon atoms involved. For 4H-pyrene and 6H-pyrene, this results in two inequivalent 3D structures with (locally) minimum energy. In one conformer, the aliphatic groups on either side of the molecule break the planarity in the same direction with respect to the backbone plane; in the other conformer, they break the planarity in opposite directions with respect to the plane. In the case of 2H-pyrene, there is only one conformer because the other can be considered as simply a rotation. All distinguishable conformers were considered in the calculations. For Me-pyrene, the methyl group can rotate almost freely with a relatively small barrier, making it unsuitable for \textsc{AnharmoniCaOs} calculations, but there is one well-defined lowest-energy position of the group. The structures are shown in Table~\ref{tab:structures}. 

All optimised structures and harmonic analyses performed with  \textsc{Gaussian16}, also including additional basis sets not used here, are made available on the new Cagliari-Toulouse \href{https://CosmicPAH-Qcals.oa-cagliari.inaf.it/}{Theoretical spectral database of Polycyclic Aromatic Hydrocarbons}\footnote{https://cosmicpah-qcals.oa-cagliari.inaf.it/}. The theoretical anharmonic spectra are accessible via the \href{https://cosmicpah-irdb.irap.omp.eu/}{CosmicPAH-IRDB database} with the following sample ids: \href{https://cosmicpah-irdb.irap.omp.eu/science/Pyrene/Gas/24/sample/}{id:24} and \href{https://cosmicpah-irdb.irap.omp.eu/science/Pyrene/Gas/26/sample/}{id:26} for pyrene, \href{https://cosmicpah-irdb.irap.omp.eu/science/1,2,3,6,7,8-hexahydropyrene/Gas/5/sample/}{id:5} for 6H-pyrene, \href{https://cosmicpah-irdb.irap.omp.eu/science/4,5-dihydropyrene/Gas/10/sample/}{id:10} for 2H-pyrene, and \href{https://cosmicpah-irdb.irap.omp.eu/science/4,5,9,10-tetrahydropyrene/Gas/13/sample/}{id:13} for 4H-pyrene.

\begin{table}
        \centering
        \caption{Structures of the Me-pyrene, 2H-pyrene, 4H-pyrene, and 6H-pyrene, shown with a top and side view}
        \begin{tabular}{ccc}
            \includegraphics[align=c, width=2cm]{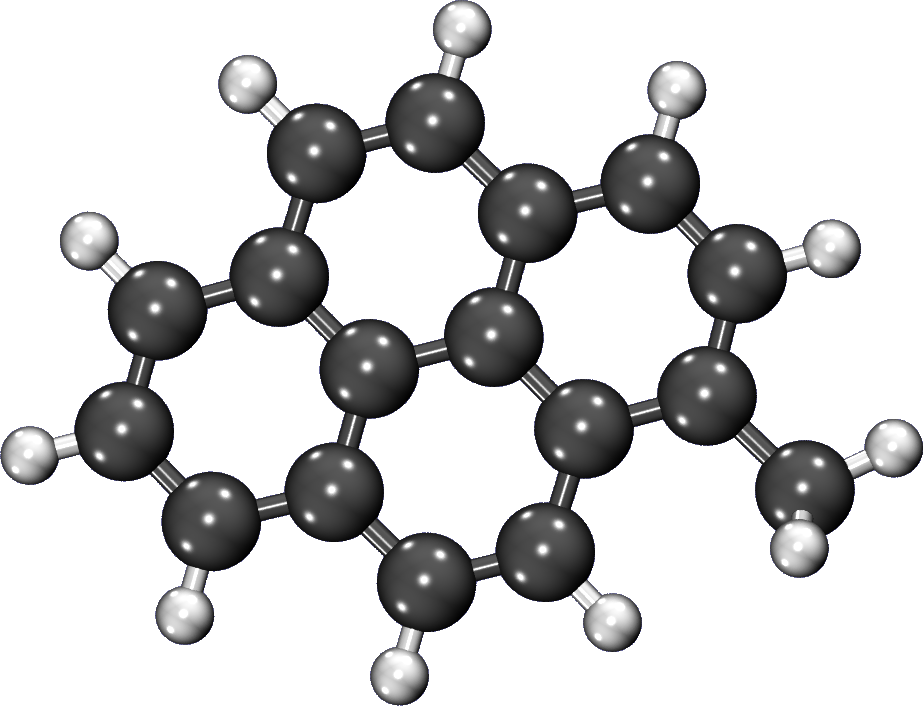} \vspace{1mm}&
            \includegraphics[align=c, width=2cm]{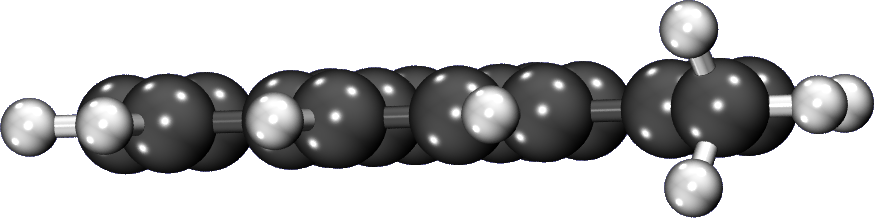} &
            Me-pyrene \\
            \includegraphics[align=c, width=2cm]{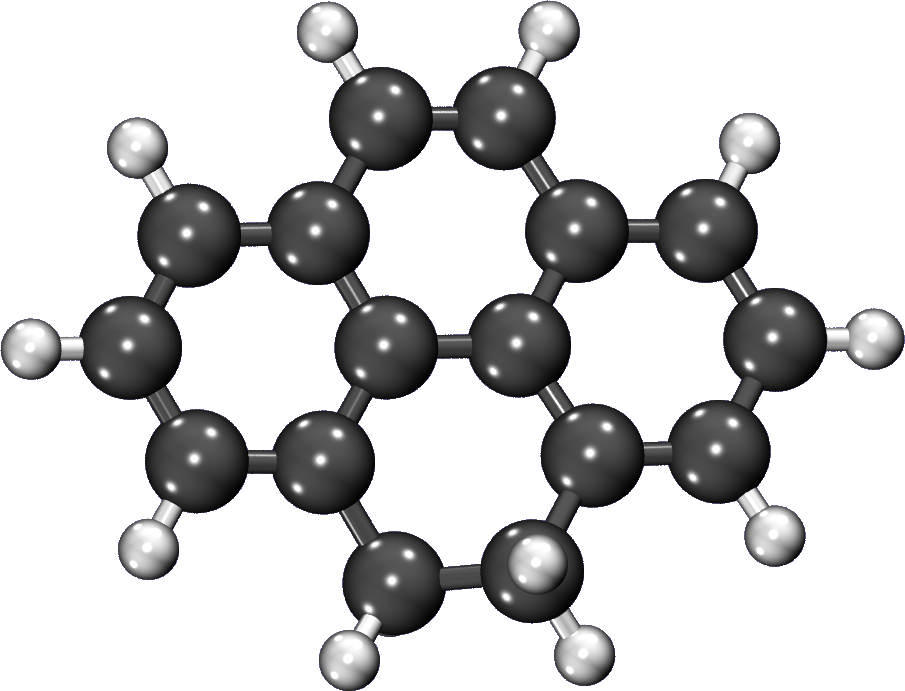}  \vspace{1mm}&
            \includegraphics[align=c, width=2cm]{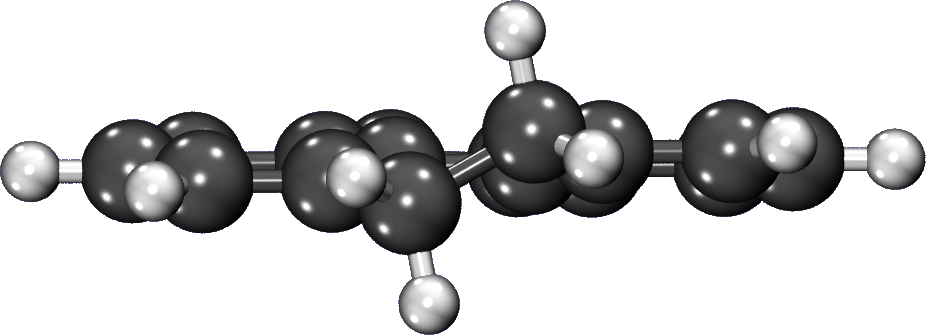} &
            2H-pyrene \\
            \includegraphics[align=c, width=2cm]{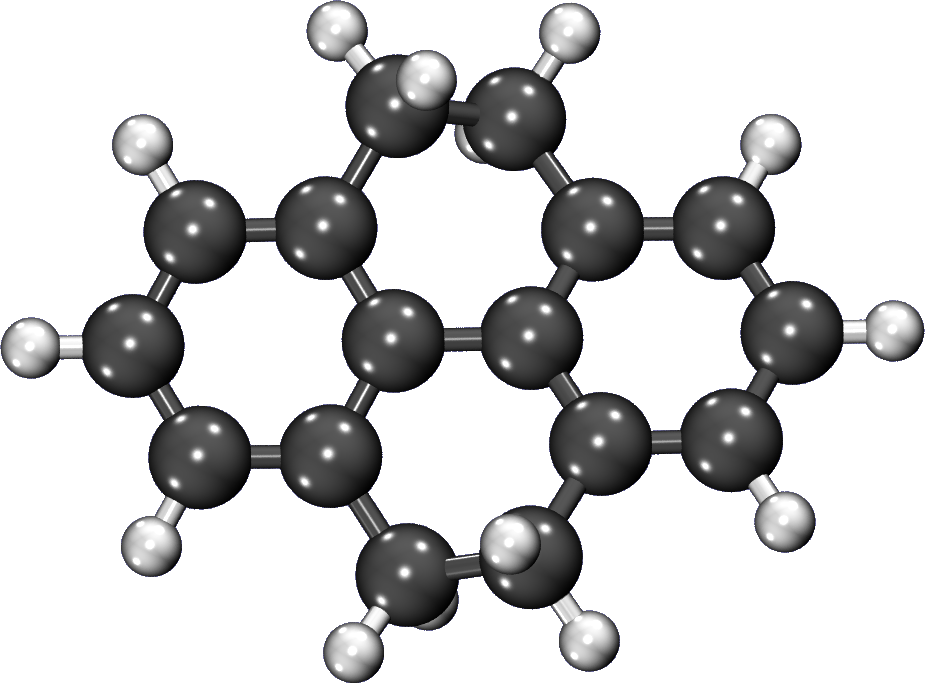}  \vspace{1mm}&
            \includegraphics[align=c, width=2cm]{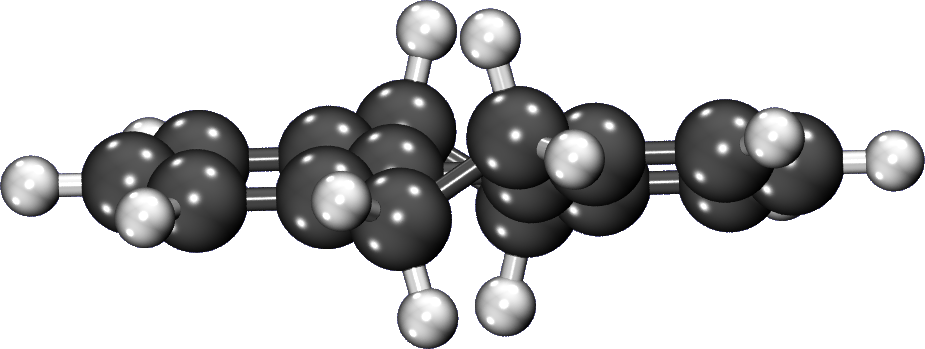} &
            4H-pyrene conf. 1\\
            \includegraphics[align=c, width=2cm]{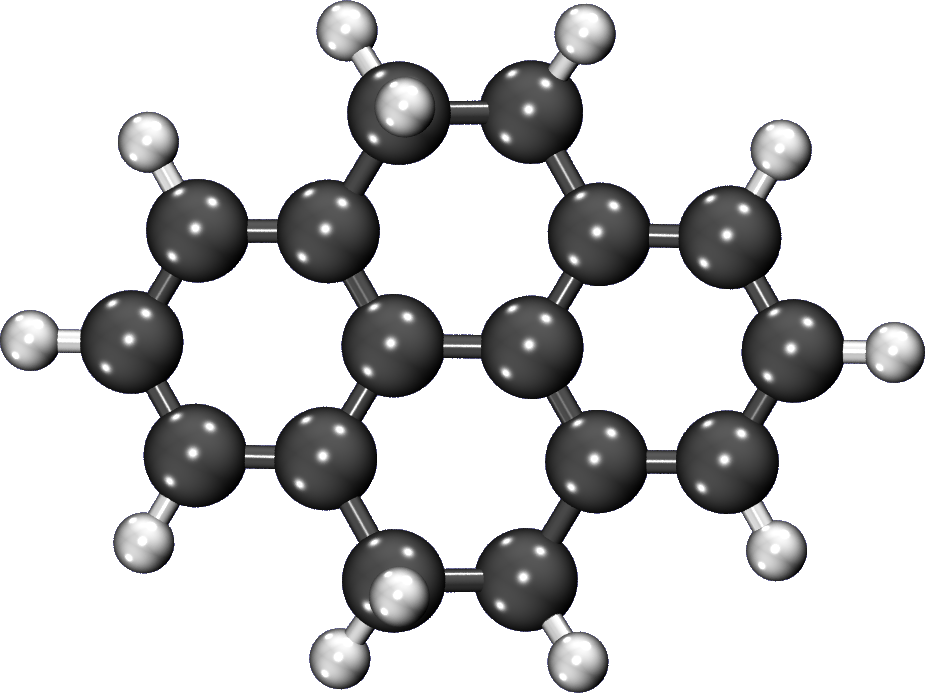}  \vspace{1mm}&
            \includegraphics[align=c, width=2cm]{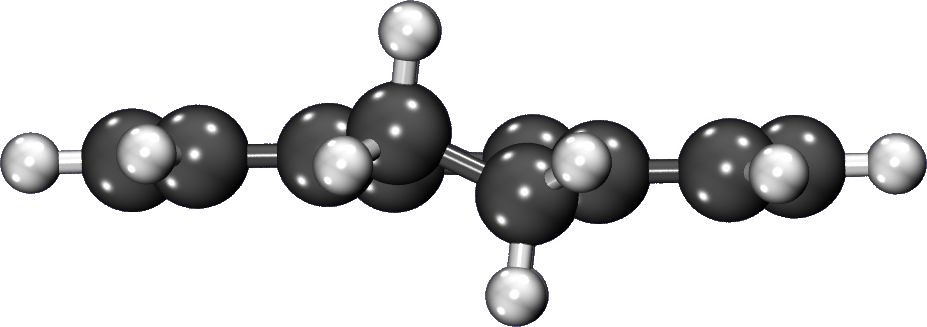} &
            4H-pyrene conf. 2\\
            \includegraphics[align=c, width=2cm]{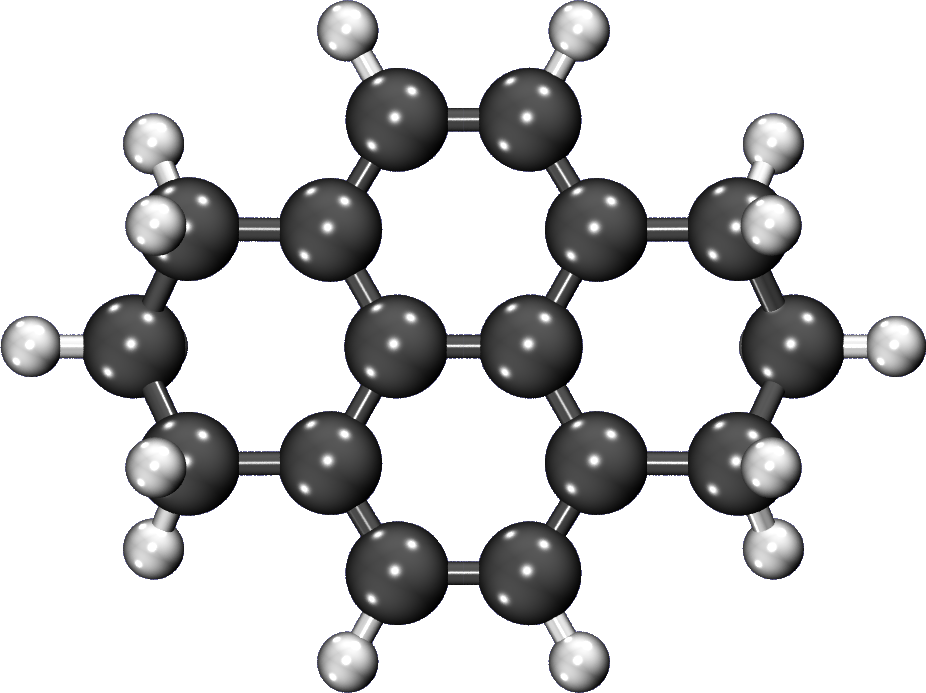}  \vspace{1mm}&
            \includegraphics[align=c, width=2cm]{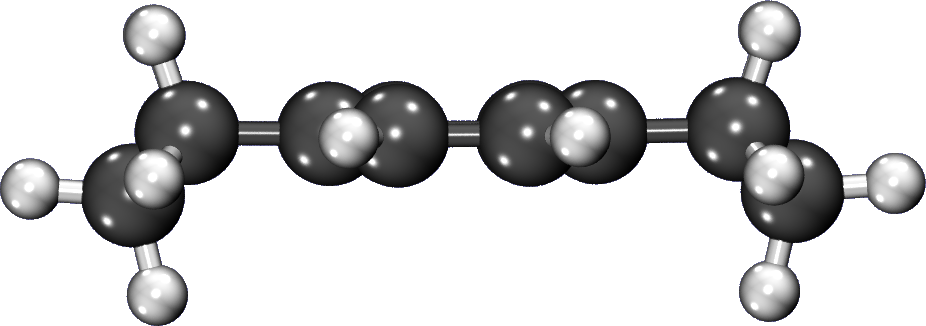} &
            6H-pyrene conf. 1\\
            \includegraphics[align=c, width=2cm]{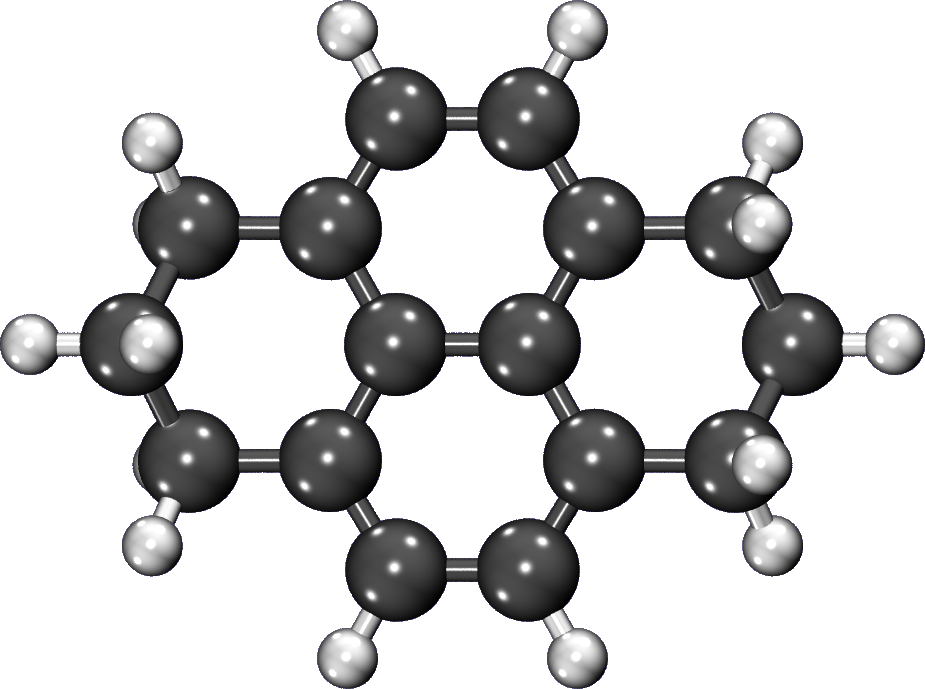} &
            \includegraphics[align=c, width=2cm]{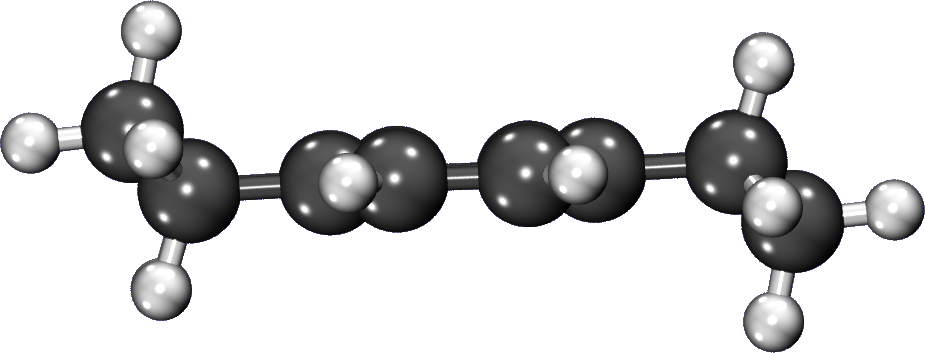} &
            6H-pyrene conf. 2\\
        \end{tabular}
        \label{tab:structures}
\end{table}

\subsection{Extraction of empirical anharmonicity parameters }
\label{sect:analysis}

  \begin{figure*}[!t]
   \begin{center}
  \includegraphics[scale=.6, trim={0.5cm 0cm 0.5cm 0}, clip]{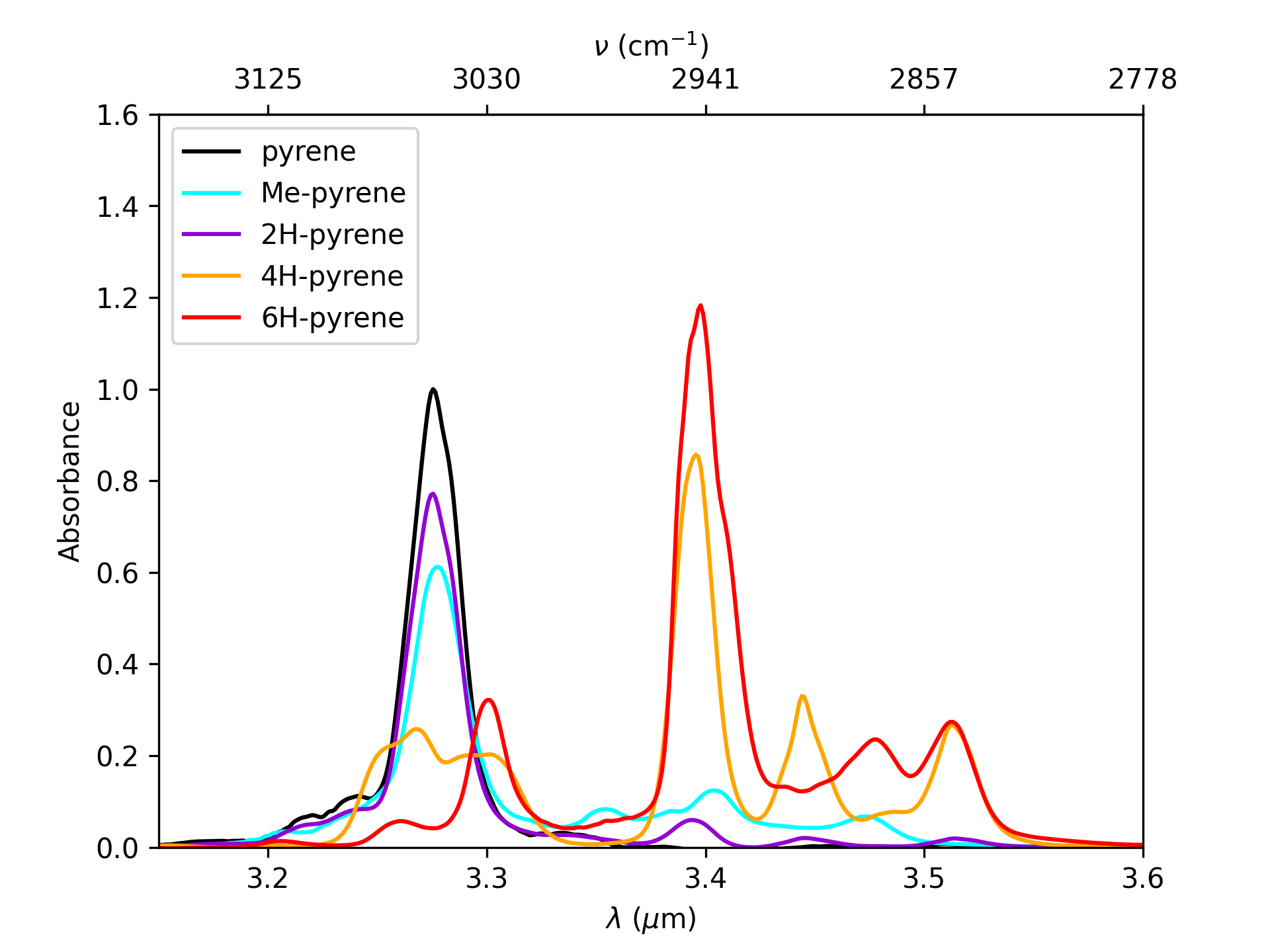}
  \includegraphics[scale=.6, trim={0.5cm 0cm 0.5cm 0}, clip]{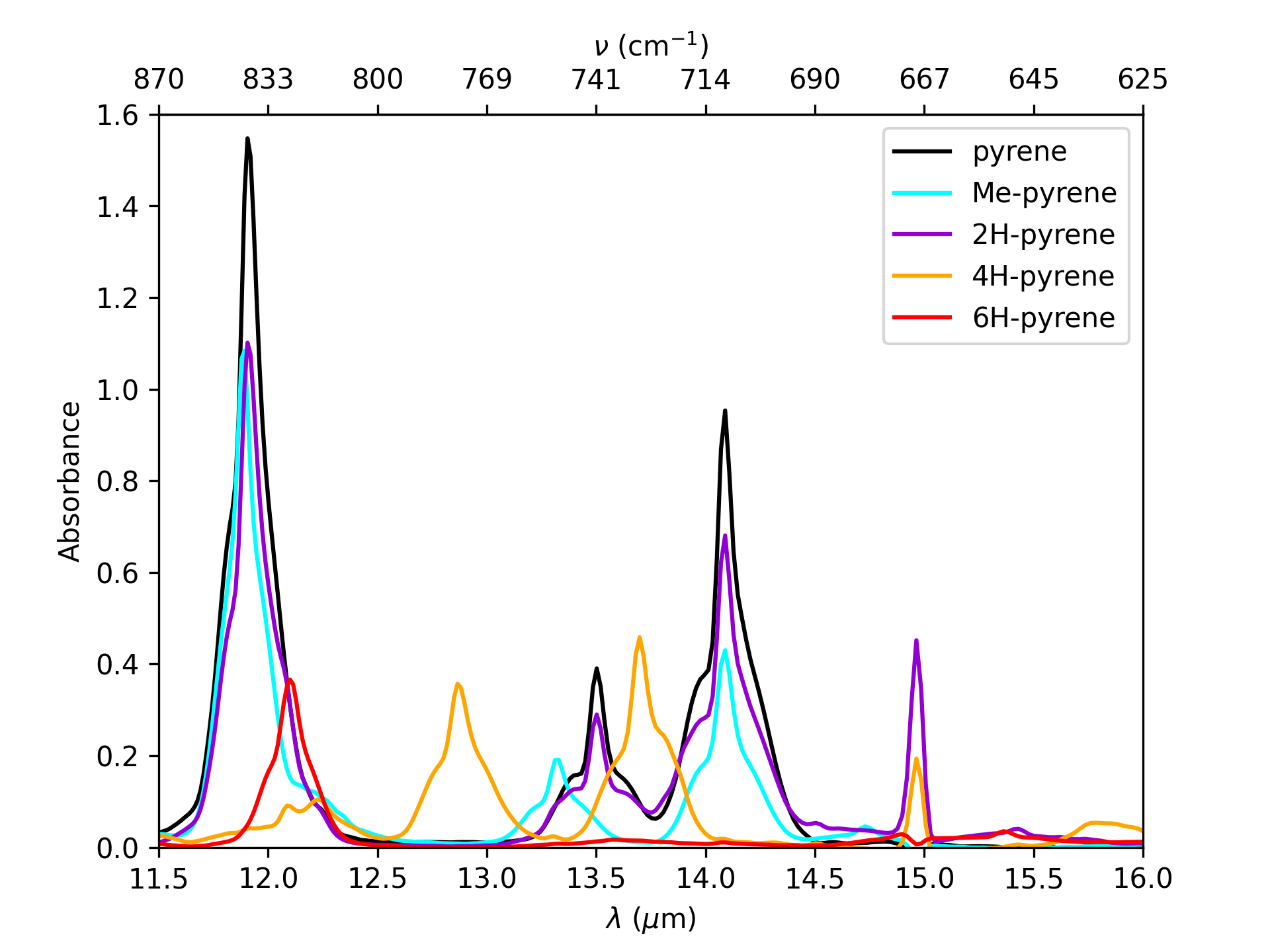}
  \caption{Infrared spectra of pyrene, hydrogenated (2H-, 4H-, and 6H-)pyrene and methylated pyrene in the gas phase at 473~K. Left: CH stretching mode spectral region. Right: CH out-of-plane bending mode region. The intermediate spectral range (5-11.5 $\mu$m) is shown in Fig.~\ref{fig:comp_all_species_5_11mic}. The spectra were first normalised to their total area and then scaled so that the intensity of the aromatic CH stretching band of pyrene equaled 1. }
   \label{fig:comp_all_species}%
  \end{center}
    \end{figure*}

The evolution of the position and width of the bands as a function of temperature can be quantified using empirical anharmonicity parameters \citep{joblin1995, chakraborty2019}. In order to facilitate the extraction of these parameters, we developed an interactive tool, \textit{cosmicPAHmfit}, written in Python and accessible via the \href{https://cosmic-pah.irap.omp.eu}{Cosmic PAH portal}\footnote{https://cosmic-pah.irap.omp.eu}. This tool allows the user to perform a multi-component fit of selected bands and to determine empirical anharmonicity parameters from polynomial functions that describe how the band positions and widths evolve with temperature. The bands are fitted with one or more pseudo-Voigt functions, that is, a linear combination of a Gaussian and a Lorenztian profile, which can be asymmetric \citep[][]{stancik2008}. The parameters for the fit are the position, width, and intensity of the bands, the asymmetry factor, and the relative weight of the Lorenztian and Gaussian. The user can lock some of these parameters or restrict the interval over which they can vary. The user can also interactively define the components for the fit from the spectrum measured at the lowest temperature. The program then performs the fit at each temperature by using the values providing the best spectral fit to the previous temperature as initial guesses for the parameters. At each temperature and for each band, the program determines the full width at half-maximum (FWHM) of the fit (which may include several pseudo-Voigt components) and the position, defined as the midpoint of the FWHM. This provides a table of values for the position and width resulting from the adjustments made at each temperature for each band. The anharmonicity parameters are then determined by fitting the evolution with the temperature of the peak positions and widths using a degree polynomial ranging from one to three over a temperature range selected by the user. 

We sought to minimise subjective choices when performing the multi-component fit. All bands were thus fitted using a single pseudo-Voigt component without asymmetry. The relative contribution of the Lorenz and Gaussian profiles was set at the same value at all temperatures. This value was determined by making an initial fit to the band at all temperatures allowing this parameter to vary, and taking the average. Depending on the bands, the peak position, intensity, and width were either allowed to vary freely or were set to a fixed interval. For the most strongly blended bands, for which it is difficult to constrain the position and width, these values can be set as fixed. As only one component was used to fit the bands, the FWHM is simply the FWHM of the pseudo-Voigt function. The shape of the bands is affected by rotational broadening, as discussed in previous work \citep{joblin1995}. This contribution should be removed for astrophysical applications \citep[see e.g.][]{pech2002}. We therefore built additional datasets in which rotational broadening was estimated at best and subtracted.
The procedure is detailed in Appendix~\ref{sect:rot_broadening}.

\subsection{Emission model}
\label{sect:model}
 For comparison with astronomical data, the cooling cascades of PAHs submitted to the UV radiation field need to be calculated and it needs to be considered that each IR photon is emitted at a given temperature. This leads to a specific band position and width. This type of approach was previously used in \citet{pech2002}, who calculated the IR emission rates based on the thermal approximation. We used the microcanical approach and our Monte Carlo kinetic code \citep{joblin2002} to calculate the evolution of the internal energy over time and quantify the cooling cascade thus. The input parameters of the Monte Carlo model were the energy of the absorbed UV photon and the list of modes and associated IR band intensities. This list was constructed using the frequency list appropriately scaled from our harmonic DFT calculations and the intensities derived from our experimental measurements.
The rates of spontaneous emission were calculated using the harmonic approximation, in which the (v-1)$\leftarrow$v transition is given by

\begin{equation}
k_{IR}^\mathrm{v, v-1}(\nu_{i})=A^\mathrm{v, v-1}_{i}P_{i}^\mathrm{v}
\label{Eq:kIR}
\end{equation}
where $A^\mathrm{v, v-1}_{i}=\mathrm{v}A^\mathrm{1, 0}_{i}$ is the Einstein coefficient for the transition and $P_{i}^\mathrm{v}=\rho_{r}(U-\mathrm{v}h\nu_{i})/\rho(U)$ is the probability to find the system in the level v of the $\nu_{i}$ frequency; $\rho(U)$ is the total density of vibrational states at the energy $U$, and $\rho_{r}$ is the density of states excluding the emitting mode $\nu_{i}$.

If we were to directly take anharmonicity into account in the Monte Carlo model of the emission cascade in detail, we would need extensive calculations. At this stage, we included corrections that can be derived from the decoupled anharmonic Morse oscillators. This was included in the calculation of the density of states using a direct count for states as given in the Stein-Rabinovitch extension formalism (see Fig.~\ref{fig:T_vsU}). We also modified Eq.~\ref{Eq:kIR} following Eq. (3) in \citet{Stockett2025}. However, these corrections led to negligible changes in the simulated profiles of the fundamental bands, for which the primary effects of anharmonicity are indeed accounted for by the empirical anharmonicity parameters derived from spectral analysis. Additional bands that can appear due to anharmonicity, such as overtones and combination bands, are minor. They are well separated from the fundamentals unless they are involved in close resonances, and in that case, they were included in the spectral profile.

The Monte Carlo code simulated 10000 cascades, which is enough to accumulate vast statistics. It recorded all the emitted photons, their energy, and the excitation of the molecule when each was emitted. This was then post-processed to take anharmonicity into account using the empirical anharmonicity parameters derived from spectral analysis. In practice, it consisted of attaching the appropriate band position and width to each emitted photon, as obtained from its parametrised temperature dependence (Sect.~\ref{sect:analysis}). This resulted in a simulated anharmonic emission spectrum.

\section{Results}
\label{sect:res}

We present the IR spectra of pyrene, 2H-pyrene, 4H-pyrene, 6H-pyrene, and Me-pyrene measured in the gas phase between 373~K and 673~K and their anharmonic spectrum at 0 K (with the exception of Me-pyrene, which is not suitable for this type of calculation) in Sect.~\ref{Sect:res:description} .
The temperature evolution of the spectra and the characterisation of the anharmonicity of the bands are presented in  Sect.~\ref{Sect:res:T_evol}. 

\subsection{Spectral diversity}
\label{Sect:res:description}

The CH$_{\rm{arom}}$ stretching modes of pyrene and the four decorated pyrene molecules peak in the range 3.2 -- 3.35~$\mu$m (Fig.~\ref{fig:comp_all_species}). The comparison of the spectra of pyrene, 2H-pyrene, and Me-pyrene shows that the methyl group and the two additional hydrogen atoms do not appreciably affect the peak position, width, and shape of the CH$_{\rm{arom}}$ stretching band. 
In contrast, the CH$_{\rm{arom}}$ stretching band of 4H-pyrene and 6H-pyrene is split into two components with a similar intensity for 4H-pyrene and with a different intensity for 6H-pyrene, and the 3.300~$\mu$m component is strongest (Table~\ref{Table:3mic_bands}). These differences are related to the structural changes induced by the additional hydrogen atoms and the methyl group, which break planar symmetry and cause more bands to become IR-active. 

\begin{table}[!t]
\caption {Position of the peak maximum of the aromatic and aliphatic CH stretching bands at 473 K of the studied species. } \begin{center}
\begin{tabular}{c c c c c}
\hline 
\hline 
& \multicolumn{2}{c}{ Band peak (473K) } & Relative  & Assignment \\ 
            		&   ($\mu$m)  &  (cm$^{-1}$) & intensity & \\ 
\hline 
\hline 
Pyrene  & 3.276 & 3052.5 & - & CH$_{\rm{arom}}$ \\ 
\hline 
\hline 
\multirow{4}{*}{2H-Pyrene}  & 3.275 & 3053.0 & 13.0 & CH$_{\rm{arom}}$\\  
\cline{2-5} 
 					          & 3.393 & 2947.0 & 1 & ip asym. CH$_{\rm{ali}}$ \\ 
 					          & 3.445 & 2902.6 & 0.34 &  oop CH ali.    \\
 					          & 3.514 & 2845.7 & 0.32 & oop sym. CH$_{\rm{ali}}$  \\
\hline 
\hline 
\multirow{5}{*}{4H-Pyrene}  & 3.268 & 3059.8 & 0.30 & CH$_{\rm{arom}}$  \\
 					        & 3.301 & 3029.0 & 0.23 & CH$_{\rm{arom}}$  \\
\cline{2-5} 
					        & 3.396 & 2945.0 & 1 & ip asym. CH$_{\rm{ali}}$ \\
 					          & 3.444 & 2903.6 & 0.38 & oop CH$_{\rm{ali}}$    \\
 					          & 3.512 & 2846.7 & 0.31 &  oop sym. CH$_{\rm{ali}}$  \\
\hline 
\hline 
\multirow{4}{*}{6H-Pyrene}  & 3.261 & 3066.5  & 0.05 & CH$_{\rm{arom}}$  \\
 					        & 3.300 & 3029.9  & 0.27 & CH$_{\rm{arom}}$ \\
 \cline{2-5} 
						      & 3.398 & 2943.1  & 1 & ip asym. CH$_{\rm{ali}}$  \\
 					          & 3.478 & 2875.6  & 0.20 & oop CH$_{\rm{ali}}$  \\
 					          & 3.513 & 2846.7  & 0.23 & oop sym. CH$_{\rm{ali}}$  \\

\hline 
\hline 
\multirow{3}{*}{Me-pyrene}  & 3.277 & 3051.1  & 4.9 & CH$_{\rm{arom}}$ \\
\cline{2-5} 
 					          & 3.355 & 2980.7 & 0.67 & asym. CH$_3$ \\
 					          & 3.403 & 2938.3 & 1 & asym. CH$_{\rm{ali}}$ \\
 					          & 3.473 & 2879.5 & 0.54 & sym. CH$_3$  \\
 \cline{2-5} 
\hline
\hline 
\end{tabular} 
\tablefoot{The peak intensities are given relative to the 3.4 $\mu$m band. The band ratios, for example, 3.4/3.3 $\mu$m, can be derived from the intensities listed in Table~\ref{Table:band_intensity}. The overall assignments, in-plane asymmetric (ip-asym.), symmetric (sym.) or asymmetric (asym.) out-of-plane (oop), are from \cite{maltseva2018}, \cite{russo2014}, and \cite{mackie2018}. } 
\label{Table:3mic_bands}
\end{center}
\end{table}

\begin{figure}[!t]
\centering
\includegraphics[scale=.55]{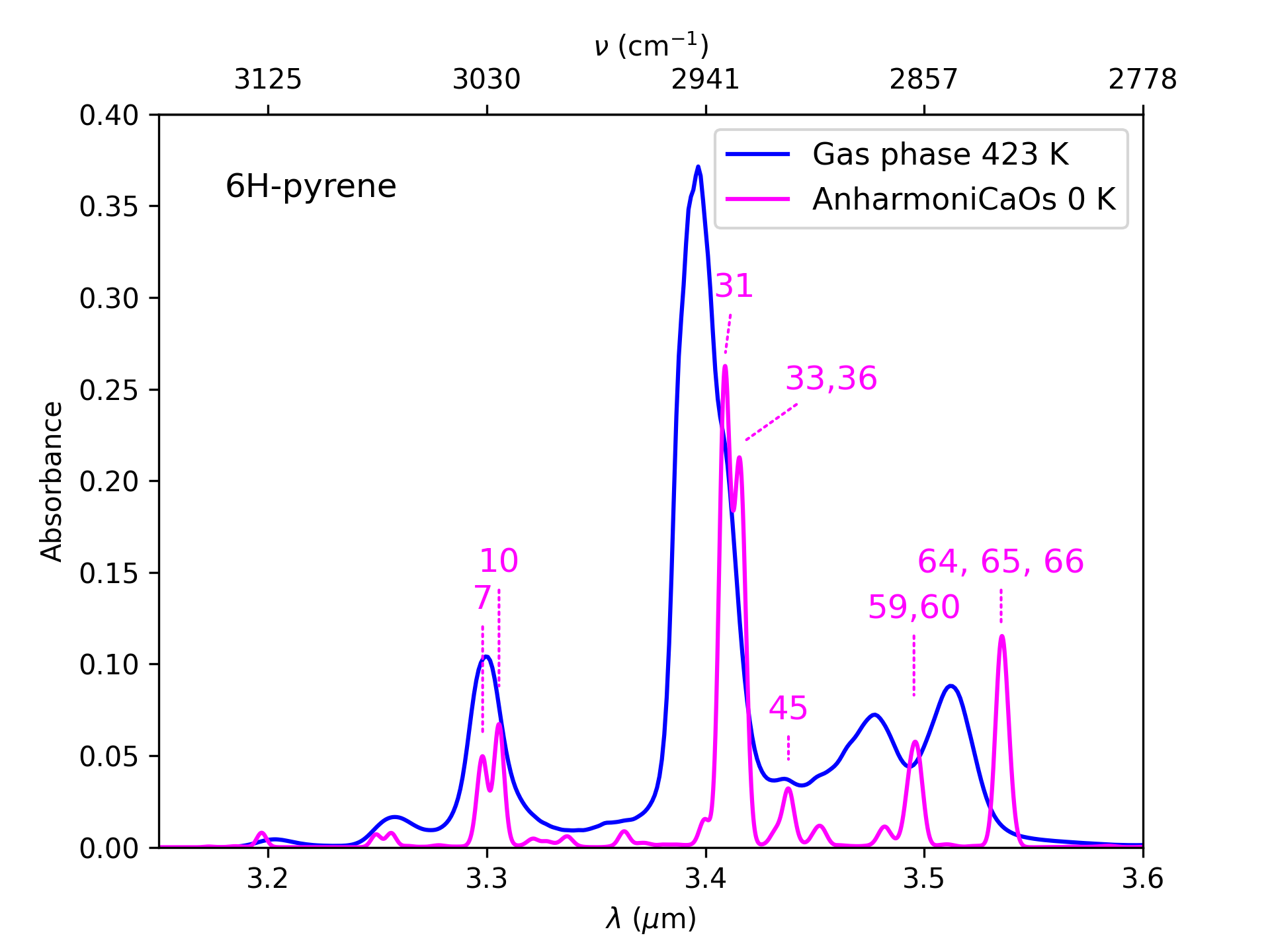}
\caption{Comparison of the 6H-pyrene spectrum in the gas phase ({\it in blue}) with theoretical anharmonic spectra at 0 K ({\it in magenta}). The theoretical spectra are unscaled and have a Gaussian profile with an FWHM of 5 cm$^{-1}$. The bands are labelled according to the first column of Table~\ref{tab:h6pyre_extract_table} for conformer 1. This figure, including the assignments in the table, can be interactively examined on the Cosmic PAHs site at \href{https://cosmicpah-irdb.irap.omp.eu/science/1,2,3,6,7,8-hexahydropyrene/Gas/3/sample/?overplot-tab\#/3/5/5/15/3140/2800/1200/500/423}{this link}.
The differences with the computed anharmonic spectrum for conformer 2 are negligible in this spectral range. The comparison for 2H- and 4H-pyrene in the 3.1-3.6 $\mu$m range and in the 5-15 $\mu$m range  for the three species is shown in Figs.~\ref{fig:3microns_exp_vs_theo_2H_4H} and ~\ref{fig:5-15microns_exp_vs_theo}. }
\label{fig:3microns_exp_vs_theo}%
\end{figure}

All the decorated pyrene species show three main groups of CH$_{\rm{ali}}$ stretching modes in the 3.35 -- 3.55~$\mu$m range (Fig.~\ref{fig:comp_all_species}). The position of the band at $\sim$ 3.39 -- 3.40~$\mu$m, attributed to the CH$_{\rm{ali}}$ asymmetric in-plane stretching vibration, is remarkably similar for 2H-, 4H-, and 6H-pyrene and is only slightly redshifted for Me-pyrene (Table~\ref{Table:3mic_bands}). This feature thus does not seem to depend strongly on the exact structure and nature of the molecule. On the other hand, the peak positions of the two out-of-plane CH$_{\rm{ali}}$ stretching bands vary more strongly with the molecular structure. These bands are similar for 2H- and 4H-pyrene. The out-of-plane symmetric CH$_{\rm{ali}}$ stretching band peaks at a longer wavelength for 6H-pyrene. For Me-pyrene, the two bands peak at distinct positions compared to hydrogenated pyrenes. The intensity of the CH$_{\rm{ali}}$ stretching bands increases with the number of additional hydrogen atoms and dominates the CH stretching region for 4H- and 6H-pyrene, while the CH$_{\rm{arom}}$ band becomes less intense. As a consequence, the relative intensity of the CH$_{\rm{ali}}$ bands with respect to that of CH$_{\rm{arom}}$ increases with the number of additional aliphatic bonds.\\
The conventional assignment of these bands as 'symmetric' or 'asymmetric' refers to whether the two H atoms that are linked via aliphatic bonds to the same C atom move synchronously in the same direction or in opposite directions during their stretching motion. While this assignment is useful, we recall that it is a simplification. We can rely on the anharmonic calculations that correctly describe the main contributing components of the CH$_{\rm{ali}}$ stretching bands (Fig.~\ref{fig:3microns_exp_vs_theo} and Table~\ref{tab:d1anhbands_2800-3200} for 6H-pyrene). Aliphatic band systems result from the superposition of many individual bands, and each of them, due to resonances, is a transition to a linear combination of several harmonic states: anharmonicity causes a number of partially resolved satellite bands to appear because many nearby combination bands are activated by Fermi resonances. 

\begin{table*}
   \caption{IR-active vibrational transitions of conformer 1 of 6H-pyrene shown in Fig~\ref{fig:3microns_exp_vs_theo}, computed by \textsc{AnharmoniCaOs} from the ground vibrational state (0~K). } 
   \label{tab:d1anhbands_2800-3200}
    \begin{center}
    \begin{tabular}{c|p{10.5 cm}|c|c}
\hline 
\hline 
transition & Excited States & Energy  & Intensity \\
number &  &  (cm$^{-1}$) & (km/mol)  \B\\
\hline
7	& $\nu_{87}$ (0.62), $\nu_{68} + \nu_{71}$ (0.15), $\nu_{19}+\nu_{87}$ (0.05)&	3032.06	&	20.038 \T \\
10	&	$\nu_{89}$ (0.65), $\nu_{68}+\nu_{72}$ (0.15), $\nu_{64}+\nu_{72}$ (0.06), $\nu_{19}+\nu_{89}$ (0.06)	&	3025.20		&35.441 \\
19	&	$\nu_{68}+\nu_{71}$ (0.60), $\nu_{87}$ (0.16), $\nu_{19}+\nu_{68}+\nu_{71}$ (0.08), $\nu_{64}+\nu_{71}$ (0.06)&	2973.74		& 4.529 \\
29	&	$\nu_{80}$ (0.36), $2\nu_{69}$ (0.20), $2\nu_{70}$ (0.19), $\nu_{66}+\nu_{70}$ (0.06), $\nu_{67}+\nu_{69}$ (0.05)&	2941.49	&	7.080 \\
31	&	$\nu_{85}$ (0.56), $\nu_{83}$ (0.35)	&	2933.64	&	133.484 \\
33	&	$\nu_{84}$ (0.51), $\nu_{86}$ (0.38)&	2928.77		&46.356 \\
36	&	$\nu_{82}$ (0.87), $\nu_{19}+\nu_{82}$ (0.07)	&	2926.93	&	68.377 \\
45	&	$\nu_{66}+\nu_{70}$ (0.20), $\nu_{78}$ (0.13), $\nu_{67}+\nu_{69}$ (0.12), $2\nu_{69}$ (0.11), $\nu_{5}+\nu_{52}+\nu_{67}$ (0.10)&	2908.74		&15.645 \\
59	&	$\nu_{79}$ (0.47), $\nu_{69}+\nu_{70}$ (0.23), $\nu_{66}+\nu_{69}$ (0.10), $\nu_{67}+\nu_{70}$ (0.06), $\nu_{57}+\nu_{58}$ (0.05)		&2862.48		&15.836 \\
60	&	$\nu_{80}$ (0.52), $2\nu_{69}$ (0.12), $2\nu_{70}$ (0.11), $\nu_{66}+\nu_{70}$ (0.09), $\nu_{67}+\nu_{69}$ (0.06)	&2859.51	&22.490 \\
64	&	$\nu_{77}$ (0.53), $\nu_{64}+\nu_{65}$ (0.30), $\nu_{66}+\nu_{67}$ (0.06), $\nu_{19}+\nu_{77}$ (0.05)		&2828.86		&32.251 \\
65	&	$\nu_{78}$ (0.52), $2\nu_{64}$ (0.17), $2\nu_{65}$ (0.13), $\nu_{19}+\nu_{78}$ (0.05)&	2828.75	&	20.281 \\
66	&	$\nu_{75}$ (0.57), $\nu_{64}+\nu_{67}$ (0.16), $\nu_{65}+\nu_{66}$ (0.14), $\nu_{19}+\nu_{75}$ (0.05)	&2826.35	&	15.586  \B\\
\hline 
\hline 
    \label{tab:h6pyre_extract_table}
    \end{tabular}
   \tablefoot{Only the transitions contributing 5\% at least of the integrated band intensity in this spectral interval are listed. The definitions of the harmonic normal modes used in the table can be found at \href{https://cosmicpah-qcals.oa-cagliari.inaf.it/database/vibrationalanalysesharmonic/196/490/}{this link}. } 
   \end{center}
    \end{table*}

The 5 -- 20~$\mu$m spectral region exhibits the bending vibrational modes of the CH$_{\rm{arom}}$ and CH$_{\rm{ali}}$ bonds together with CC stretching modes, combination modes, and skeletal vibrations (Figs.~\ref{fig:comp_all_species} and~\ref{fig:comp_all_species_5_11mic}). The aromatic CC stretching and CH bending modes are affected by the presence of extra hydrogen atoms and the methyl group, which modify the aromaticity of the molecules. The 6.2~$\mu$m band attributed to C=C stretch aromatic vibration does not vary strongly with the addition of hydrogen atoms or the methyl group to pyrene (\ref{fig:comp_all_species_5_11mic}).

Additional bands due to the bending vibrations of the extra hydrogen atoms and CH$_3$ groups are expected in the spectra of the decorated species. They fall in the 5-7~$\mu$m region, where contamination by gas-phase water present in the cell complicates the analysis. Despite that, in the spectra of the three hydrogenated pyrene species, we observe a band at $\sim$ 6.9~$\mu$m that can be attributed to the CH$_2$ scissoring vibrational mode. This band is most intense in 4H-pyrene and weakest in 2H-pyrene. In addition, 6H-pyrene has relatively intense bands at 7.11,  7.43, 7.72, and 7.91~$\mu$m that are dominated at the harmonic level by the bending modes of the CH$_2$ groups. In the spectrum of Me-pyrene, we observe bands at $\sim$ 6.67, 6.9, and 7.7~$\mu$m that are associated with CH$_3$ bending vibrations.\\
\indent Concerning the intense out-of-plane vibrational modes, the bands of pyrene at 13.5 and 14.1~$\mu$m, which are attributed to trio CH$_{\rm{arom}}$, are absent from the spectrum of 6H-pyrene as this species contains no trio CH bonds. In the spectra of 2H-, 4H-, and Me-pyrene, these bands are blue-shifted as a result of the presence of additional hydrogen atoms and of the methyl group adjacent to the trio bonds. The CH$_{\rm{arom}}$ out-of-plane vibrational band observed in pyrene at 11.9~$\mu$m is mostly due to the vibration of duo hydrogen atoms. It is red-shifted and much weaker for 6H-pyrene and 4H-pyrene than for the other species.

Our anharmonic calculations at 0~K for 2H-, 4H-, and 6H-pyrene are overall consistent with our high-temperature gas-phase spectra (Figs. ~\ref{fig:3microns_exp_vs_theo}, ~\ref{fig:3microns_exp_vs_theo_2H_4H}, and ~\ref{fig:5-15microns_exp_vs_theo}). For 6H-pyrene, our calculations are consistent with the low-temperature experimental spectrum in \citet{maltseva2018}. All bands are unambiguously identified without empirical correction factors. A theoretical anharmonic spectrum of 6H-pyrene was also previously published by \citet{mackie2018}. It appears to agree well with our spectrum, with some slight differences that are presumably due to our inclusion of more explicit resonances, which resulted in a somewhat more accurate description of the detailed spectral structure around the C-H stretches. \\
Our B3LYP-D3/6-311G(d,p) harmonic spectra are also consistent with those shown in Fig.~18 of \citet{yang2020}, which were computed at a very similar level, and with those in the latest release of the NASA Ames PAH database \citep{Ricca2026}. The latter only contains one unspecified conformer of 4H- and 6H-pyrene. In our comparison with experimental spectra, we averaged the IR spectra of the two conformers because they are most likely to be both present; they are separated only by a tiny energy difference. The IR spectra of the two conformers are very similar, and this results in a slight broadening of some bands in the averaged spectra. \\

\subsection{Spectral evolution with temperature}
\label{Sect:res:T_evol}

\begin{figure*}[!ht]
\centering
\includegraphics[scale=.6, trim={1cm 0.3cm 0cm 0.3cm}, clip]{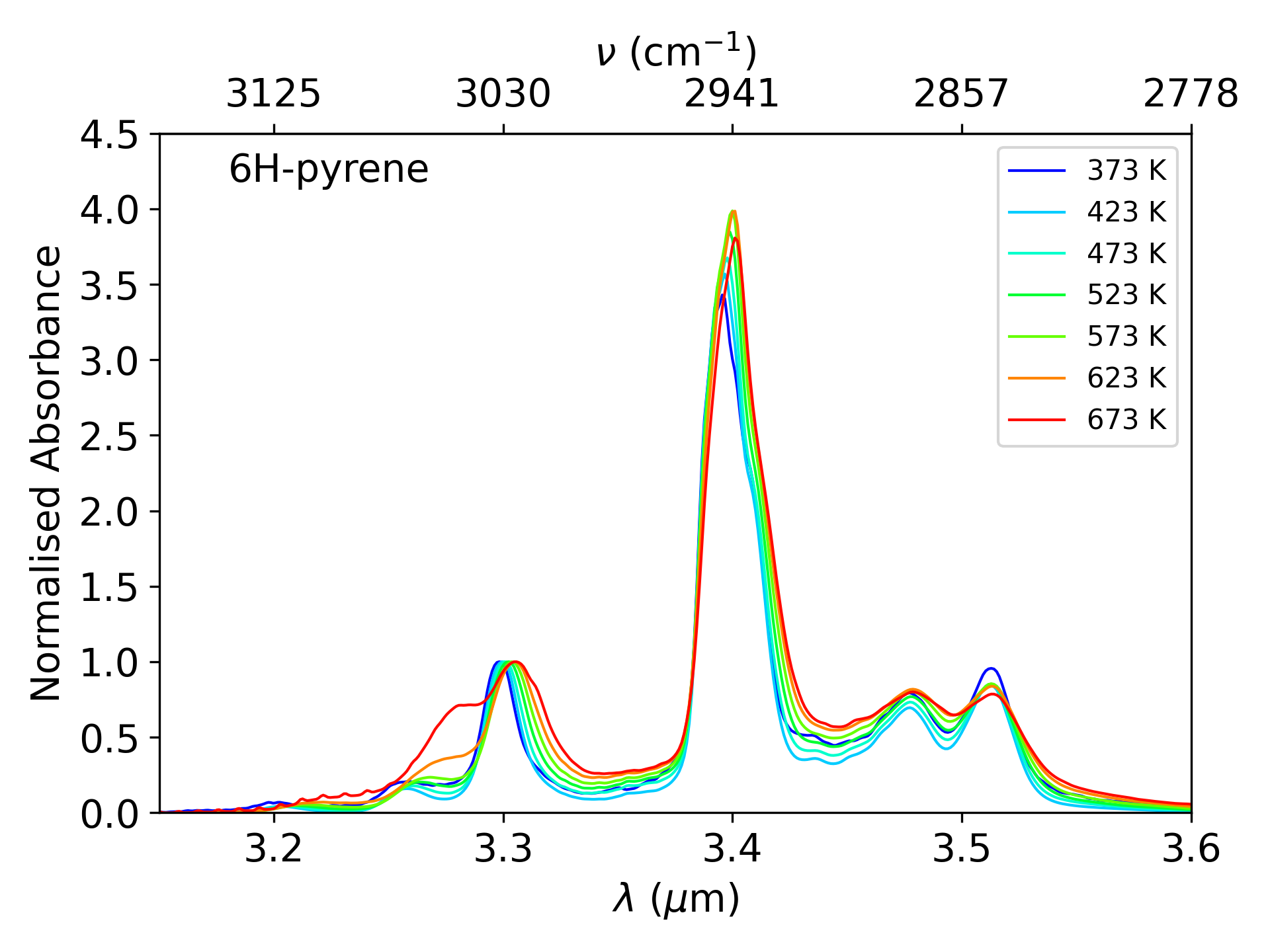}
\includegraphics[scale=.6, trim={1cm 0.3cm 0cm 0.3cm}, clip]{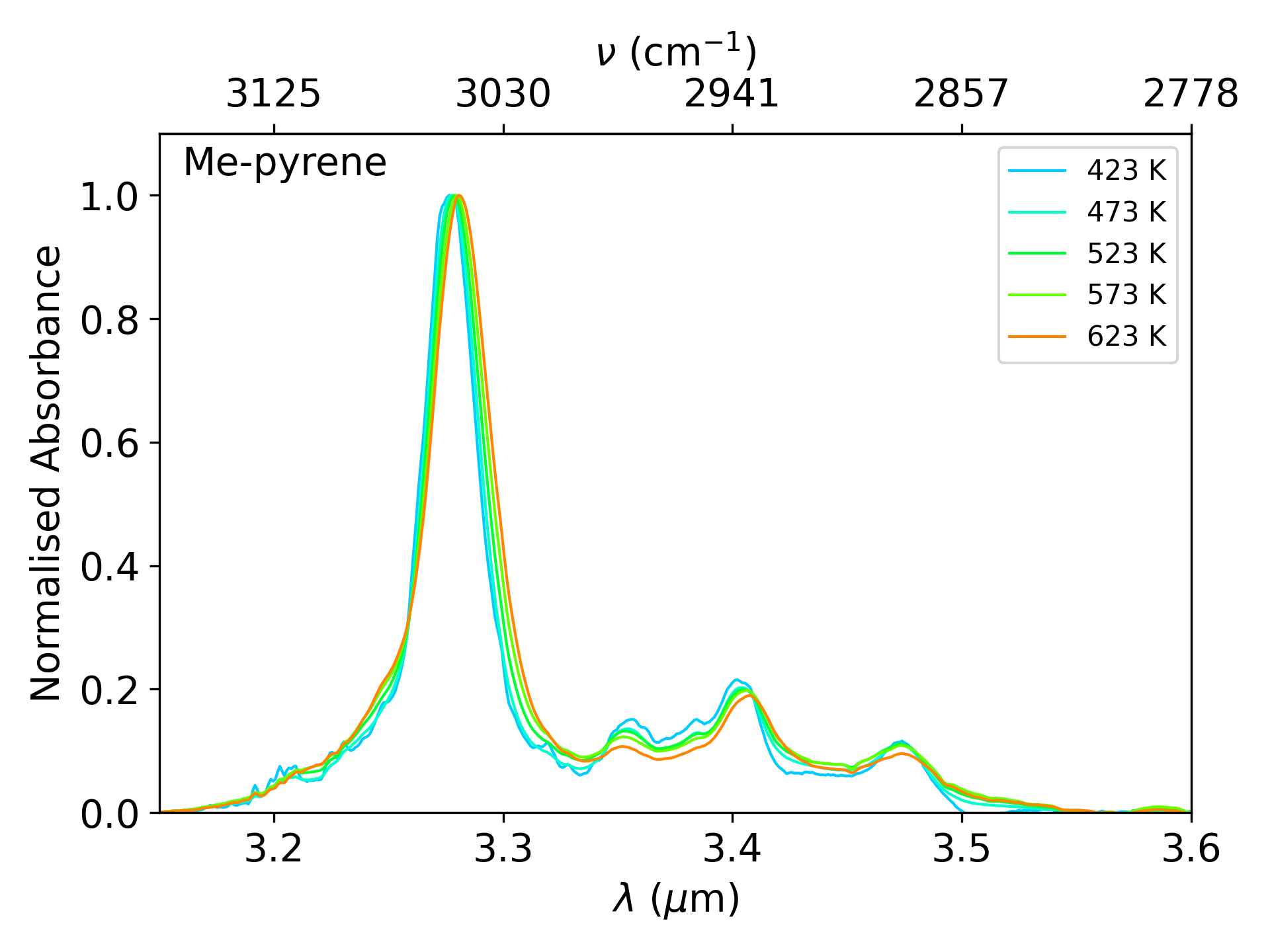}
\caption{Infrared spectra of 6H-pyrene and Me-pyrene at various temperatures in the CH stretching mode spectral region. The spectra are normalised to the aromatic CH stretching band. The spectra of pyrene, 2H-, and 4H-pyrene are shown Fig.~\ref{fig:3microns_vs_T_norm_2H_4H}.}
\label{fig:3microns_vs_T_norm}
\end{figure*}

A first variation we observed in the spectra with temperature concerns the band intensities. They initially increased as the vapour pressure rose with temperature, resulting in an increasing number of molecules in the gas phase. Above a certain temperature, which varied depending on the species, the integrated band intensities began to decrease: above 473 K for pyrene, 2H-pyrene, 6H-pyrene, and Me-pyrene, and above 423 K for 4H-pyrene. This evolution can be seen for each species in the \href{https://cosmicpah-irdb.irap.omp.eu/}{CosmicPAH-IRDB database}.  

The observed decrease cannot be attributed to intrinsic spectral effects but is instead due to our experimental conditions. It can arise from condensation onto the colder parts of the cell, particularly near the valve, or from thermal alteration of the molecules. The latter effect is evident from the decrease in the relative intensity of the aliphatic bands compared to the aromatic bands for 2H- and 4H-pyrene (Figs.~\ref{fig:3microns_vs_T_norm_2H_4H} and ~\ref{fig:ali_aro_ratio}). This shows that aliphatic bonds are lost and aromatic bonds are formed, leading to pyrene as the temperature increases. This behaviour is clearly confirmed for 4H-pyrene by the disappearance of the 12.9 and 13.7 $\mu$m bands, accompanied by the appearance of bands at 11.9, 13.0, and 14.1 $\mu$m, which are characteristic of pyrene (Fig.~\ref{fig:4hpyr-to-pyr}). The aliphatic-to-aromatic band ratio of 6H-pyrene decreases more gradually and at higher temperatures than for 2H- and 4H-pyrene, indicating that 6H-pyrene is more thermally resistant. This is confirmed by the thermal evolution of the 12.1 $\mu$m band of 6H-pyrene, which only decreases weakly (Fig.~\ref{fig:4hpyr-to-pyr}). In addition, there is no indication of a transformation of 6H-pyrene into pyrene (Fig.~\ref{fig:4hpyr-to-pyr}). This might indicate that 6H-pyrene fragments via hydrocarbon rather than hydrogen loss, which is known for the cation \citep{marciniak2021, stockett2021}. Therefore, tracing 6H-pyrene loss by aromatic CH band might just provide a limited view of the thermal alteration of 6H-pyrene. 
The behaviour of Me-pyrene is similar to that of 6H-pyrene in the 423-573~K range: its aliphatic-to-aromatic band ratio remains constant with temperature, even though the integrated intensities of the aliphatic and aromatic bands decrease. This suggests an additional loss channel for Me-pyrene, similar to that of pyrene and potentially 6H-pyrene. One possible explanation is polymerisation, leading to larger species that can condense out of the gas phase due to their high mass \citep{joblin1992PhDT,joblin1997}.\\

As expected from intrinsic anharmonic effects \citep{joblin1995, chakraborty2019}, there are clear variations in the band position and width with temperature (Figs.~\ref{fig:3microns_vs_T_norm} and ~\ref{fig:3microns_vs_T_norm_2H_4H} for the CH stretching region).
To parametrise these variations, and thus, to empirically quantify anharmonicity in simple terms for modelling purposes, we proceeded as described in Sect.~\ref{sect:analysis}. Rotational envelopes with P, Q and R branches are clearly observed in the case of the low-frequency bands  (Fig.~\ref{fig:rot_broadening_3comp}). We estimate the rotational broadening to be in the range of 5.8\,-\,6.8 cm$^{-1}$ from 423 to 673~K (see Appendix~\ref{sect:rot_broadening} for the method). As a first approximation, we assumed that these values apply to all bands and species because their rotational constants are similar. Rotational broadening is not expected to significantly affect the band positions because the rotational envelopes are very symmetric for these species. 

The derived anharmonicity parameters characterising the evolution with temperature of the band positions and widths (including subtraction of rotational broadening) are presented in Table~\ref{Table:anha_coeff} for several bands of 6H-pyrene and Me-pyrene, the most stable molecules in our experimental conditions. The evolution of the peak positions and widths with temperature along with the fits we used to derive the anharmonicity parameters are shown in Fig~\ref{fig:pos_anha_h6pyr} for 6H-pyrene. The corresponding data for Me-pyrene can be retrieved from the \href{https://cosmicpah-irdb.irap.omp.eu/}{CosmicPAH-IRDB database}. Except for some CH$_{\rm{ali}}$ stretching bands, all the analysed bands are red-shifted when temperature increases. The position of the CH$_{\rm{arom}}$ stretching band of both species shows stronger temperature-induced variation in position ($\mid{\chi}_1^{'}\mid \ge 2 \times 10^{-2}$) than the CH$_{\rm{ali}}$ bands ($\mid{\chi}_1^{'}\mid \leq 2 \times 10^{-2}$). Most bands broaden with increasing temperature, although a few, generally weak and blended, appear to narrow in the fits.

\begin{table*}[!t]
\caption {Empirical anharmonicity parameters derived for the bands of 6H-pyrene and Me-pyrene, corrected for rotational broadening.}
\label{Table:anha_coeff}
\begin{center}
\begin{tabular}{c c c c c c c c c}
\hline 
\hline 
        Species & RatioLG & \multicolumn{2}{c}{${\lambda}_{peak}$(523K)} & \multicolumn{2}{c}{Position (cm$^{-1}$)} &  \multicolumn{2}{c}{Width (cm$^{-1}$)} & T range  \\ 
            	&         &      $(\mu$m)  &  (cm$^{-1})$                    & ${\chi}_0^{'}$   &  ${\chi}_1^{'}$ ($\times$ 10$^{-2}$)     &  ${\chi}_0^{"}$   &  ${\chi}_1^{"}$ ($\times$ 10$^{-2}$)  & (K)  \\ 
            	&         &        &                    & (cm$^{-1}$)   &  (cm$^{-1}$ K$^{-1}$)    & (cm$^{-1}$)  &  (cm$^{-1}$ K$^{-1}$)  &    \T\B \\ 
\hline
\hline
\multirow{4}{*}{Pyrene} & 0.61 & 3.276 & 3052.5 & 3062.5$\pm$0.3 & -2.16$\pm$0.05 & 10.0$\pm$1.3 & 1.59$\pm$0.25 &  423-673 \T\\
\cline{2-9} 
& 1    & 11.92 & 838.9 & 846.3$\pm$0.1 & -1.39$\pm$0.02 & 2.56$\pm$0.31 &  0.77$\pm$0.6 &  \multirow{3}{*}{423-673}  \T\\
& 0.41 & 13.51 & 740.2& 746.2$\pm$0.1 & -1.20$\pm$0.02 & 1.1$\pm$0.58 & 0.06$\pm$0.01 & \T\\
& 0.41 & 14.09 & 709.7 & 714.2$\pm$0.12 & -0.91$\pm$0.02 & -3.4$\pm$0.89 & 1.45$\pm$0.16 &   \T\\
\hline
\hline
\multirow{17}{*}{6H-Pyrene} &\multirow{2}{*}{0.9$^{(\#3)}$} & \multirow{2}{*}{3.303} & \multirow{2}{*}{3028.0} & 3041.1$\pm$0.9 & -2.49$\pm$0.18 & -7.46$\pm$0.17 & 4.09$\pm$0.03 & 423-523 \T\\
           &                     &                        &                         & 3040.2$\pm$1.8  & -2.35$\pm$0.3  & -29.26$\pm$6.8  & 8.20$\pm$1.19 & 523-623  \T\\

           &\multirow{2}{*}{0.15$^{(\#4)}$}&\multirow{2}{*}{3.398} & \multirow{2}{*}{2942.7} & 2947.1$\pm$0.0 & -0.85$\pm$0.00 & 14.90$\pm$3.15&-0.05$\pm$0.66 & 423-523  \T\\
           &                     &                       &                         & 2948.7$\pm$0.1 &-1.15$\pm$0.01 & 10.2$\pm$1.09&0.84$\pm$0.18& 523-673  \T\\

& 0$^{(\#7)}$ & 3.429 & 2916.3 & 2916.3$\pm$0.0 & 0.0$\pm$0.0 & 28.40$\pm$0.88 & -0.34$\pm$0.16 & 423-623  \T\\

  &\multirow{2}{*}{0.4$^{(\#5)}$} & \multirow{2}{*}{3.475} & \multirow{2}{*}{2877.3}  & 2879.6$\pm$0.5 & -0.52$\pm$0.11 & 13.25$\pm$9.34 & 2.11$\pm$1.97 & 423-523  \T\\
  &                     &                        &                          & 2875.7$\pm$0.5 & 0.22$\pm$0.08 & 8.79$\pm$1.02 & 2.83$\pm$0.17 & 523-673  \T\\

&\multirow{2}{*}{0.2$^{(\#6)}$} & \multirow{2}{*}{3.514} & \multirow{2}{*}{2846.1} & 2850.0$\pm$0.6 &-0.76$\pm$0.13 & 2.73$\pm$1.39&2.29 $\pm$ 0.29 & 423-523  \T\\
&                     &                        &                         & 2851.9$\pm$0.8 &-1.09$\pm$0.13 & -14.5$\pm$0.96 & 5.58$\pm$0.16 & 523-673  \T\\
\cline{2-9} 
  & 0.15 & 6.262 & 1597.0 & 1612.7$\pm$0.53 & -3.04$\pm$0.10 & 0.09$\pm$0.66&2.15$\pm$0.12 & 423-673  \T\\

 &\multirow{2}{*}{0} & \multirow{2}{*}{6.902} & \multirow{2}{*}{1448.8} & 1452.5$\pm$0.8 & -0.70$\pm$0.17 & 18.2$\pm$1.09 & 0.17$\pm$0.23 & 423-523  \T\\
 &                   &                        &                         & 1458.8$\pm$2.2 & -1.80$\pm$0.35 & 11.2$\pm$0.80 & 1.44$\pm$0.13 & 573-673  \T\\

 &\multirow{2}{*}{0} & \multirow{2}{*}{6.902} & \multirow{2}{*}{1448.8} & 1418.4$\pm$0.3 & -2.70$\pm$0.06 & -1.5$\pm$2.4 & 3.47$\pm$0.51 & 423-523  \T\\
 &                   &                        &                         & 1415.6$\pm$1.4 & -2.2$\pm$2.8 & -2.3$\pm$0.06 & 3.68$\pm$0.51 & 573-673  \T\\

& \multirow{2}{*}{1} & \multirow{2}{*}{7.915} & \multirow{2}{*}{1263.5} & 1270.3$\pm$0.3 & -1.30$\pm$0.06 & -6.29$\pm$2.13 & 3.85$\pm$0.41 & 473-573  \T\\
&                    &                        &                         & 1274.9$\pm$1.1 & -2.10$\pm$0.17 & -13.44$\pm$0.15 & 5.12$\pm$0.02 & 573-673  \T\\

& \multirow{2}{*}{0.5} & \multirow{2}{*}{12.108} & \multirow{2}{*}{825.9} & 832.2$\pm$0.4 & -1.20$\pm$0.08 & 4.07$\pm$1.80 &  0.74$\pm$0.36 & 423-573 \T\\
&                      &                         &                        & 829.5$\pm$0.8 & -0.75$\pm$0.13 & -27.23$\pm$3.45 & 6.21$\pm$0.55 & 573-673 \T\\
\hline
\hline
\multirow{7}{*}{Methyl-Pyrene} & 0.95$^{(\#1)}$ & 3.279 & 3049.9 & 3061.6$\pm$0.2 & -2.23$\pm$0.04 & 5.80 $\pm$0.96 & 2.80$\pm$0.18  & \multirow{6}{*}{473-623} \T\\
                               & 0.1$^{(\#2)}$  & 3.354 & 2981.1 & 2981.7$\pm$.19 & -0.11$\pm$0.35  & 23.91$\pm$4.75 & -1.68$\pm$0.86 &  \T\\
                               & 0.2$^{(\#3)}$  & 3.382 & 2956.9 & 2951.3$\pm$1.9 & 1.23$\pm$0.34  & 14.56 $\pm$1.35 & -0.19$\pm$0.25  &  \T\\
                               & 0.3$^{(\#4)}$  & 3.405 & 2936.8 & 2944.2$\pm$1.1 & -1.39$\pm$0.20 & 0.27 $\pm$2.73 & 2.71$\pm$0.50  &  \T\\
                               & 0.0$^{(\#6)}$  & 3.438 & 2908.4 & 2920.9$\pm$2.9 & -1.99$\pm$0.52 & 40.30 $\pm$1.57 & -3.03$\pm$0.28  &  \T\\
                               & 0.3$^{(\#5)}$  & 3.475 & 2877.8 & 2883.6$\pm$1.4 & -0.98$\pm$0.26 & 0.13$\pm$1.37 & 3.81$\pm$0.25  &   \T\\
\cline{2-9} 
                               & 0.8  & 11.89 & 840.8  & 846.3$\pm$0.1  & -1.06$\pm$0.02 & -0.85$\pm$1.29 & 1.49$\pm$0.24  &  \multirow{3}{*}{423-623} \T\\
                               & 0.9  & 13.33 & 750.1  & 756.0$\pm$0.1  & -1.14$\pm$0.03 &  3.55$\pm$0.91 & 0.45$\pm$0.17  &  \T\\
                               & 0.75 & 14.10 & 709.3  & 715.2$\pm$0.1  & -1.12$\pm$0.01 & 1.36$\pm$0.97  & 0.77$\pm$0.18  &   \T\\
\hline
\hline 
\end{tabular} 
\tablefoot{ 
The second column indicates the band shape of each component; the number in parentheses corresponds to the component label used in the database and in Figs.~\ref{fig:decomp6H} and ~\ref{fig:decomp-mpyr}. The third and fourth columns list the band positions obtained from the multi-component fit of the 523 K spectrum. Columns 5 to 8 report the empirically derived anharmonicity parameters (together with the fitting uncertainty) for the band positions and widths, assuming linear anharmonicity laws ( $\nu(T) = {\chi}_0^{'} + {\chi}_1^{'} T$ and $\Delta\nu(T) = {\chi}_0^" + {\chi}_1^" T$ ), over the temperature ranges specified in column nine. Components 7 of 6H-pyrene and components 2 and 6 of Me-pyrene are strongly blended. Consequently, the derived parameters for these bands should be considered with caution. Specifically, the negative values of ${\chi}_1^"$ result in unphysical band widths. In our emission model, we therefore refrained from extrapolating the band widths beyond the measured range.}
\end{center}
\end{table*}

\section{Comparison with JWST spectra}
\label{sect:astro}

\subsection{Comparison of measured laboratory spectra with observations}
\label{Sect:res:comp_obs}

\begin{figure}[!ht]
   \centering
  \includegraphics[scale=0.6, trim={0cm 1.2cm 0cm 0.8cm}, clip]{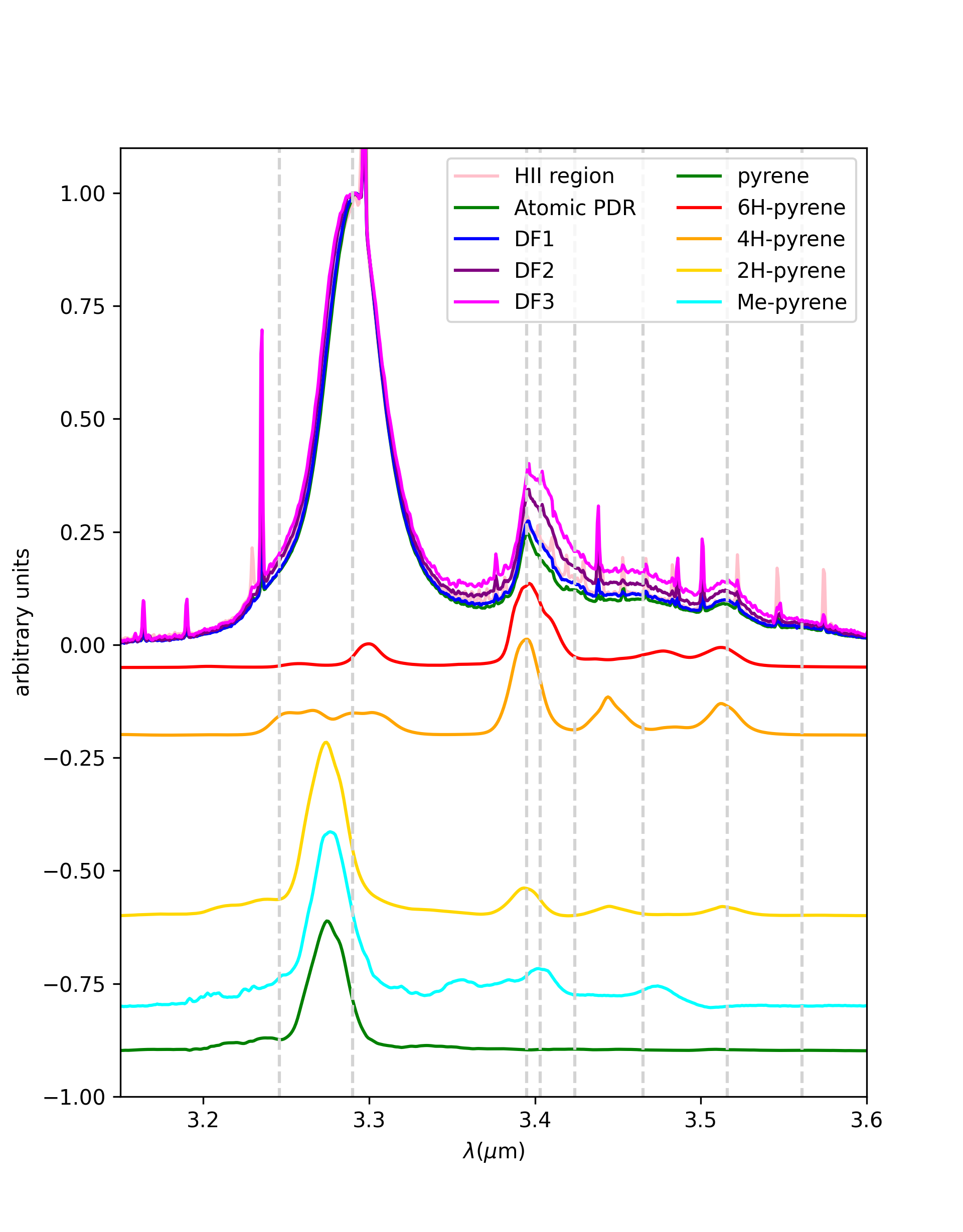}
   \caption{Comparison of the spectra of 2H-pyrene, 4H-pyrene, 6H-pyrene, and Me-pyrene (473~K; this work) with the JWST observations of the Orion PDR. The JWST spectra are the template spectra of the HII region, the atomic PDR, and the three dissociation fronts (DF1, DF2, and DF3). The dashed light grey lines indicate the observed band position from Table A.1 of \cite{chown2024}. } 
              \label{fig:astro:compallspecies}%
    \end{figure}

Figure~\ref{fig:astro:compallspecies} shows the comparison of selected experimental spectra at 473~K with Orion Bar spectra observed with the JWST in the spectral region of the CH stretching modes. We used the NIRSPEC spectra from \cite{vandeputte2025}, from which we subtracted a spline continuum. 
The laboratory spectra fall into two categories: 2H-pyrene and methyl-pyrene mainly contribute to the 3.3 $\mu$m band (similar to pyrene), with minor additional contributions to the 3.35–3.55 $\mu$m range. In contrast, more heavily hydrogenated species primarily contribute to the 3.4 $\mu$m band and may account for the 3.51 $\mu$m band, unlike methyl-pyrene, which shows no features at this wavelength. Weaker bands, which vary in position among species, may collectively contribute to the broad 3.4–3.5 $\mu$m massif or plateau. At this specific temperature, the comparison with observations suggests that the species shown in Fig.~\ref{fig:astro:compallspecies} would contribute to the blue sides of the 3.3 and 3.4~$\mu$m bands. However, this direct comparison does not account for the full temperature effects associated with  cooling cascades following UV photon absorption. Further comparison in band position and profile requires a simulation of these cascades and must take band anharmonicity into account in the simulated IR emission spectra.

\subsection{Comparison of simulated spectra with observations}
\label{Sect:res:model_obs}

We present the results of our emission model for 6H- and Me-pyrene, which are representative of the two categories explained above. The empirical anharmonicity parameters used for the modelling are provided in Table~\ref{Table:anha_coeff}. In our model, we selected the main components for which corresponding band intensities are given in the second column of Table~\ref{Table:band_intensity}.  A number of other weaker bands are expected to contribute to the emission spectrum and might contribute to the continuum and the plateau in addition to the bands. For Me-pyrene, the anharmonicity of the seven relatively weak bands between 960 and 1600 cm$^{-1}$ (listed in Table~\ref{Table:band_intensity}) has not been characterised. For these bands, we used fixed positions and widths, as specified in the first column and footnote of Table~\ref{Table:band_intensity}, respectively.  \\

  \begin{figure*}[!th]
   \centering
  \includegraphics[scale=.55, trim={0.cm 0.cm 0cm 0.cm}, clip]{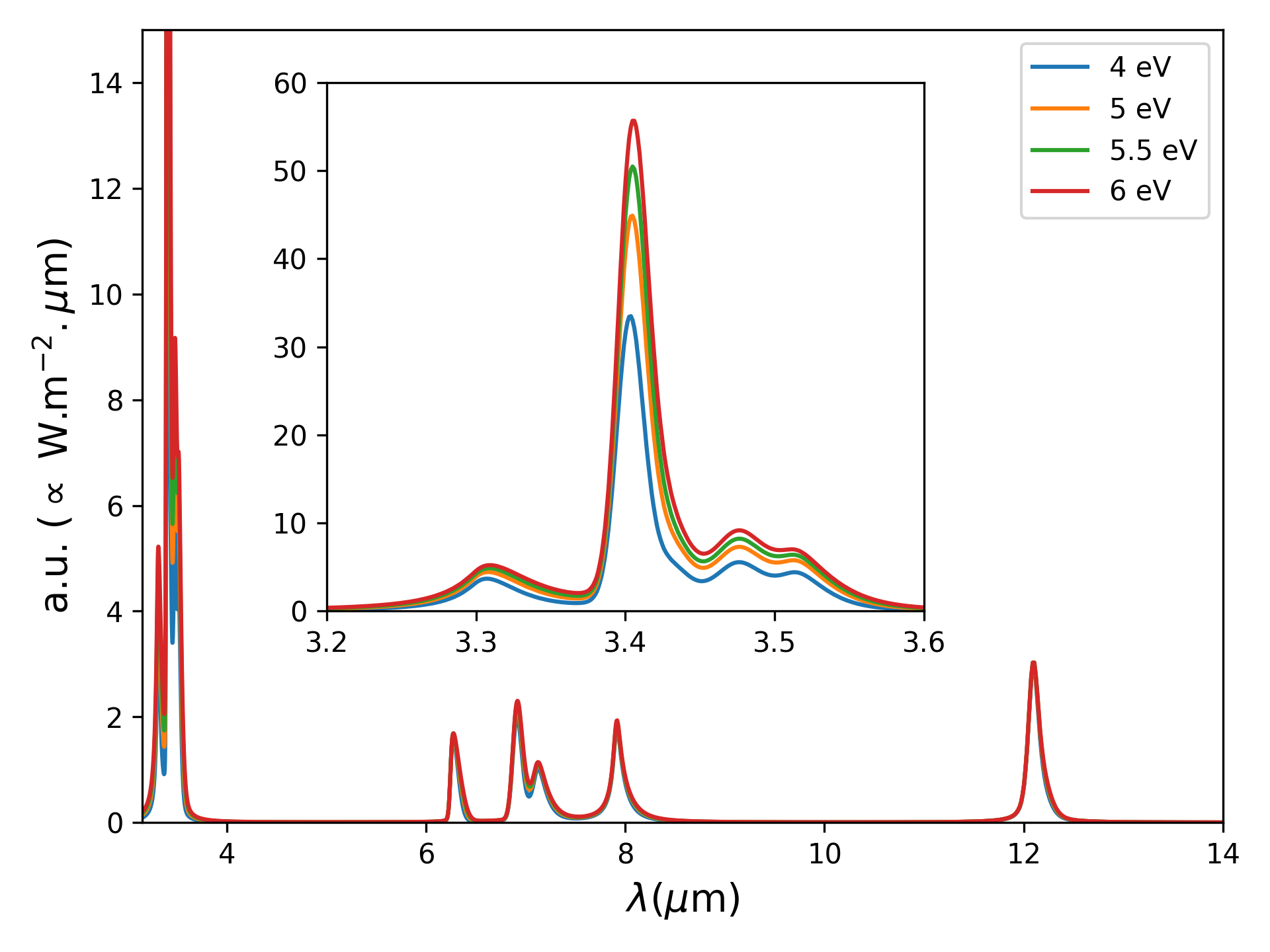}
  \includegraphics[scale=.55, trim={0.cm 0.cm 0cm 0.cm}, clip]{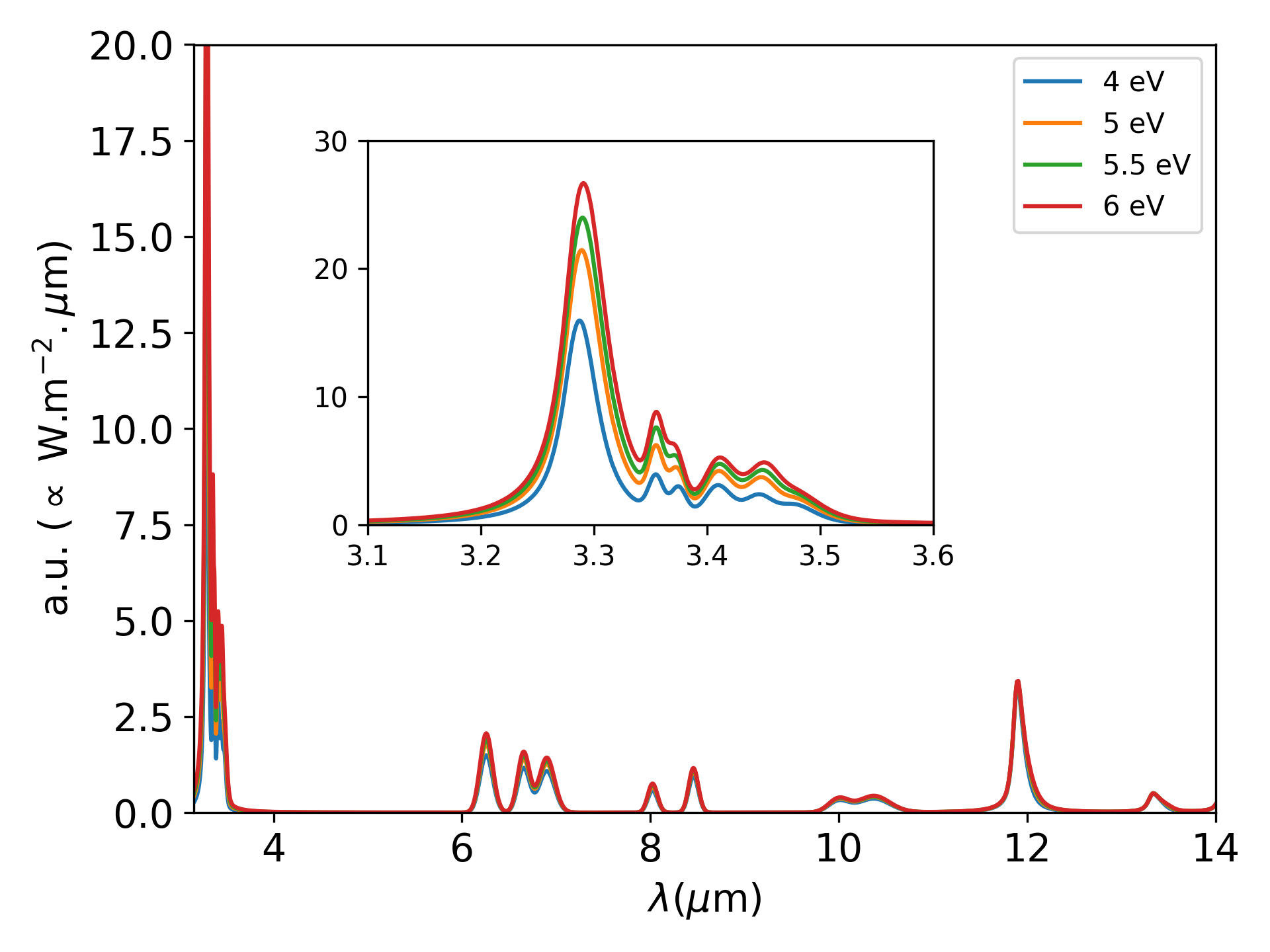}
  \caption{Emission spectrum of 6H-pyrene {\it (left panel)} and Me-pyrene {\it (right panel)}, modelled for the absorption of photons of energy of 4, 5, 5.5, and 6\,eV. }
              \label{fig:model:emission}%
    \end{figure*}

The modelling requires consideration of energies for the UV photons. Photo-absorption measurements for 6H-pyrene and Me-pyrene were recorded in Ne matrices \citep[][Joblin \& Salama, unpublished data]{halasinski2005}. In addition, laboratory experiments using UV synchrotron light showed that the 6H-pyrene cation dissociates at $\sim$5 eV \citep{stockett2021}, while the Me-pyrene cation dissociates at $\sim$6 eV \citep{west2018}; we assumed similar dissociation energies for their neutral counterparts. The photo-absorption spectrum of 6H-pyrene exhibits a strong vibronic band between 5.5 and 5.7~eV (expected to lead to dissociation) and an about ten times weaker band between 3.8 and 4.5~eV. For Me-pyrene, there are two bands in the 3.7-4.8~eV range and a strong band between 5.1 and 5.5~eV (which does not lead to dissociation). Time-dependent DFT (TD-DFT) calculations suggest that this band contains two transitions (see the calculated electronic spectra in the \href{https://CosmicPAH-Qcals.oa-cagliari.inaf.it/}{Theoretical spectral database of Polycyclic Aromatic Hydrocarbons}). We therefore simulated emission spectra for absorbed photon energies of 4, 5, 5.5, and 6~eV. For comparison with astronomical spectra, we considered synthetic spectra corresponding to excitation at energies of 4~eV for 6H-pyrene and 6~eV for Me-pyrene.

Figure~\ref{fig:model:emission} shows that the bands at short wavelengths are affected most by the energy of the absorbed UV photon. Their intensity decreases significantly with decreasing photon energy, whereas the intensity of the bands above 10~$\mu$m remains unchanged. This is related to the fact that the emission in the 3\,$\mu$m range occurs at the highest temperatures during the cooling cascade \citep[see Fig.~3 in][]{mulas2006}.

\begin{figure*}[!t]
   \centering
  \includegraphics[scale=.37, trim={0.2cm 0.cm 0cm 0.cm}, clip]{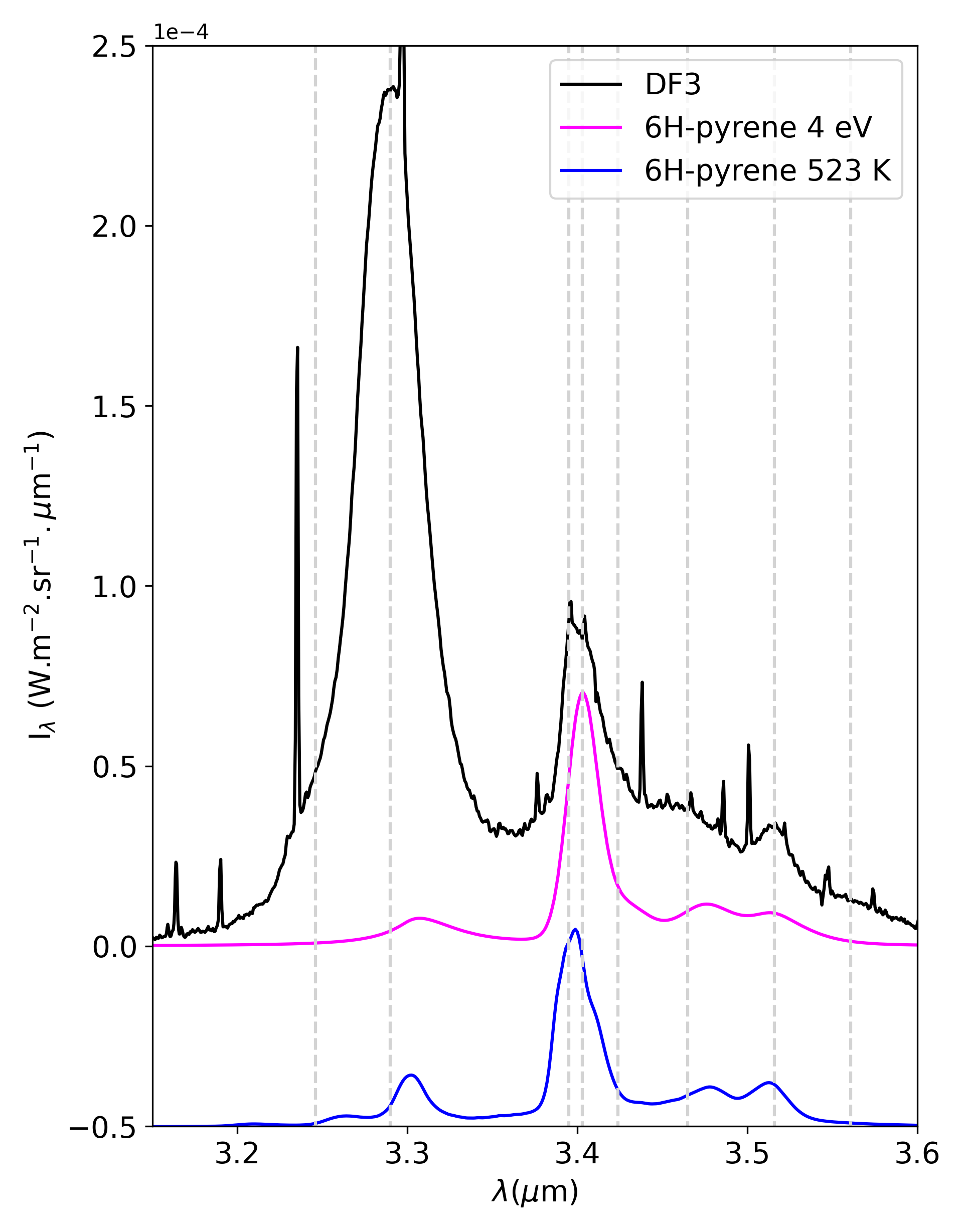}
   \includegraphics[scale=.37, trim={0.2cm 0.cm 0cm 0.cm}, clip]{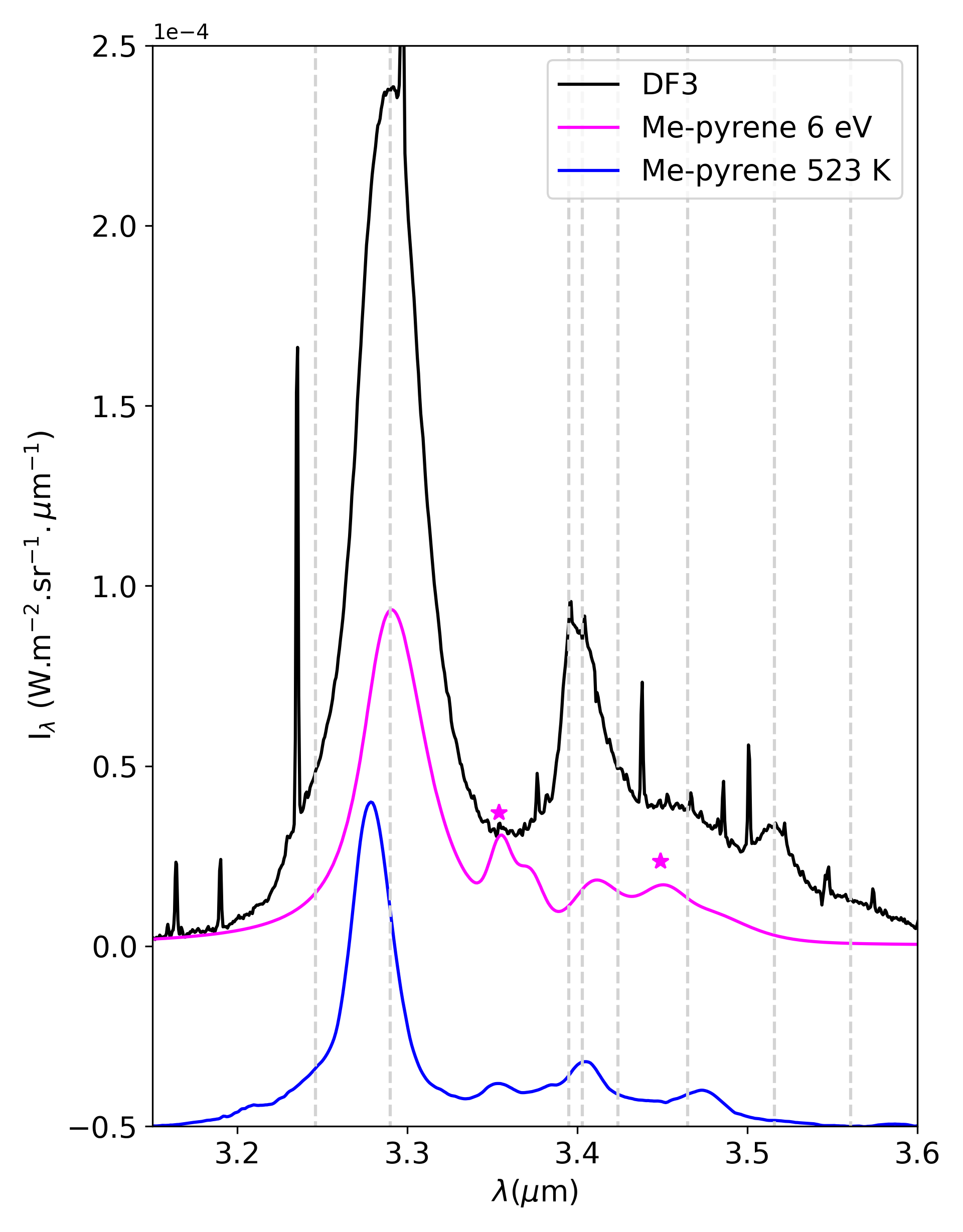}
  \includegraphics[scale=0.37, trim={0.2cm 0.cm 0cm 0.cm}, clip]{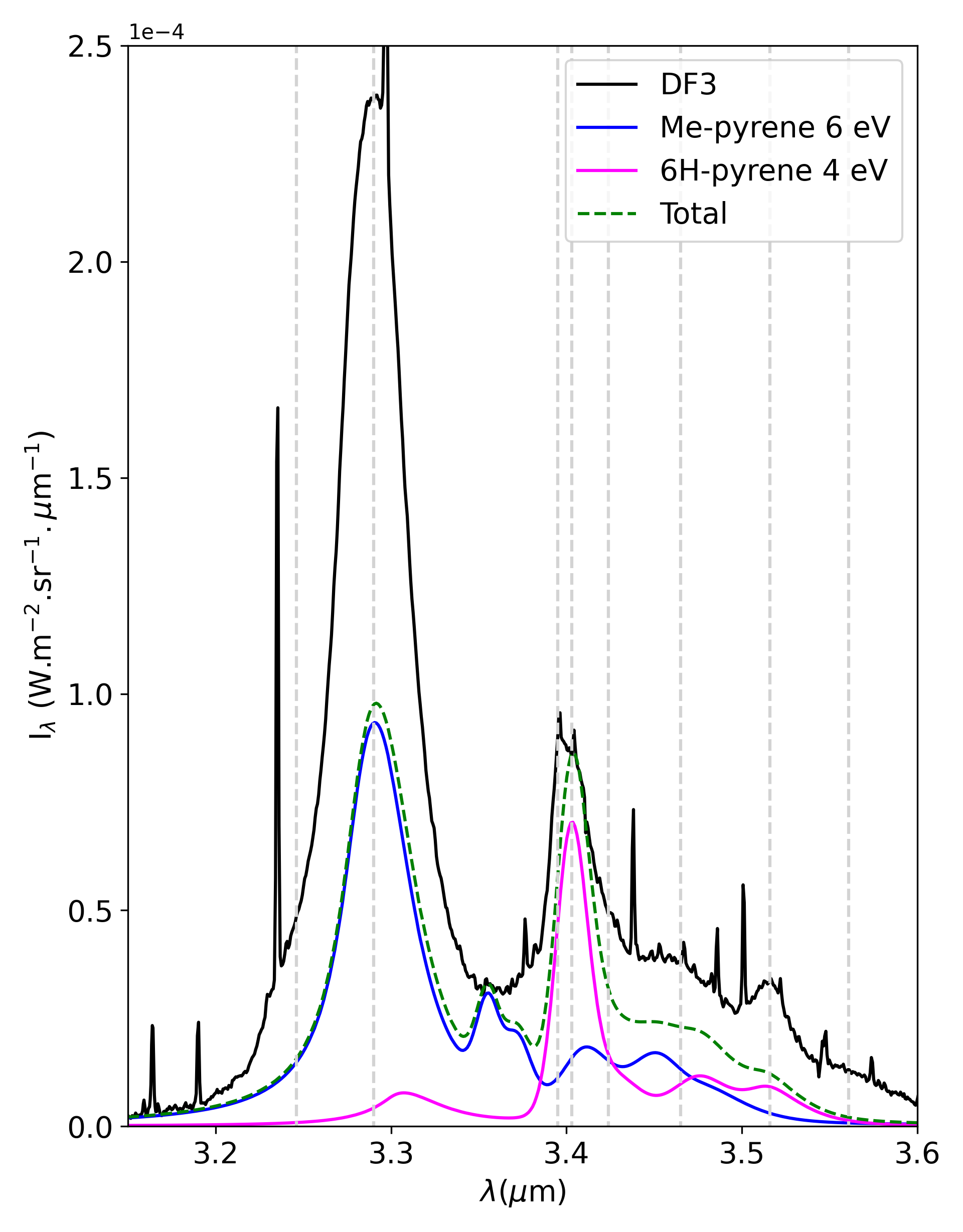}
 \caption{Comparison of the JWST observations of the Orion PDR \citep[DF3;][]{vandeputte2025} with experimental and modelled spectra of 6H-pyrene and Me-pyrene (this work). {\it Left panel:} Emission spectrum of 6H-pyrene modelled after the absorption of a 4~eV UV photon and arbitrary scaling by 2.1$\times10^{-6}$. {\it Middle panel:} Emission spectrum of Me-pyrene modelled after the absorption of a 6~eV UV photon and arbitrary scaling by 3.5$\times10^{-6}$.  {\it Right panel :} Comparison of the observations with the sum of the modelled 6H- and Me-pyrene. The experimental spectra are those measured at 523 K. The bands marked with an asterisk are not well modelled (see text; see Fig.~\ref{fig:astro:comp6hpyr_all} for the comparison on the spectral range from 3 to 14\,$\mu$m).}
              \label{fig:astro:emission}
    \end{figure*}

The comparison of the simulated spectra with the observed spectrum at the dissociation front DF3 of the Orion Bar is presented in Fig.~\ref{fig:astro:emission} for the CH stretch region and in Fig.~\ref{fig:astro:comp6hpyr_all} for longer wavelengths. In the CH stretch region, the two strongest bands are the CH$_{\rm{ali}}$ band of 6H-pyrene at 3.4\,$\mu$m  and the CH$_{\rm{arom}}$ band of Me-pyrene at 3.3\,$\mu$m. In both cases, the simulated spectra appear to be more consistent with the observed 3.3 and 3.4\,$\mu$m bands with respect to the laboratory spectra measured at 523~K (Fig.~\ref{fig:astro:emission}). This shows the importance of an accurate cooling model that incorporates anharmonic effects. The bands at longer wavelengths are far weaker, and most of them are undetectable in the observations. The bands at $\sim$ 10 and 14.1\,$\mu$m of Me-pyrene might be detected in principle because they fall outside strong AIBs. They are not observed in the DF3 spectrum, and they thus constrain the maximum amount of Me-pyrene that could be present in this object. The bands at 6.68\,$\mu$m (Me-pyrene) and at 6.9\,$\mu$m (Me- and 6H-pyrene) might correspond to two weak bands in the observations. The band at 6.9 $\mu$m appears as a potential tracer of aliphatic CH bonds. However, it is complicated to interpret the 6.9/3.3 $\mu$m ratio in observations because it depends on the excitation conditions, such as the energy of the absorbed UV photon. Based on our simulated spectra, we determined values of 1.08 and 0.19 for the 6.9/3.3 $\mu$m integrated band intensity ratio of 6H-pyrene (4 eV) and Me-pyrene (6 eV), respectively. \\

 \begin{figure}[!t]
   \centering
  \includegraphics[width=\hsize, trim={0.0cm 0.cm 0cm 0.cm}, clip]{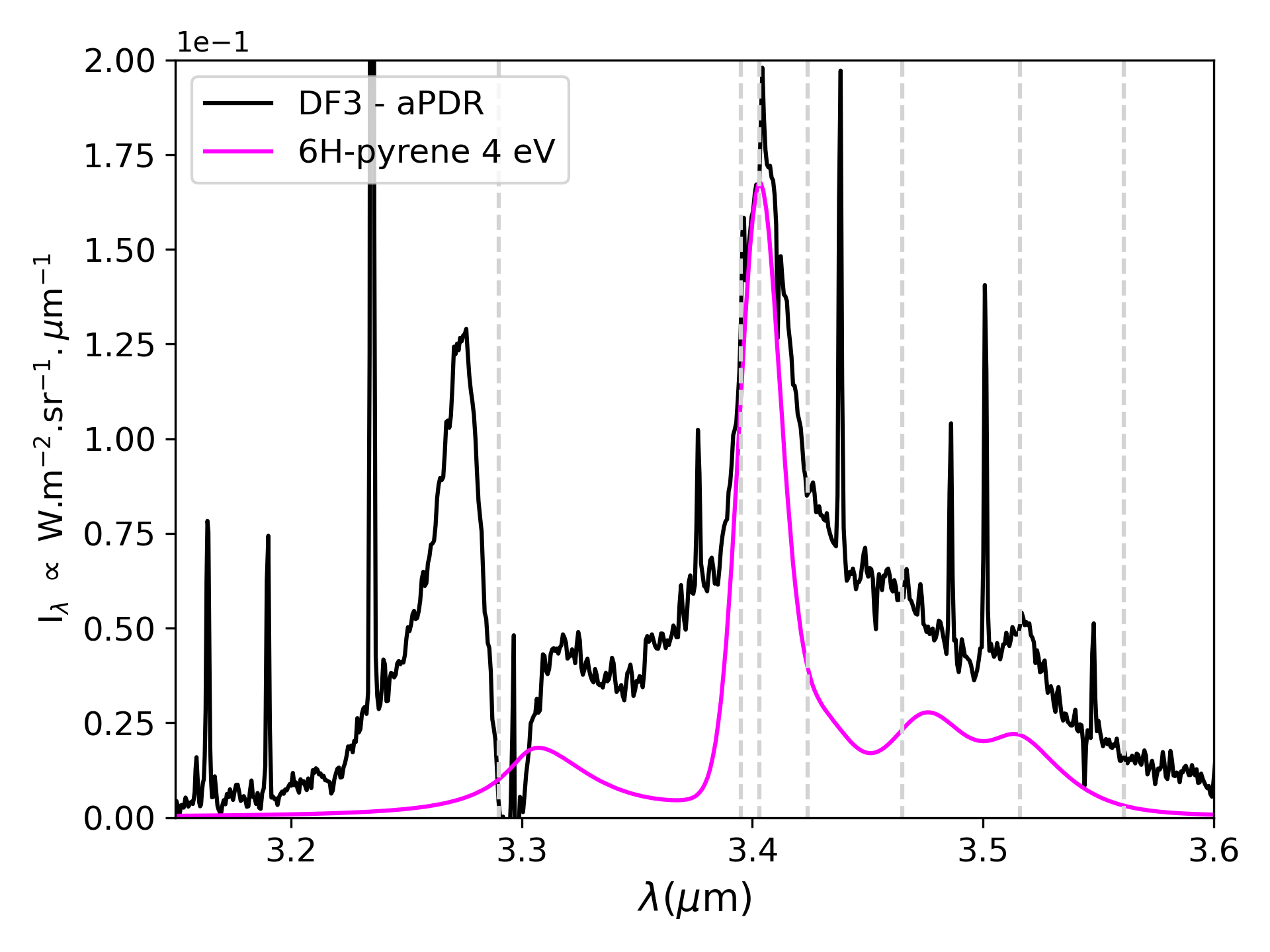}
   \includegraphics[width=\hsize, trim={0.0cm 0.cm 0cm 0.cm}, clip]{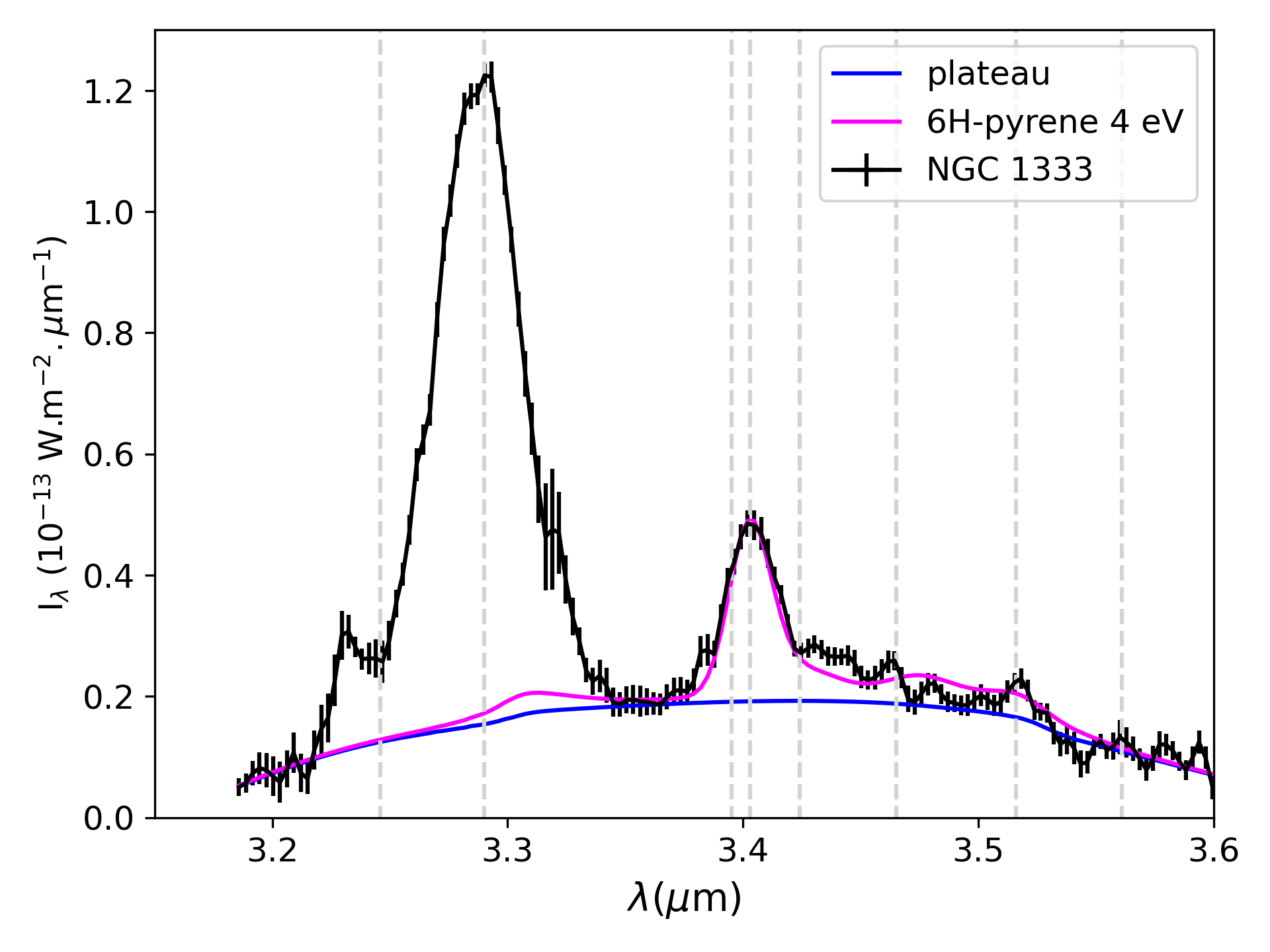}
\caption{Comparison of the simulated emission spectrum of 6H-pyrene upon 4~eV photon excitation (magenta lines) with observations. {\it Upper panel:} Comparison with the Orion PDR observed with the JWST \citep{peeters2024,vandeputte2025}; the simulated spectrum is compared with the difference of the normalised DF3 and aPDR spectra. {\it Lower panel:} Comparisons with NGC~1333  ground-based observations \citep{joblin1996b}; the simulated spectrum is represented on the top of the continuum and plateau defined in Fig.\,3 of \cite{joblin1996b}.}
\label{fig:astro2:emission}
    \end{figure}

\subsection{Discussion of the accuracy of the method}
\label{Sect:res:discuss}

We significantly enhanced the quality of synthetic PAH spectra for comparison with observational data. This improvement primarily relied on the incorporation of empirical anharmonicity functions, which quantify the evolution of band positions and widths as a function of temperature. These functions were derived from the analysis of gas-phase experimental spectra, which are the most detailed data obtained to date.  Our method built upon pioneering studies \citep{joblin1995, pech2002}, with advancements in band analysis, extraction of anharmonicity parameters, and data dissemination through the CosmicPAH-IR database. However, several factors remain inherently limiting under our experimental conditions and cannot be overcome. The first is the temperature range over which the study can be conducted. This range is constrained by the vapour pressure of the studied species (PAHs are refractory compounds), their thermal stability (as evidenced in this study by 4H- and 2H-pyrene), and the thermal resistance of the oven components (up to $\sim$900~K, but with a risk of damaging the diamond window seals). While \cite{farouki2025} recently demonstrated that a new experimental setup can achieve slightly higher temperatures, the current limitations still necessitate extrapolating the anharmonicity functions towards the high temperatures involved in the emission of the 3\,$\mu$m band (see Fig.~\ref{fig:T_vsU}). Linear extrapolations are consistent with molecular dynamics simulations \citep{chakraborty2021}.
A similar question arises for low temperatures, although this has a very minor impact here (Fig.~\ref{fig:histo_T}). In our simulations, we therefore avoided extrapolation at low temperatures and instead used the band positions and widths corresponding to the lowest measured temperature.  For the band positions, this approach is indeed consistent with measurements at low temperature reported in our previous study of pyrene \citep{chakraborty2019}.

The band analysis also faces several complications. Under our experimental conditions, the rotational temperature is in equilibrium with the vibrational temperature, unlike in space. This necessitates subtracting the contribution of rotational broadening from the band widths. While we employed a simplified method for this purpose, more sophisticated approaches \citep[e.g. \textsc{PGOPHER}, ][]{Western2015} 
could be explored in future work. Another challenge in band analysis involves the selection of parameters for the multi-component fit. The fit results depend on user-defined conditions, particularly the number of components included. This aspect will be discussed in detail in a separate work \citep{debentzmann2026}. Our approach prioritizes methodological consistency to optimise comparisons between different species. \\

\section{Astrophysical implications}
\label{Sect:res:astro}

Laboratory studies \citep{bernstein1996,joblin1996b,steglich2013, sandford2013, maltseva2018} and calculations, both harmonic and anharmonic  \citep{yang2013, mackie2018, yang2020}, demonstrated that hydrogenated and methylated PAHs can contribute to the 3.4\,$\mu$m band. Observationally, \cite{sloan1997} first noted that the 3.4\,$\mu$m band consists of two components at 3.395 and 3.405\,$\mu$m, with the latter spatially correlating with H$_2$ emission. This was recently confirmed by PDRs4All observations \citep{peeters2024}, which refined the positions of the two main components to 3.395 and 3.403\,$\mu$m and identified an additional weaker component at 3.424\,$\mu$m. Our study shows that methylated pyrene does not provide a major contribution to the 3.4 $\mu$m band, 
whereas 6H-pyrene can account for the observed 3.403~$\mu$m band at an excitation energy of $\sim$4~eV, which balances photo-absorption and photo-stability (Sect.\,\ref{Sect:res:model_obs}). For a detailed spectral assignment, we present in Fig.\,\ref{fig:astro2:emission} additional comparisons with astronomical spectra. The top spectrum is a difference spectrum between the DF3 and atomic positions in Orion, which isolates a cleaner 3.403~$\mu$m band. The bottom spectrum, selected from the literature, corresponds to ground-based observations of NGC~1333 \citep{joblin1996b}. It has the peculiarity of containing only the 3.403\,$\mu$m component, which provides further confidence that this emission feature is independent of the 3.395\,$\mu$m component. Moreover, it demonstrates that our modelled emission spectrum matches the observations with high accuracy. We note that detailed modelling does not yield any bands that could confirm or refute this assignment when comparing with Orion Bar observations.

This finding reshapes our understanding of the use of the 3.4~$\mu$m band in the analysis of astronomical observations. First, it is important to note that only the 3.403\,$\mu$m component is confirmed to be linked to aliphatic content, while the origin of the 3.395\,$\mu$m component remains unidentified. Consequently, using the 3.4-to-3.3~$\mu$m band ratio to determine the superhydrogenation degree of AIB carriers (as done in previous studies) is necessarily approximate. For example, \cite{yang2020} derived a superhydrogenation degree of $\leq$2\% for neutral species, which are expected to dominate the emission in the 3~$\mu$m range. Additionally, the 3.403\,$\mu$m band is associated with shielded regions \citep{schroetter2024,thatte2026}, which aligns with the need to minimise the destruction of fragile aliphatic bonds. In contrast, the 3.3\,$\mu$m band may be dominated by contributions from regions with higher-energy UV photons. Regular PAHs, including pyrene, can indeed withstand higher excitation conditions, with their VUV excitation energy threshold generally increasing with size \citep{zhen2016, west2018, wenzel2020}. Thus, using the 3.4-to-3.3~$\mu$m band ratio as a tracer of the aliphatic content of PAHs can result in a rough approximation.


\cite{bernstein1996} studied the spectroscopy of several superhydrogenated PAHs, ranging from naphthalene-like (C$_{10}$H$_{10}$, C$_{10}$H$_{12}$, C$_{10}$H$_{16}$) to coronene-like (C$_{24}$H$_{24}$) species in Ar matrices at 12\,K. Based on these measurements, the authors concluded that 6H-pyrene provided the best match to the 3.4~$\mu$m band in the Orion Bar. Our results advance this conclusion by providing a detailed match to the observed band profile. Using the derived anharmonicity function, we predict that the band of 4H-pyrene (although it was not modelled here) lies between the two main components of the observed 3.4~$\mu$m band. While it remains to be demonstrated whether 6H-pyrene is the only candidate satisfying spectral constrains, its robustness is supported by excitation conditions: in regions that are only penetrated by mild UV photons, 6H-pyrene emission can be triggered by efficient absorption at $\sim$4\,eV in a non-dissociative regime. Pyrene is expected to be abundant in the gas-phase in the dark cold molecular cloud TMC-1, as evidenced by the detection of its cyano derivatives \citep{wenzel2024a,wenzel2024b}. In PDRs, 6H-pyrene could be a direct product of the evaporation of eVSGs \citep{pilleri2015}, with further gas-phase chemistry also possible. However, the dissociation of 6H-pyrene is not expected to produce pyrene derivatives that could reform 6H-pyrene via hydrogenation because 6H-pyrene, or at least its cation, dissociates by losing hydrocarbons \citep{marciniak2021, stockett2021}. Thus, if exposed to dissociative photons, 6H-pyrene is unlikely to be reformed through gas-phase chemistry. Its IR emission would therefore only be observed in regions in which the conditions are optimised for both its production and its survival.

\section{Conclusions}
\label{Sect:concl}

We recorded IR spectra of gas-phase hydrogenated and methylated pyrene species in the temperature range 373-673~K. This allowed us to determine empirical anharmonicity functions that quantify the evolution of the band positions and widths with temperature. These spectral functions were incorporated into our emission model to derive emission spectra at internal energies compatible with the photophysics of these species. This approach is expected to yield more accurate simulated emission spectra than theoretical anharmonic IR methods, which necessarily rely on approximations to address the complexity of the problem \citep{chen2018,mackie2021,mackie2022}. The comparison with observed spectra of PDRs such as the Orion Bar and NGC~1333 showed that
\begin{itemize}
    \item the emission of 6H-pyrene might account for the 3.403\,$\mu$m component and contribute to the 3.3--3.6\,$\mu$m plateau,
    \item methyl-pyrene is also expected to contribute to the plateau, with its main contribution at 3.3\,$\mu$m;
    \item in both cases, when the modelled spectra were scaled to match the observed spectra in the C-H stretch spectral region, no strong unaccounted-for features appeared elsewhere.
\end{itemize}

The 3.4\,$\mu$m band is a common component in the AIB spectrum, and this work has implications for the use of this band in current JWST observations as a tracer of physical and chemical conditions in a range of environments, from galactic PDRs \citep{peeters2024,misselt2025} to protoplanetary disks \citep{bouteraon2019} and galaxies in the near- and far-Universe \citep{mentzer2026,thatte2026}.

\section*{Data availability}
Data from this article are publicly-available through the \href{https://cosmicpah-irdb.irap.omp.eu/}{CosmicPAH-IRDB database} (https://cosmicpah-irdb.irap.omp.eu/) and the \href{https://CosmicPAH-Qcals.oa-cagliari.inaf.it/}{Theoretical spectral database of Polycyclic Aromatic Hydrocarbons} (https://CosmicPAH-Qcals.oa-cagliari.inaf.it/). The \textit{cosmicPAHmfit} tool can be downloaded via the \href{https://cosmic-pah.irap.omp.eu}{Cosmic PAH portal}  (https://cosmic-pah.irap.omp.eu) tool tab. 
 
\begin{acknowledgements}
 This work was supported by the European Research Council under the European Union’s Seventh Framework Programme (FP/2007-2013) ERC-2013-SyG, Grant agreement N$^{o}$610256 NANOCOSMOS, the CNES (APR LAIBrary/JWST), the Université of Toulouse, the Thematic Action “Physique et Chimie du Milieu Interstellaire” (PCMI) of the INSU Programme National “Astro”, with contributions from CNRS Physique, CNRS Chimie, CEA, and CNES.  The authors gratefully acknowledge Anthony Bonnamy and Loïc Noguès for their technical support on the ESPOIRS setup of the Nanograin platform. The authors acknowledge support from the Observatoire Virtuel Grand Sud Ouest (OVGSO, https://https://ov-gso.irap.omp.eu/) funded by INSU and Observatoire Midi-Pyrénées (OMP).

\end{acknowledgements}

\bibliographystyle{aa}
\bibliography{bib.bib}

\begin{appendix}
\nolinenumbers

\section{Correction for rotational broadening} 
\label{sect:rot_broadening}

\begin{figure}[!h]
   \centering
  \includegraphics[scale=0.75, trim={1.5cm 0.3cm 0.5cm 0.3cm}, clip]{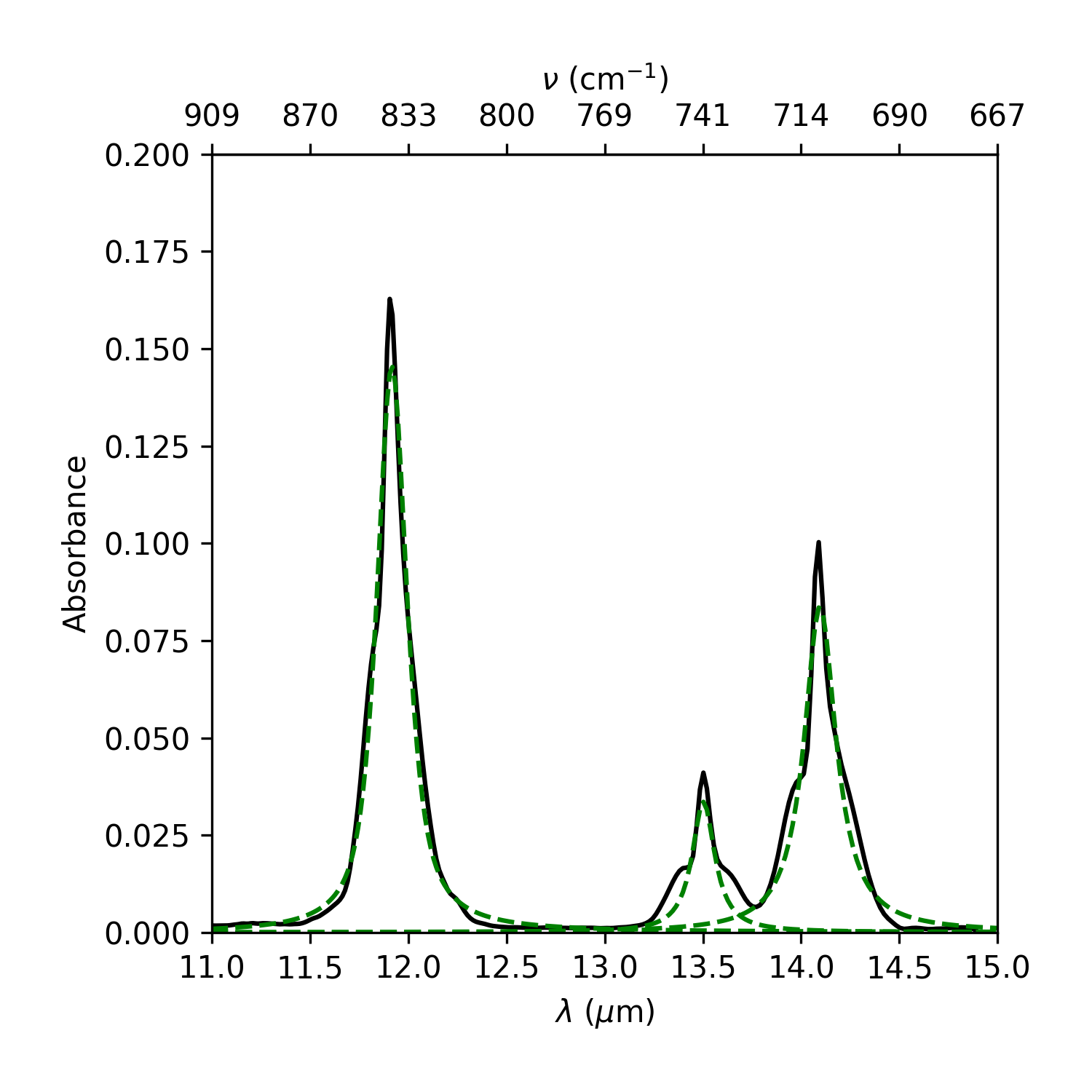}
  \caption{Absorption spectrum of pyrene at 473~K. The three bands, which show a rotational structure with three branches similar to the P, Q and R branches, with the Q branch being the middle component, are fitted with only one components.  }
              \label{fig:rot_broadening_1comp}%
    \end{figure}

\begin{figure}[!h]
   \centering
  \includegraphics[scale=0.75, trim={1.5cm 0.3cm 0.5cm 0.3cm}, clip]{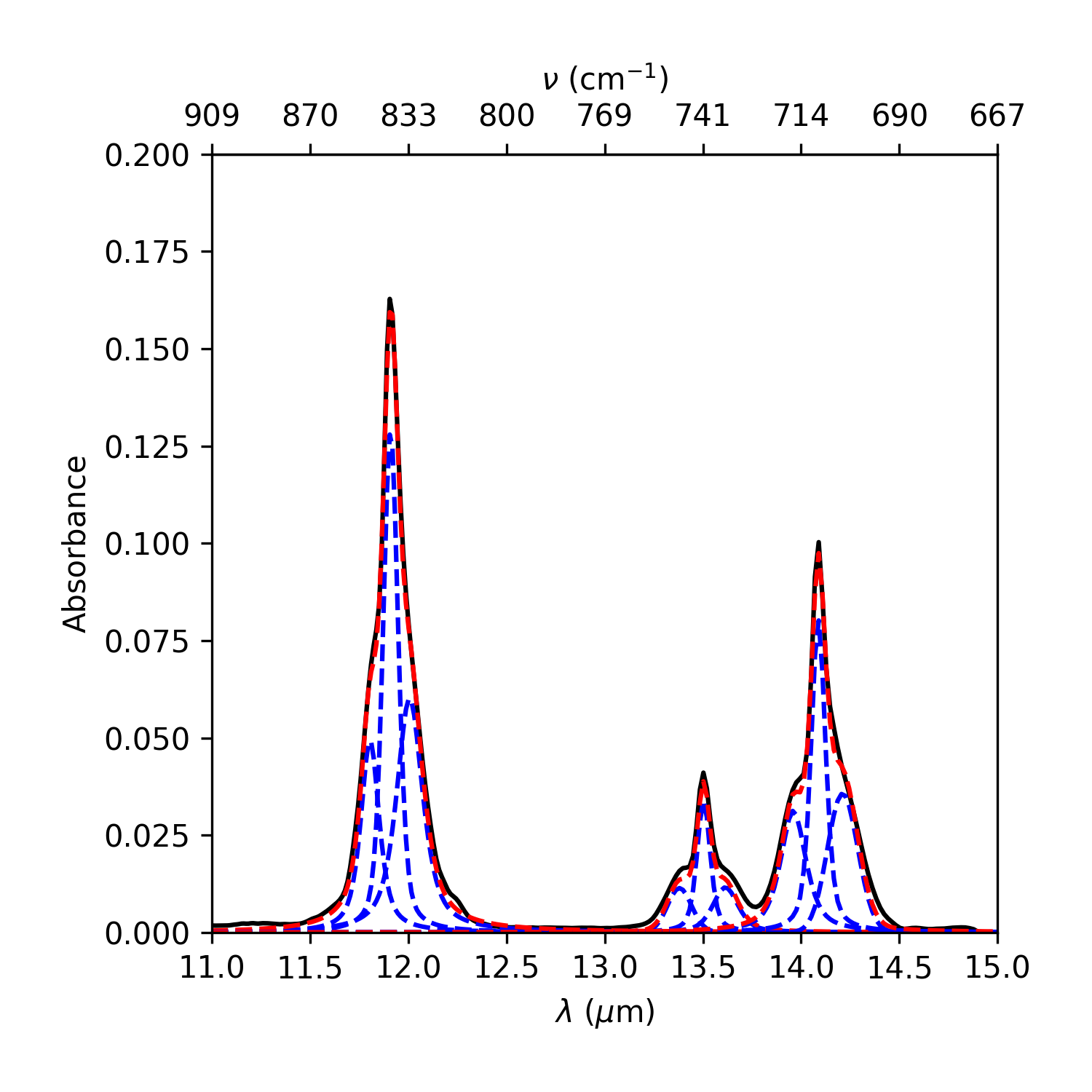}
   \caption{Absorption spectrum of pyrene at 473~K. The three bands, which show a rotational structure with three branches similar to the P, Q and R branches, with the Q branch being the middle component, are fitted with three components. }
              \label{fig:rot_broadening_3comp}%
    \end{figure}

In the experimental conditions of this study, the molecules are hot and the bands are broadened by rotation. For the determination of the bandwidth anharmonicity parameters, we have corrected the bands for the rotational broadening. This was estimated from bands at low frequency for which a rotational structure with the three P-Q-R branches is observed. These bands were fitted first with a single component (Fig.~\ref{fig:rot_broadening_1comp}) and second with three sub-components (Fig.~\ref{fig:rot_broadening_3comp}) corresponding to the P-Q-R branches . The rotational broadening is defined as the difference between the FWHM of the one-component fit and that of the Q-branch, i.e. the middle components. The rotational broadening was estimated at each temperature for the three pyrene bands at 710, 740 and 840~ cm$^{-1}$, and then averaged. The values are 5.9, 6.1, 6.7, 6.6, .6.5, 6.9~cm$^{-1}$ for 423, 473, 523, 573, 623, 673~K, respectively, close to the theoretical estimation of 4.5\,-\,5.8.~ cm$^{-1}$ for the 840~cm$^{-1}$ band of pyrene \citep{chakraborty2021}.  These rotational broadening values were used to correct all pyrene, 6H-pyrene and methyl-pyrene bands for which the P-Q-R branches are not observed. Where the P-Q-R branch is visible, the Q branch adjustment is adopted as the rotationally corrected band.

\newpage

\section{Emission model} 

\begin{figure}[!h]
\centering
\includegraphics[scale=.55]{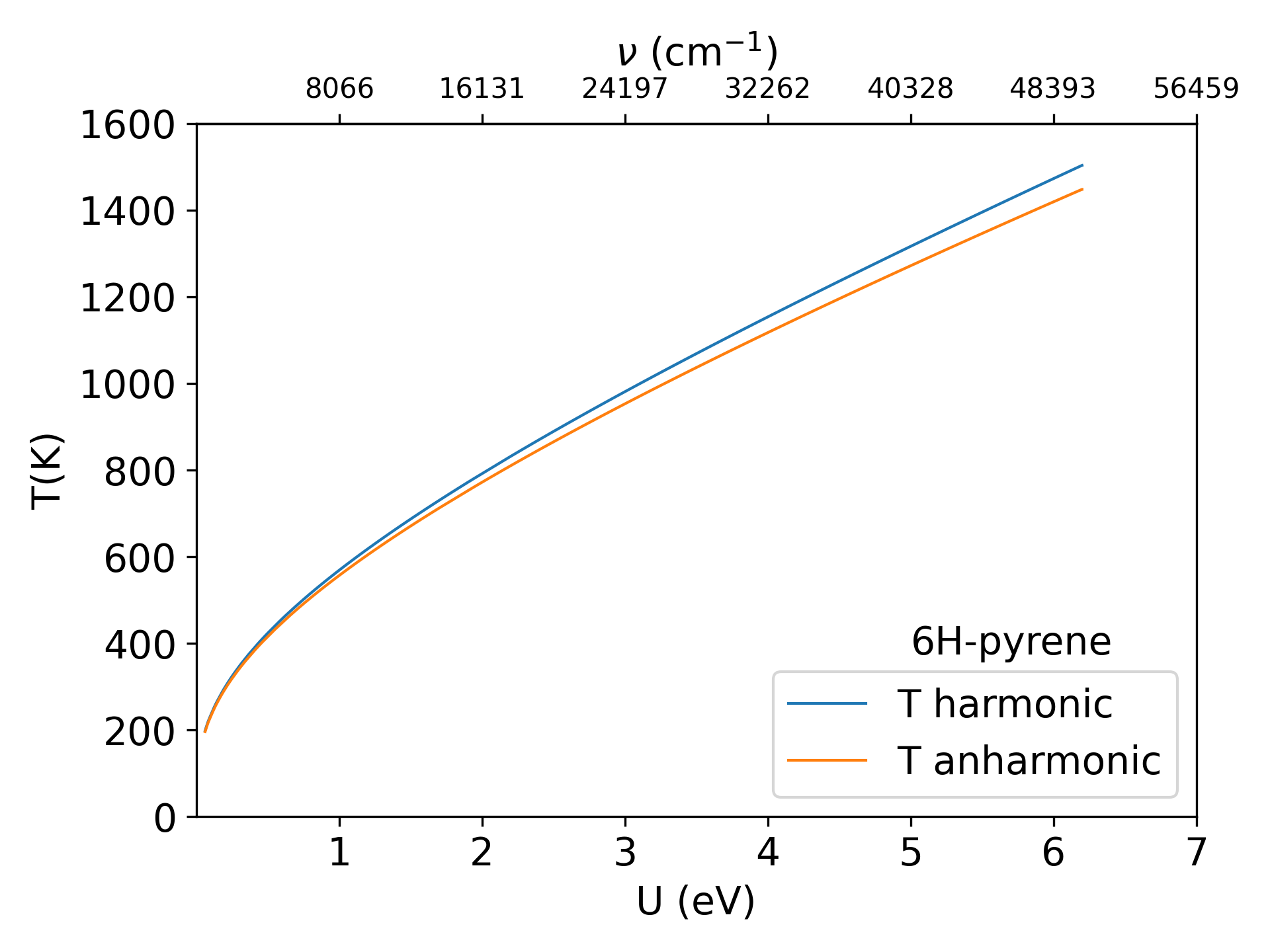}
\caption{ Correspondence between the internal energy U and the microcanonical temperature T, derived using harmonic and anharmonic state counts to compute the density of states, as described in Sect.~\ref{sect:model}.}
\label{fig:T_vsU}
\end{figure}

\begin{figure}[!h]
\centering
\includegraphics[scale=.5]{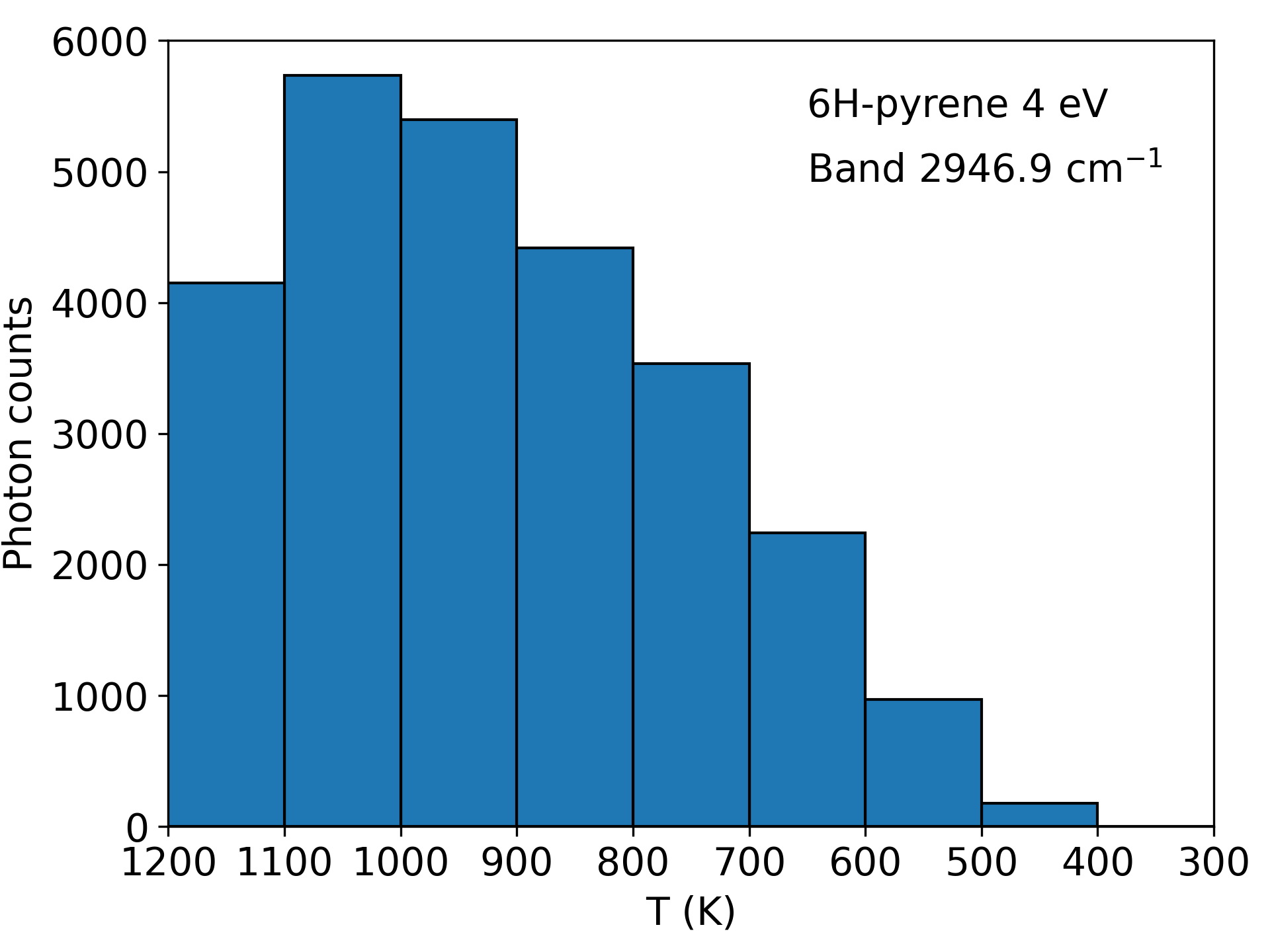}
\caption{ Temperature distribution of the photons emitted by 6H-pyrene in the 3.4~$\mu$m band after the absorption of a 4 eV photon.}
\label{fig:histo_T}
\end{figure}

\FloatBarrier
\section{ Experimental spectra in the 5-12 m range} 
\label{sect:MIR_spec}

\begin{figure}[h]
   \begin{center}
   \includegraphics[scale=.32, trim={1.5cm 0.3cm 0.5cm 0.3cm}, clip]{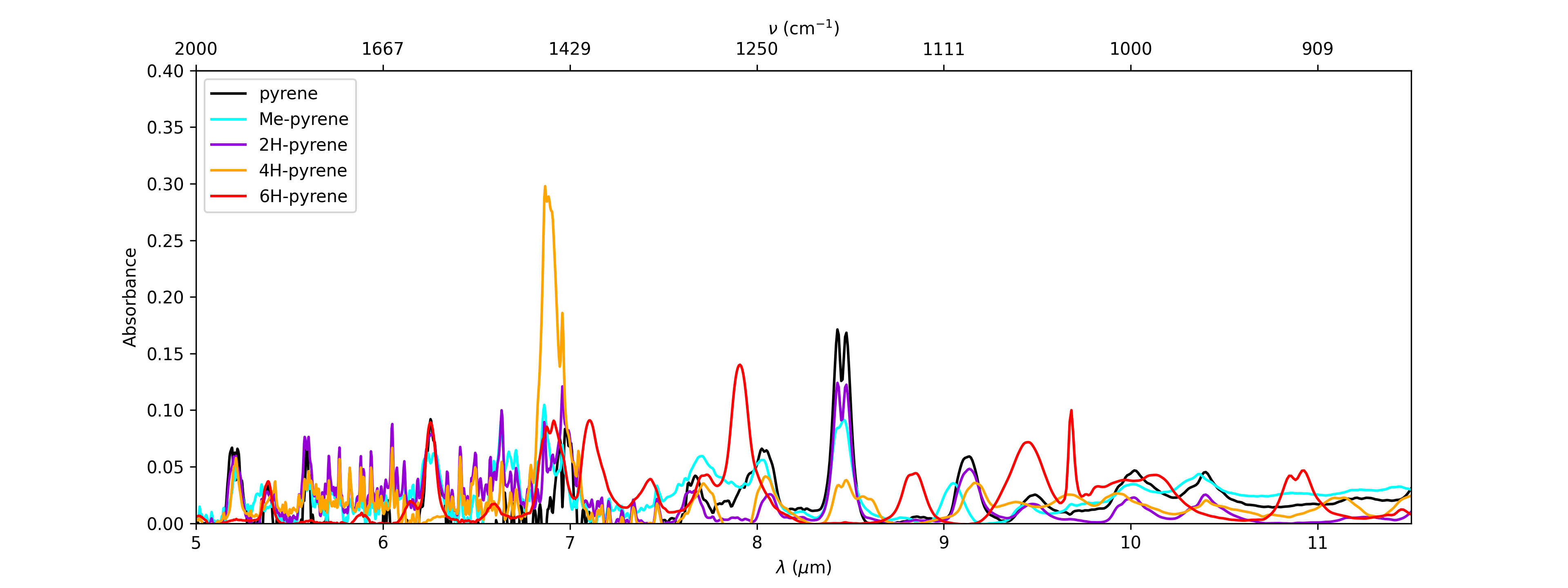}
  \caption{Comparison of the IR spectrum of pyrene, hydrogenated and methylated pyrene in gas phase, in the 5 -- 11.5$\mu$m range of the CH out-of-plane bending modes. For comparison within each others the spectra have been first normalised to their total area and second such that the intensity of the CH aromatic stretching band of pyrene is equal to 1. As the relative intensity of the CH$_{\rm{arom}}$ and CH$_{\rm{ali}}$ stretching bands varies with the temperature (see Sect~\ref{Sect:res:T_evol}), we present here the spectra measured at 473~K for all the species. }
   \label{fig:comp_all_species_5_11mic}%
  \end{center}
    \end{figure}

\section{Anharmonic theoretical calculations} 

\begin{figure}[!h]
\centering
\includegraphics[scale=.5]{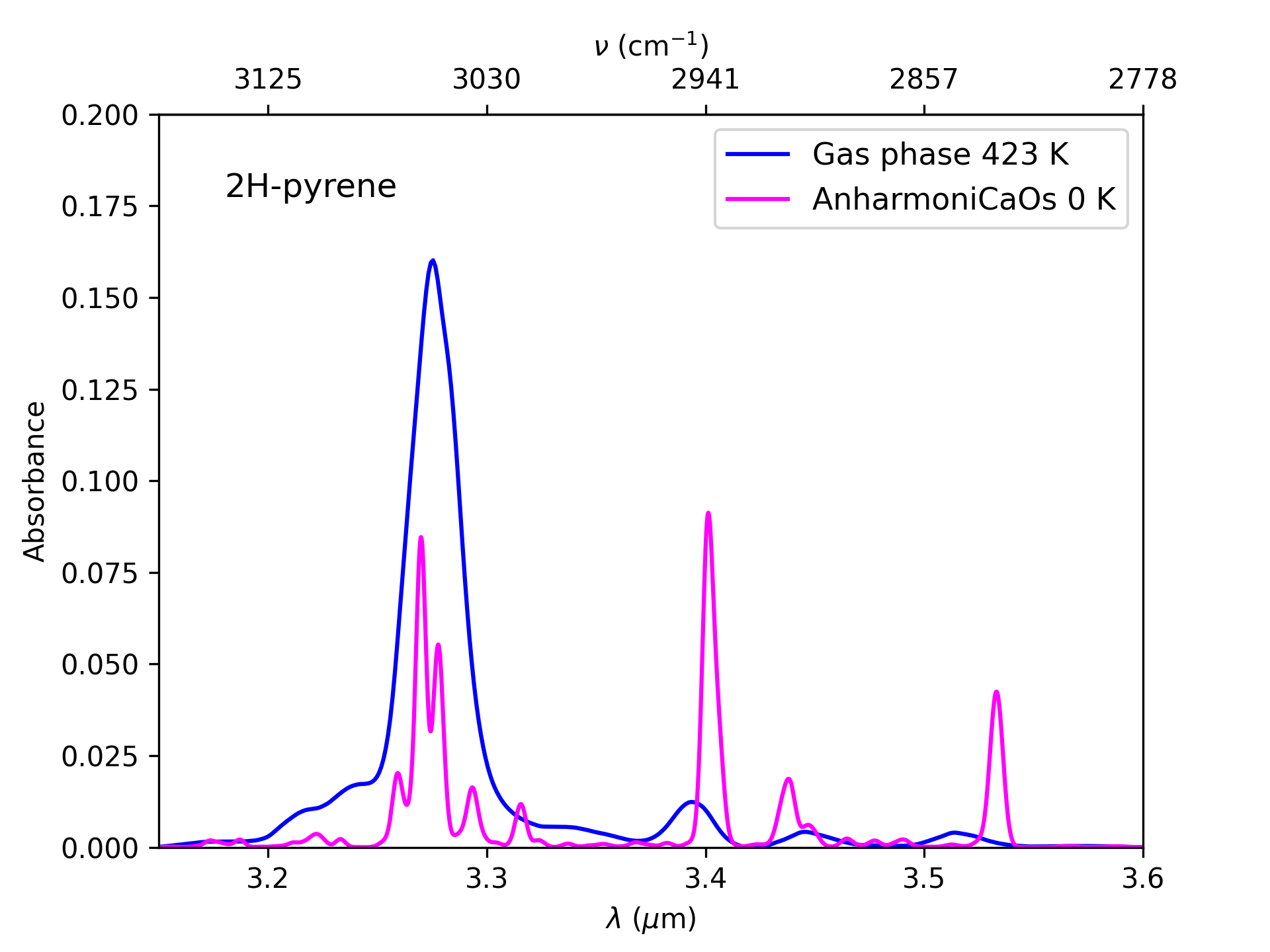}
\includegraphics[scale=.5]{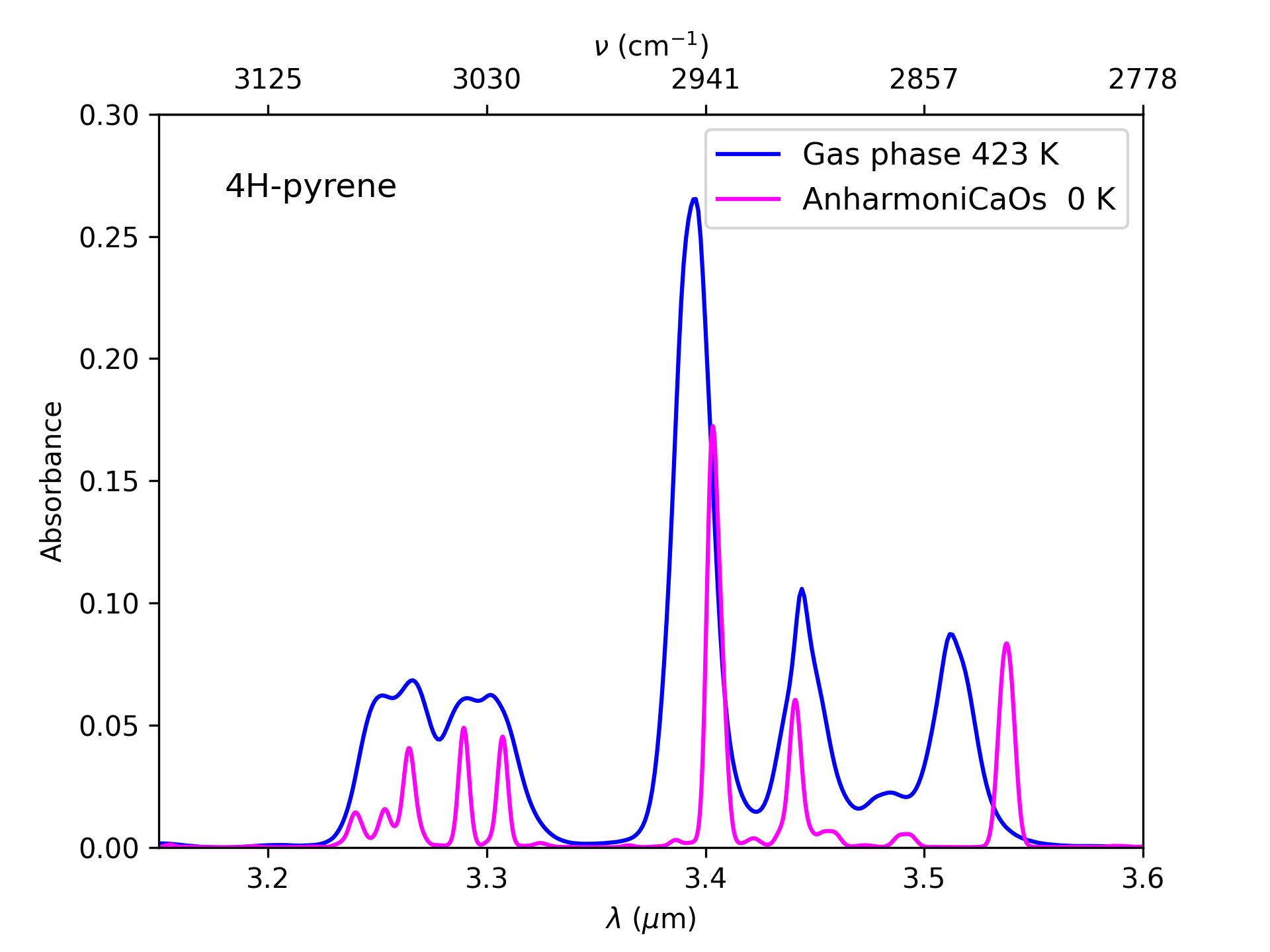}
\caption{Comparison of gas phase spectra (blue lines) with theoretical anharmonic spectra (magenta lines) for 2H-pyrene (left panel) and 4H-pyrene (right panel). The theoretical spectra are unscaled and have been convolved with a Gaussian profiles with a FWHM of 5 cm$^{-1}$. A numerical version of these figures with assignments is available on the comsicPAH-IRDB at this \href{https://cosmicpah-irdb.irap.omp.eu/science/4,5-dihydropyrene/Gas/9/sample/?overplot-tab\#/3/10/5/2/3140/2800/1200/500/423}{link} for 2H-pyrene and at this \href{https://cosmicpah-irdb.irap.omp.eu/science/4,5,9,10-tetrahydropyrene/Gas/12/sample/?overplot-tab\#/3/13/5/2/3140/2800/1200/500/423}{link} for 4H-pyrene. }
\label{fig:3microns_exp_vs_theo_2H_4H}%
\end{figure}

\begin{figure}[!h]
\centering
\includegraphics[scale=.35,trim={0.cm 0 0 0},clip]{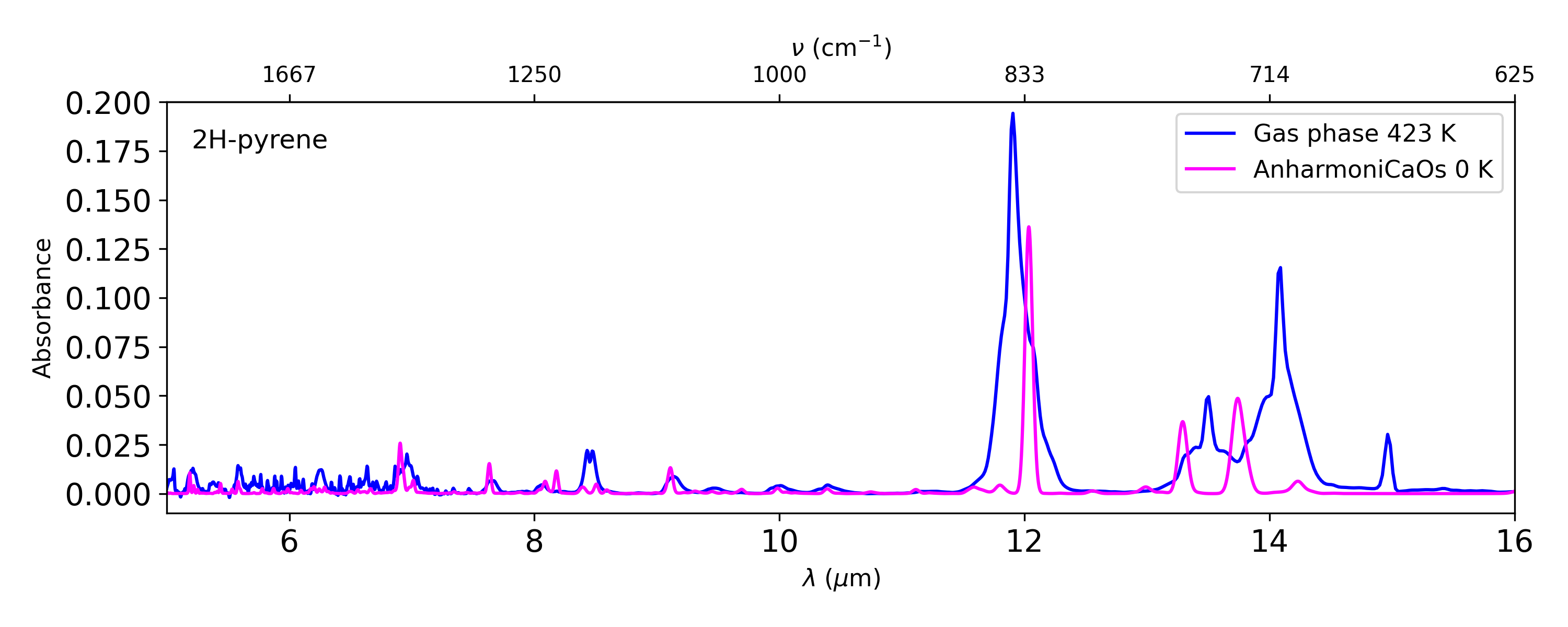}
\includegraphics[scale=.35,trim={0.cm 0 0 0},clip]{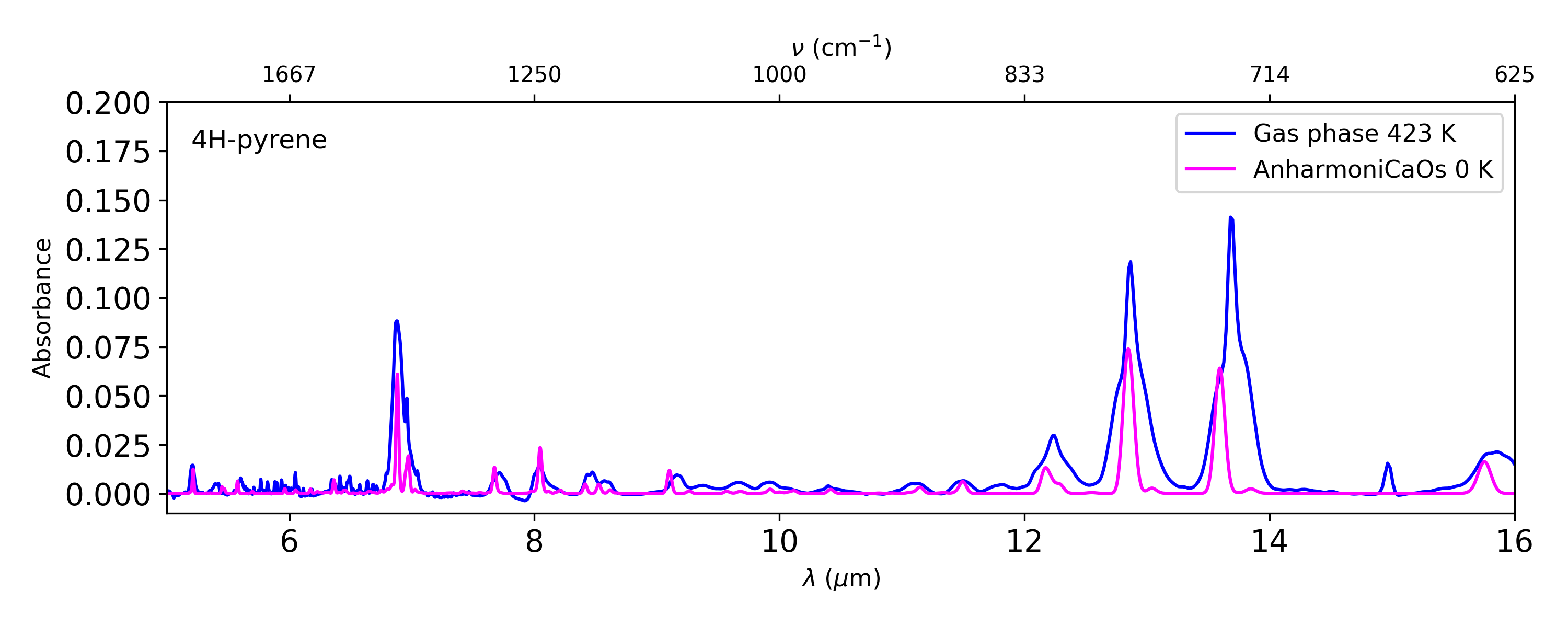}
\includegraphics[scale=.35,trim={0.cm 0 0 0},clip]{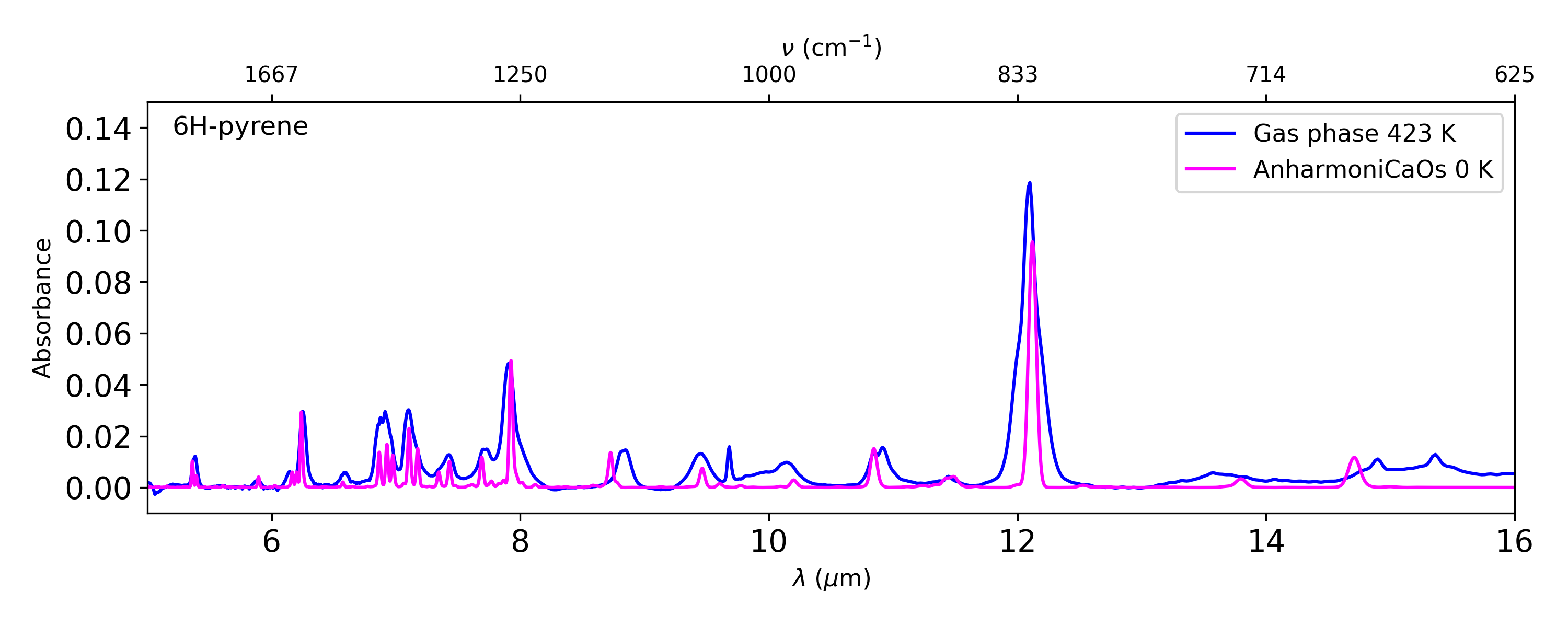}
\caption{Comparison of gas phase spectra (blue line) with theoretical anharmonic spectra (magenta line) for 2H-pyrene, 4H-pyrene and 6H-pyrene. The theoretical spectra are unscaled and have been convolved with a Gaussian profiles with a FWHM of 5 cm$^{-1}$.}
\label{fig:5-15microns_exp_vs_theo}%
\end{figure}

\FloatBarrier
\section{Spectral evolution with the temperature} 

\begin{figure}[!h]
\centering
\includegraphics[scale=.47,trim={0.4cm 0.2cm 0.4cm 0.4cm},clip]{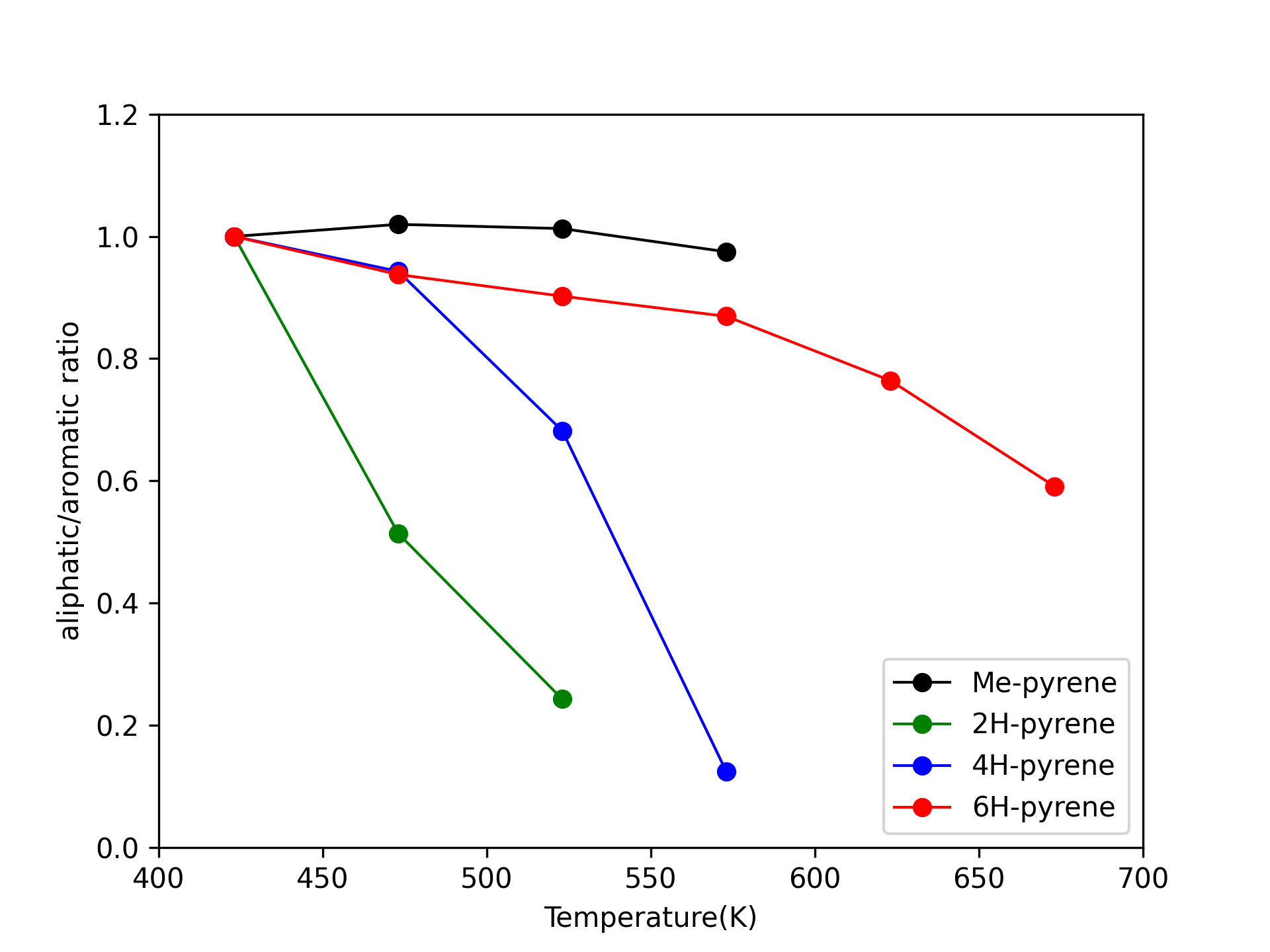}
\caption{Evolution of the aliphatic/aromatic band ratio with the temperature of the methylated and hydrogenated pyrene. The ratios were calculated from the band areas and have been normalised to 1 at 423~K. }
\label{fig:ali_aro_ratio}%
\end{figure}

\begin{figure}[!t]
\centering
\includegraphics[scale=.5]{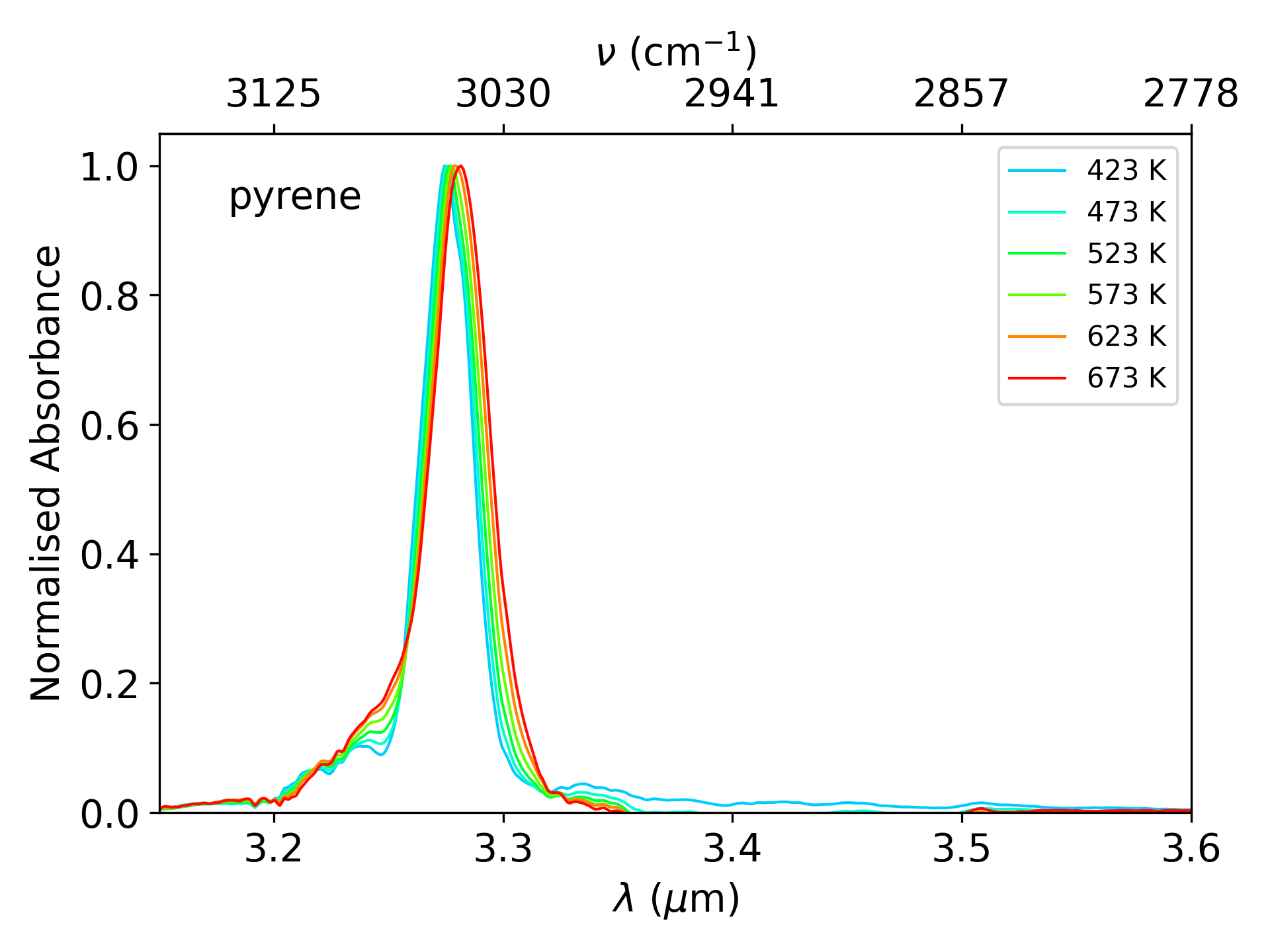}
\includegraphics[scale=.5]{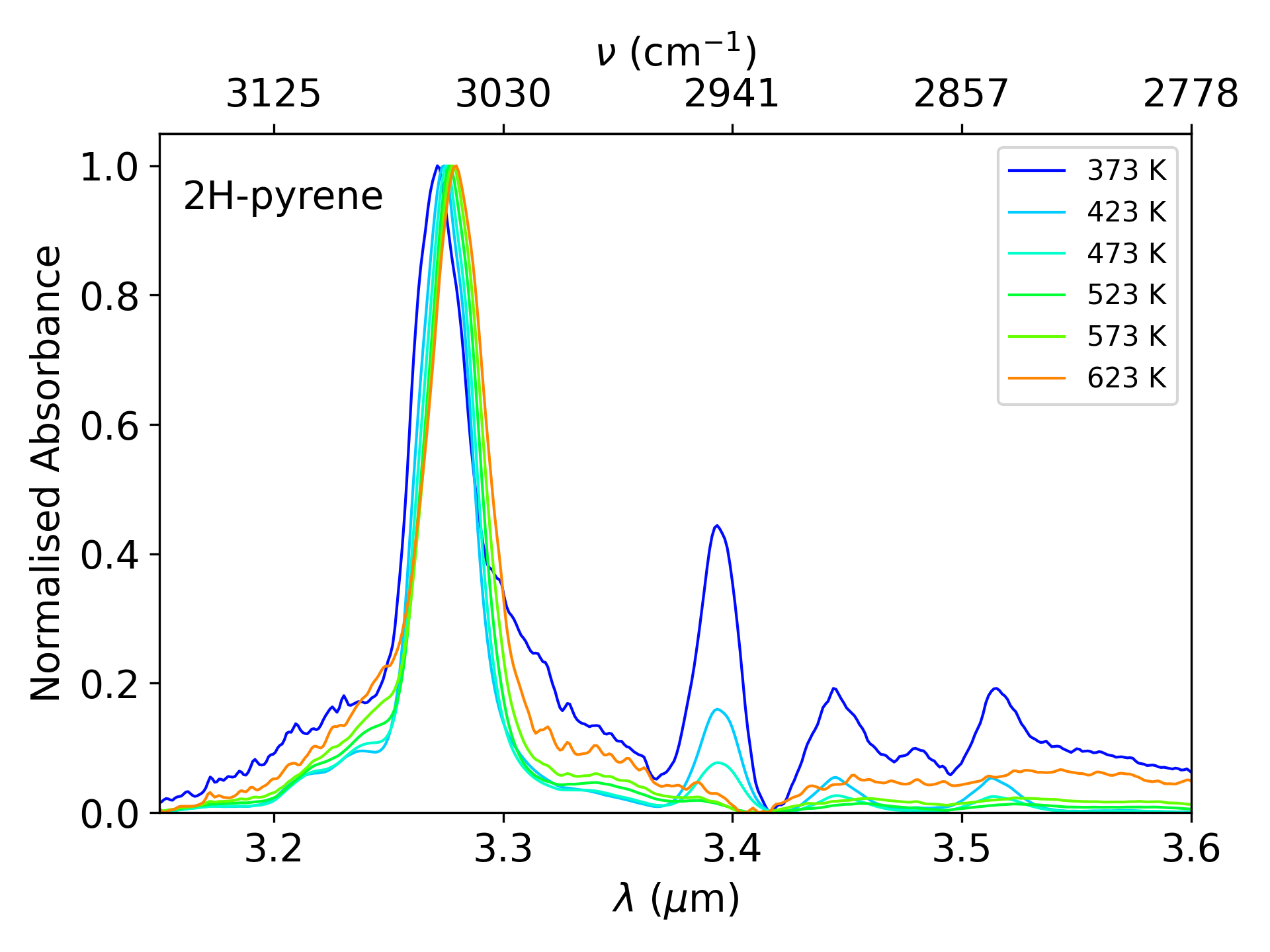}
\includegraphics[scale=.5]{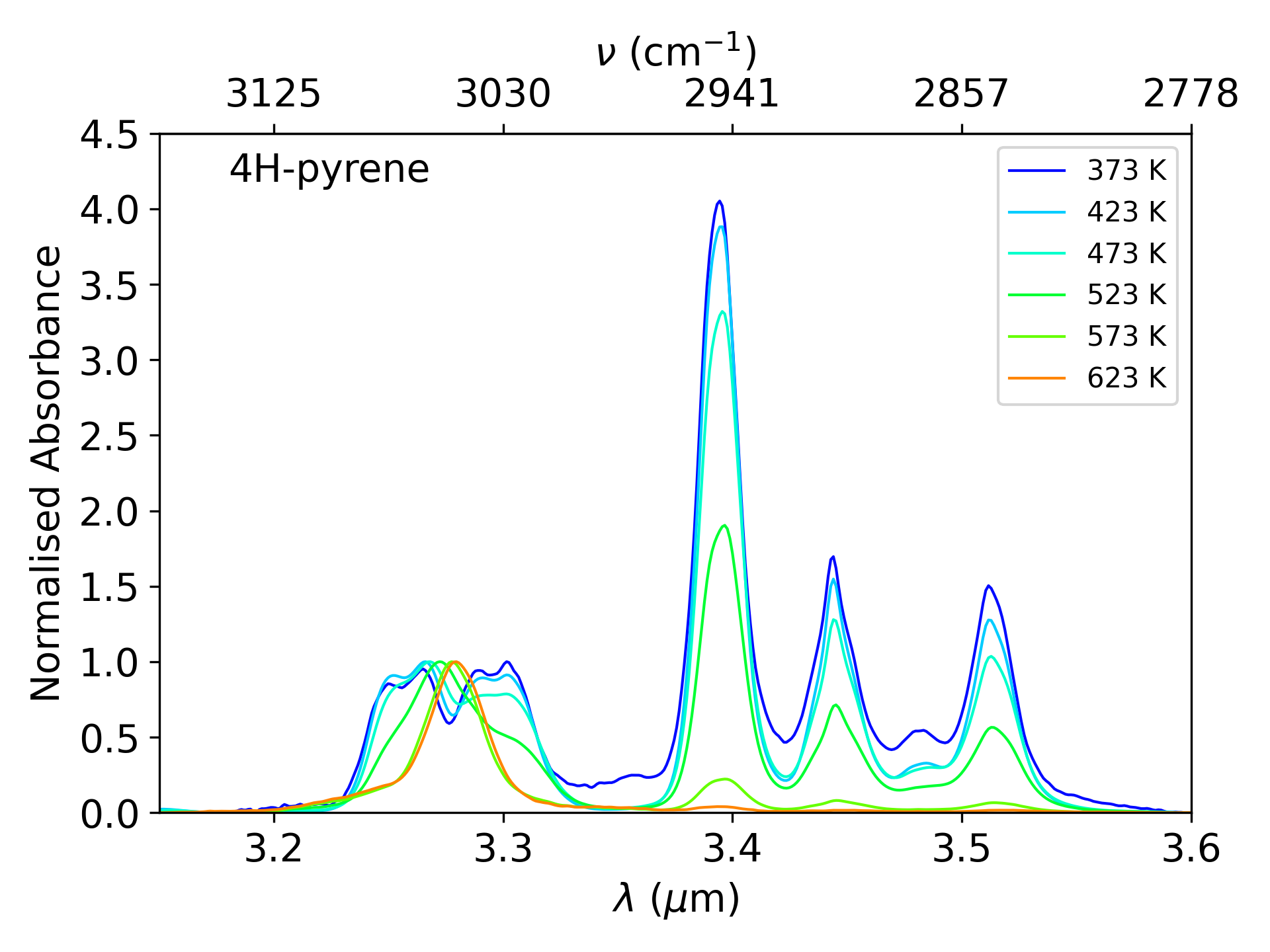}
\caption{Infrared spectrum of pyrene and 2H- and 4H-pyrene in gas phase at various temperatures in the CH stretching mode spectral region. The spectra are normalised to the maximum of the aromatic CH stretching band.  }
\label{fig:3microns_vs_T_norm_2H_4H}%
\end{figure}

\begin{figure}[!h]
\centering

\includegraphics[scale=.5, angle=0]{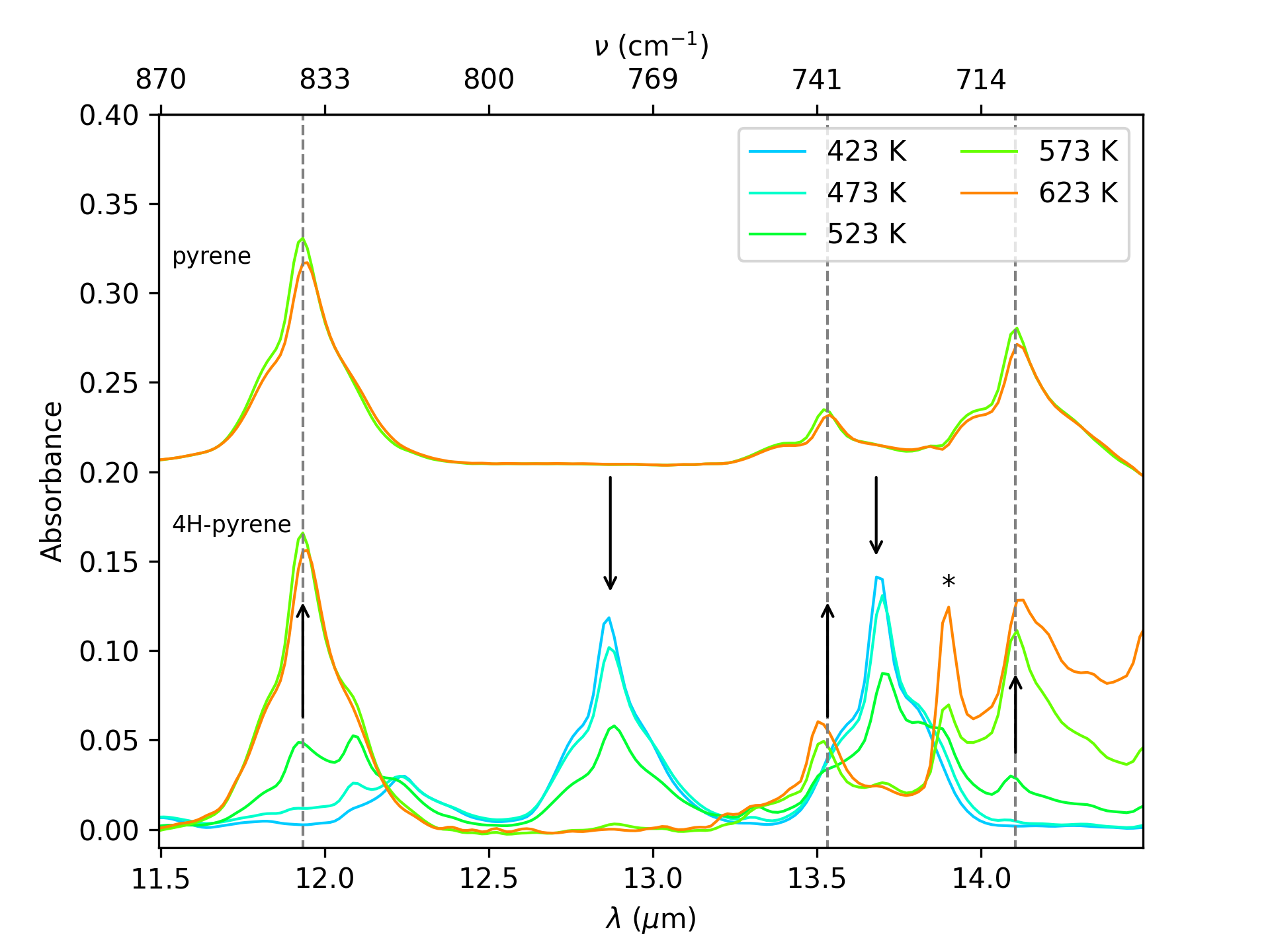}
\includegraphics[scale=.5, angle=0]{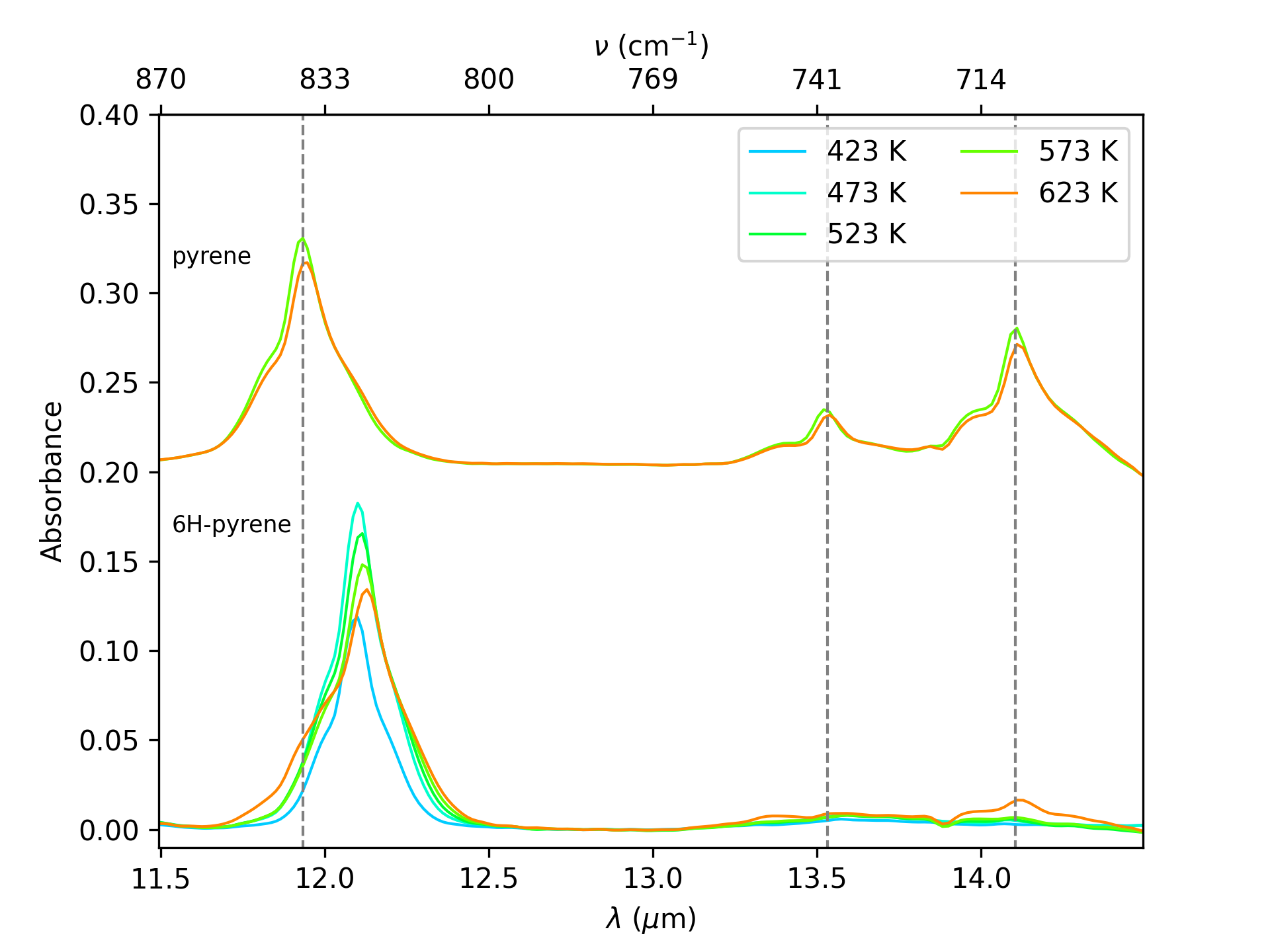}
\caption{Temperature evolution of the low frequency bands of 4H-pyrene (upper panel) and 6H-pyrene(lower panel). The grey dashed lines indicate the peak position of the pyrene bands at 709, 739 and 838 cm$^{-1}$, ie. 14.10, 13.53 and 11.93 ~$\mu$m respectively. In the upper panel, the black arrows indicate the bands that disappear (downward arrows) or appear (upward arrows) from the 4H-pyrene spectrum as the temperature increases. The band marked with a star is a band of CO$_2$.}

          \label{fig:4hpyr-to-pyr}%

\end{figure}

\FloatBarrier
\section{Deriving empirical anharmonic functions for 6H-pyrene and Me-pyrene.}
\label{sect:anha_charac}
\FloatBarrier

\begin{figure}[!th]
   \centering
  \includegraphics[scale=0.15]{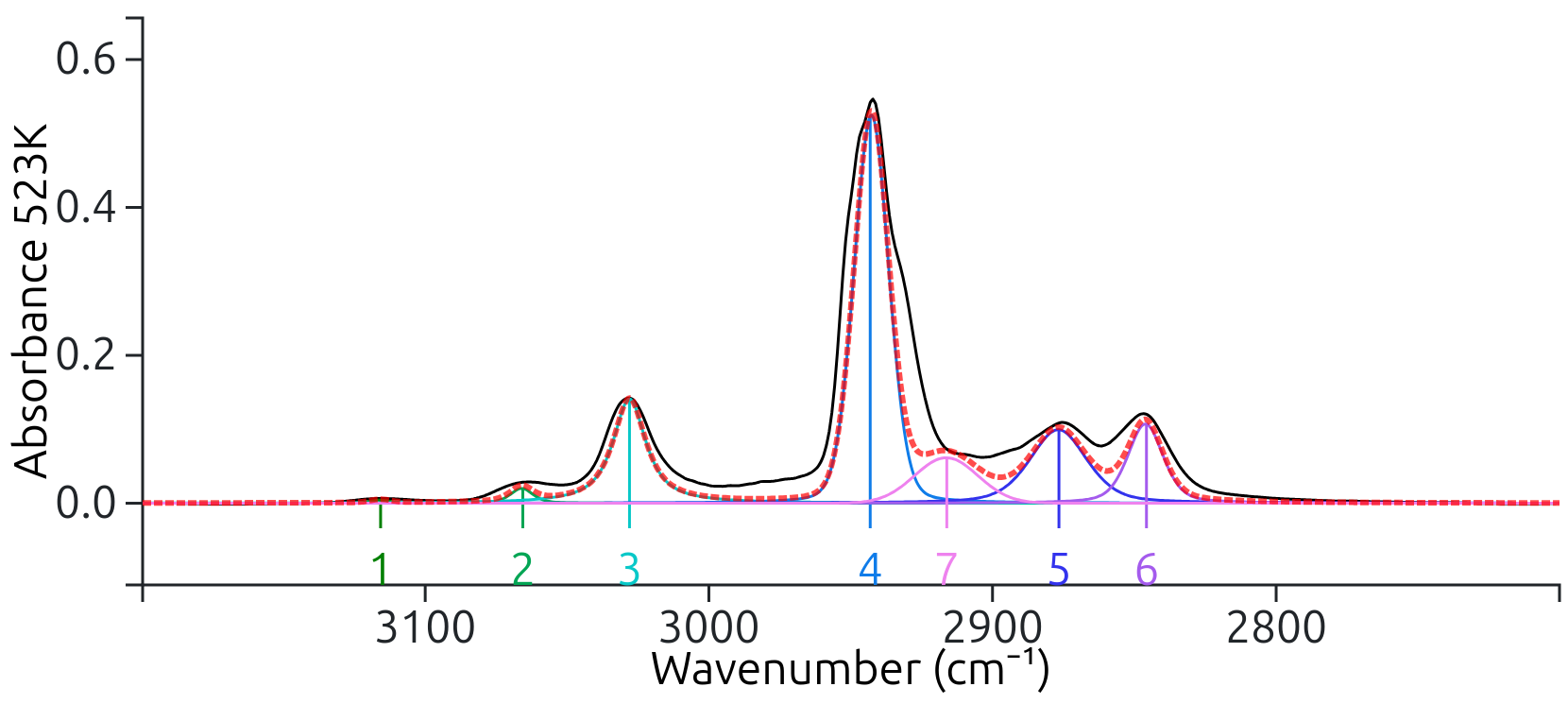}
  \caption{Spectral decomposition of the 3.3 $\mu$m  band of 6H-pyrene at 523 K. The modelled bands are narrower than the experimental ones because the rotational broadening has been subtracted.}
              \label{fig:decomp6H}%
\end{figure}

\begin{figure}[!th]
   \centering
  \includegraphics[scale=0.3]{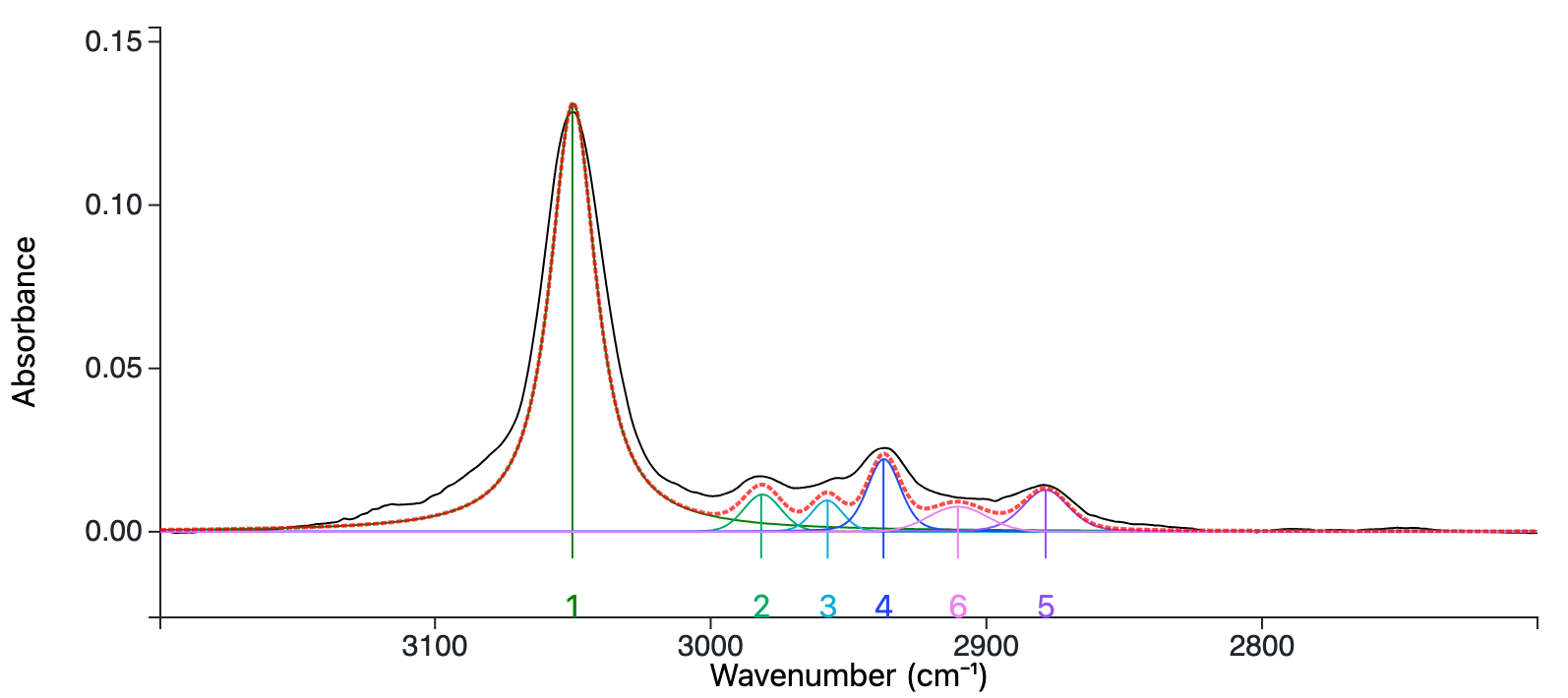}
  \caption{Spectral decomposition of the 3.3 $\mu$m band of Me-pyrene at 573 K. The modelled bands are narrower than the experimental ones because the rotational broadening has been subtracted.}
              \label{fig:decomp-mpyr}%
\end{figure}

\begin{figure*}[!th]
\begin{center}
  \includegraphics[scale=0.37, trim={0.cm 0.cm 0.cm 0cm}, clip]{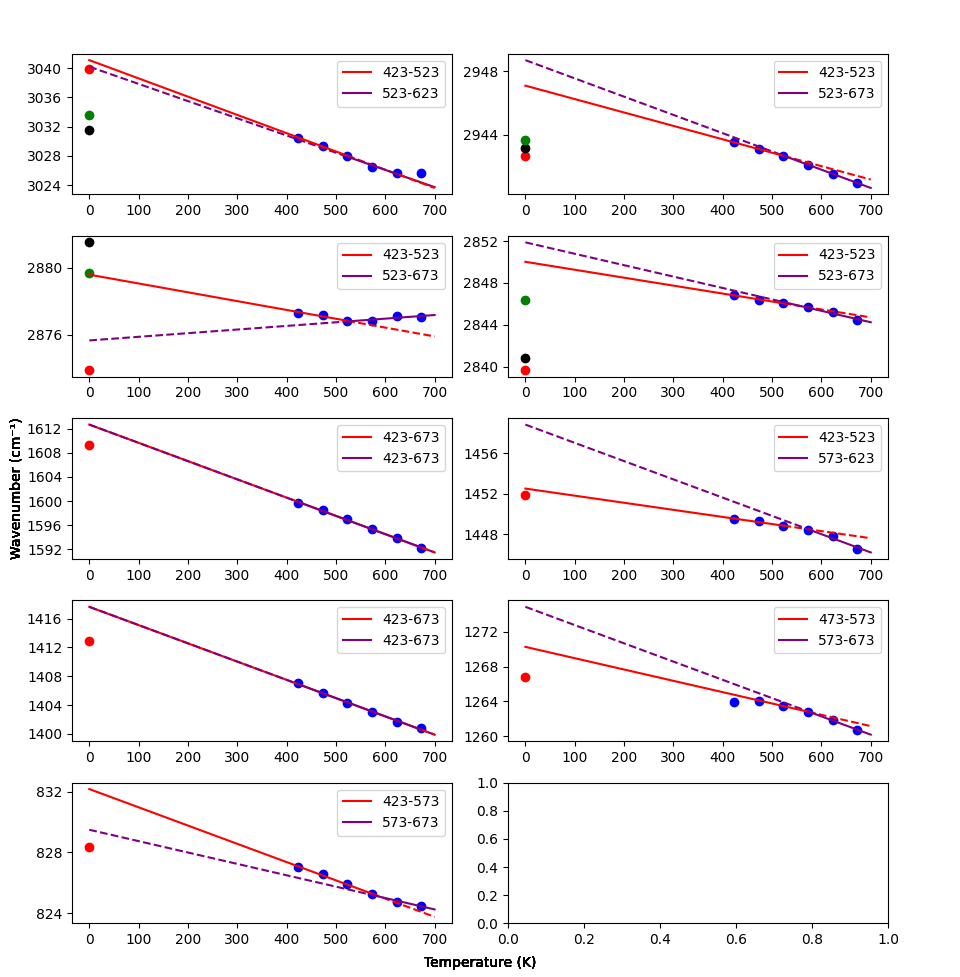}
  \includegraphics[scale=0.37, trim={0.cm 0.cm 0.cm 0cm}, clip]{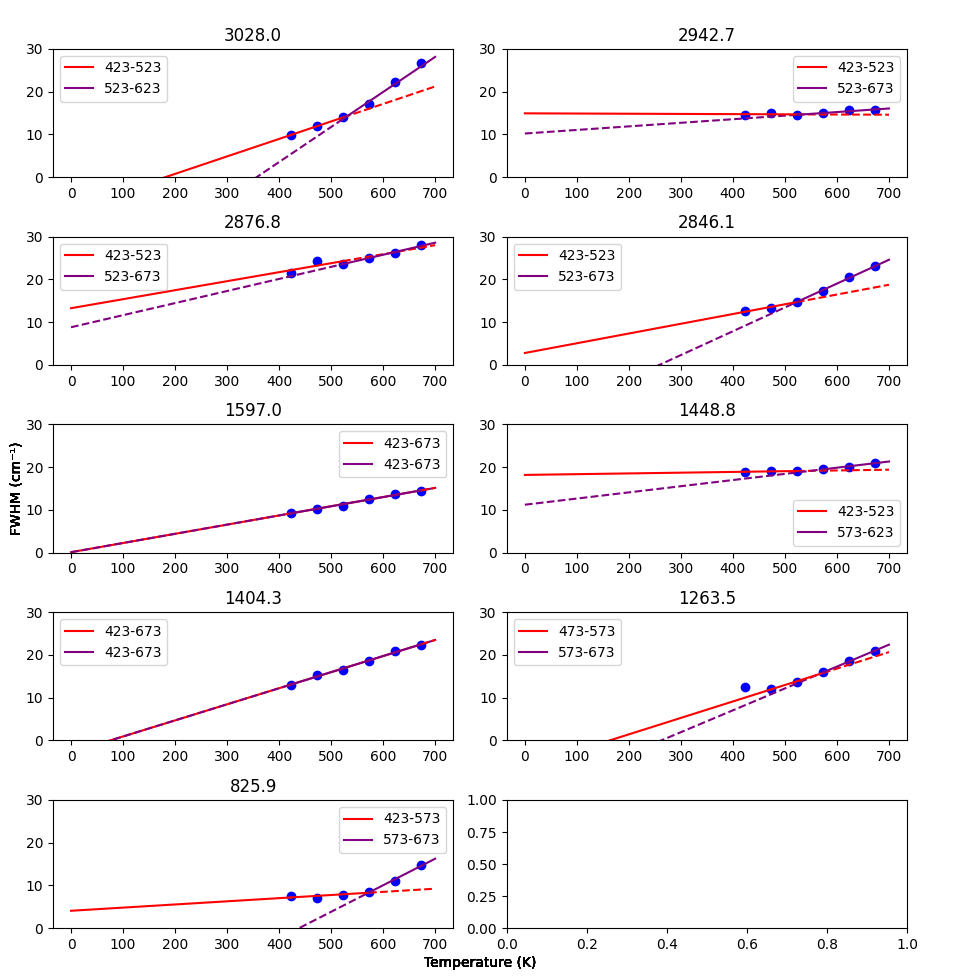}
  \caption{Band position ({\it left panel}) and width ({\it right panel}) as a function of temperature for several bands of 6H-pyrene. The data are fitted at low (red line) and high (purple line) temperatures with first order polynomials. The band positions derived from low-temperature measurements using IR-REMPI jet spectroscopy \citep[][green dots]{maltseva2018} or rare-gas matrix isolation \citep[][black dots]{sandford2013}, as well as from anharmonic calculations (corrected by a factor of 1/0.996, red dots; this work) are also shown.}
              \label{fig:pos_anha_h6pyr}%
\end{center}
\end{figure*}

Figures~\ref{fig:decomp6H} and ~\ref{fig:decomp-mpyr} show examples of spectral decomposition of the 3.3 $\mu$m band of 6H-pyrene and Me-pyrene, respectively. The temperature evolution of peak position and width of 6H-pyrene are shown in Fig.~\ref{fig:pos_anha_h6pyr} with the fits to the data used to determine the anharmonicity parameters. The data are best fit when the temperature range is split in two parts which allows to reproduced the different evolution at low and high temperature. At high temperature, the fitting function is used for the extrapolation outside the measured temperature range. At low temperature the fit takes into account the experimental data from \cite{maltseva2018} only for the bands at 3032 and 2944~ cm$^{-1}$. For other bands, the value of the band position and width at 0~K from \cite{maltseva2018} and \cite{sandford2013} and from the anharmonic calculations are shown for comparison. For these data, the bands are made up of several sub-components and the band positions are calculated by taking the intensity weighted average of the sub-components.

\onecolumn

 \FloatBarrier

 \section{Comparison of the emission model with observations} 

\begin{figure*}[h]
\centering
  \includegraphics[scale=0.66, trim={0.3cm 0.3cm 0.2cm 0.2cm}, clip]{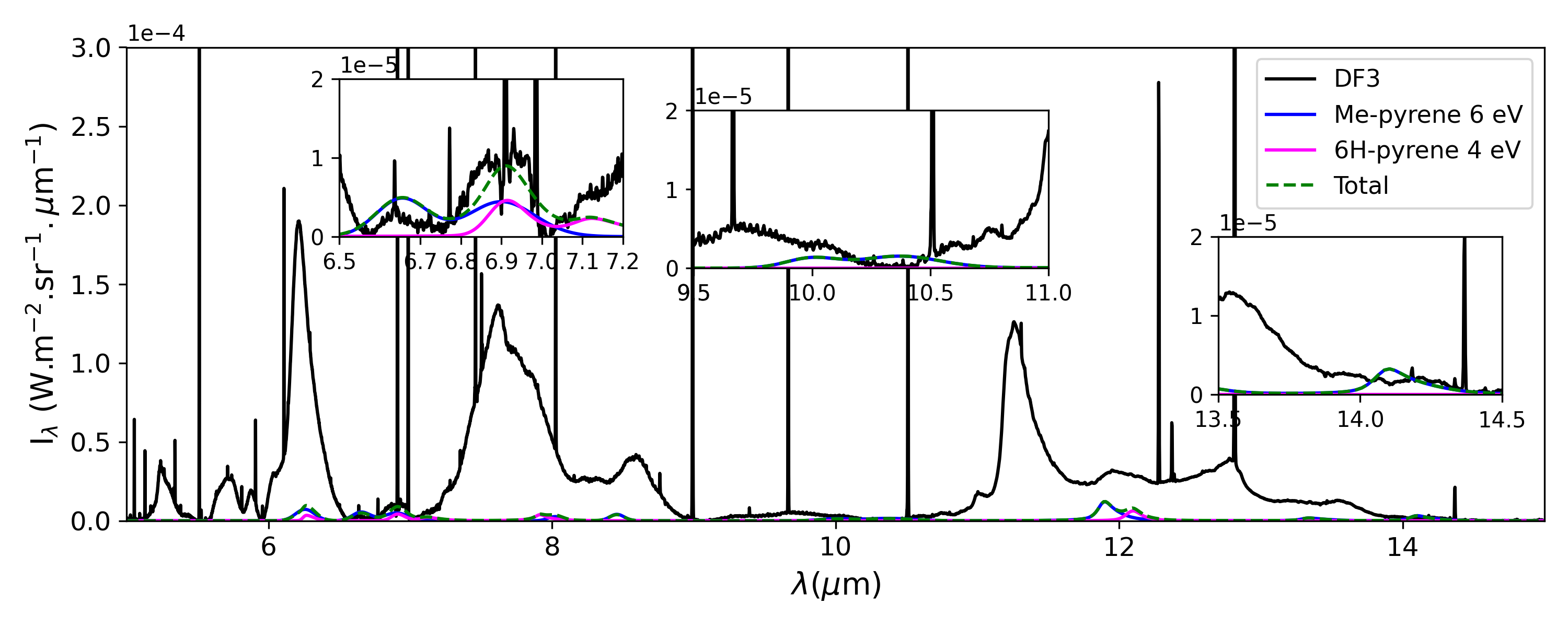}
   \caption{Comparison of the emission spectrum of 6H-pyrene and Me-pyrene with the JWST observations of the Orion PDR. The JWST spectrum is the template spectra of the third dissociation fronts from \cite{vandeputte2025} The emission spectra are scaled by 3.6$\times10^{-6}$ and 2.1$\times10^{-6}$ for Me-pyrene and 6H-pyrene, respectively. 
   }
    \label{fig:astro:comp6hpyr_all}%
   \end{figure*}

\begin{table*}
\caption {Integrated band intensities of 6H-pyrene and Me-pyrene from the experiments at 523~K and from anharmonic and harmonic calculations used as input data for the emission model. } 
\label{Table:band_intensity}
\begin{center}
\begin{tabular}{c c c| c c | c c }
\hline 
\hline 
\multicolumn{7}{c}{6H-pyrene}  \T\B\\
\hline 
\hline 
\multicolumn{3}{c|}{Experiments (523K)}  & \multicolumn{2}{c}{Anharmonic 0K} & \multicolumn{2}{c}{Harmonic} \T \B \\ 
\cline{1-7}
E$_\mathrm{peak}$ & Intensity\tablefootmark{(1)}  & E$_\mathrm{mode}$ \tablefootmark{(2)}& E$_\mathrm{peak}$ & Intensity & E$_\mathrm{peak}$ & Intensity \T \\ 
 (cm$^{-1}$) & (km.mol$^{-1}$) & (cm$^{-1}$) & (cm$^{-1}$) & (km.mol$^{-1}$)& (cm$^{-1}$) & (km.mol$^{-1}$) \B\\
\cline{1-7}
 3028.0 & 76.9 (4.43)  & 3033.5 & 3031.9+3028.6        & 64.3  & 3033.3+3050.4        & 122.7  \T\\
 2942.7 & 220.0 (12.67)  & 2946.9 &  2933.8+2927.4        & 257.3 & 2946.9+2949.9+2959   & 373.5 \\
 2916.3 & 37.5 (2.16)  & 2915.1 & 2908.8+2898.4+2895.9 & 22.47 &  2915.1              & 73.58 \\
 2876.8 & 66.5 (3.83)  & 2884.7 & 2861.6+2872.5        & 46.25 &  2884.7+2883.2       & 141.19 \\
 2846.1 & 50.7 (2.92)  & "2846.1" & 2828.3               & 68.42 &  2883.4              & 0.00      \\
 1597.0 & 18.2 (1.05)  & 1609.5  & 1603.2               & 17.34 & 1609.5               & 51.27  \\
 1448.8 & 24.5 (1.41)  & 1449.6 & 1456.4+1444.6+1432.6 & 25.11 & 1449.6+1463.6+1472.6 & 64.46  \\
 1404.3 & 24.1 (1.39)  & 1405.7 & 1395.4+1408.1        & 22.14 & 1405.7               & 64.32  \\
 1263.5 & 35.6 (2.05)  & 1261.6 & 1261.4               & 27.61 & 1306.9+1273.2        & 121.49  \\
 825.9  & 48.4 (2.80)  & 819.5 & 826.7                & 51.44 & 849.8+832.7+819.5    & 225.2  \B\\    
\hline
\hline
\multicolumn{7}{c}{Me-pyrene} \T\B\\
\hline
\hline
\multicolumn{3}{c|}{Experiments (523K)}  & \multicolumn{2}{c|}{Harmonic} & & \T \B\\ 
\cline{1-5}
\cline{1-5}
 3049.9 & 185.2 (5.47) & 3054.4 & 3042.1 to 3080.9          & 156.0  & & \T\\
 2981.1 &  9.8 (0.29) & 2994.7 & 2994.7                     & 18.6    & & \\
 2956.9 &  8.1 (0.24) & 2957.7 &  2957.7                    & 19.4       & & \\
 2936.8 &  18.6 (0.55) &  "2936.8" &  -                    & -     & & \\
 2908.4 &  9.5 (0.28) & 2912.0 &  2912.0                     & 40.0      & & \\
 2877.8 &  12.5 (0.37) & "2877.8" &  -                          & -        & & \\
 1597.5\tablefootmark{*} &  23.7 (0.70) & 1604.5 & 1595.3 + 1604.5 + 1619.5       & 19.3   & & \\
 1502.3\tablefootmark{*} &  17.3 (0.51) &  1516.1 & 1478.5 + 1493.5 + 1516.1               & 22.5    & & \\
 1448.6\tablefootmark{*} &  20.3 (0.60) & 1458.8 & 1437.8 + 1458.8 + 1463.6 & 19.4     & & \\
 1246.2\tablefootmark{*} &  6.8 (0.20) & 1247.6 & 1241.5 + 1247.6 + 1251.5        & 6.6    & & \\
 1182.3\tablefootmark{*} &  10.8 (0.32) & 1183.0 & 1183.0 + 1195.2       & 12.2   & & \\
 999.7\tablefootmark{*} &  7.5 (0.22) & 1000.6 &        1000.6             &  2.5  & & \\
 963.3\tablefootmark{*} & 14.9 (0.44) & 968.8 & 957.5 + 961.1 + 968.8 + 969.4    & 2.3    & & \\
 840.8 &  67.1 (1.98) & 844.6 & 844.6                 & 92.8    & & \\ 
 750.1 &   11.9 (0.35) & 752.0 & 752.0                 & 11.7     & & \\
 709.3 &   22.0 (0.65) & 714.0 & 714.0                       & 22.7    & & \B\\    
\hline
\hline 
\end{tabular} 
\end{center}
\tablefoot{\tablefoottext{1}{Integrated band intensities (values in parentheses, in cm$^{-1}$) were extracted from the laboratory spectra using the multi-component fit tool. The resulting intensities were normalised to the sum of the band intensities calculated at the harmonic level or, when available, at the anharmonic level.}
\tablefoottext{2} {The emission model requires an input file listing the vibrational mode frequencies and their IR intensities. To incorporate the experimental IR intensities, each value was assigned to the closest harmonic frequency. Values in quotation marks indicate cases where no sufficiently close harmonic frequency exists, requiring an adjustment to the list of mode frequencies.} \tablefoottext{*}{Bands for which no anharmonicity functions were derived and thus treated with a constant bandwidth in the model. The adopted bandwidths (in cm$^{-1}$) were: 36.18, 25.63, 16.80, 19.85, 40.25, 33.90, and 40.40 for the bands at 963.2, 999.7, 1182.3, 1246.2, 1448.6, 1502.3, and 1597.5~cm$^{-1}$, respectively.} }
\end{table*}

\end{appendix}
\end{document}